\documentclass[a4paper,fleqn]{article}

\usepackage[utf8]{inputenc}
\usepackage[T1]{fontenc}
\usepackage{amsmath,amssymb,latexsym,graphicx,hyperref,fullpage,anyfontsize,authblk}
\usepackage{newpx} 
\usepackage[dvipsnames]{xcolor}
\usepackage[
  style=phys,
  eprint=true,
  maxnames=100
]{biblatex}

\hypersetup{
  colorlinks=true,
  urlcolor=MidnightBlue,
  citecolor=MidnightBlue,
  filecolor=MidnightBlue,
  linkcolor=MidnightBlue
}

\title{
  Stationary point complexity via minimal supersymmetry breaking
}

\author{Jaron Kent-Dobias}

\affil{
  \small
  ICTP South American Institute for Fundamental Research, São Paulo, Brazil
  \emph{and}
  \\
  Instituto de Física Teórica, Universidade Estadual Paulista ``Júlio de
  Mesquita Filho'', São Paulo, Brazil
}

\begin{document}

\maketitle

\begin{abstract}
  The statistics of stationary points are a powerful way to understand
  mean-field random landscapes, and the Kac--Rice formula is a general way to
  compute them. A longstanding technical barrier to these calculations is the
  presence of the absolute value of the determinant of the Hessian. Neglecting
  the absolute value produces an elegant 2-index supersymmetric representation
  of the problem, but is often incorrect. We develop an expanded 4-index
  supersymmetric representation of the complexity problem which incorporates
  the absolute value naturally via spontaneous supersymmetry breaking along a
  particular superspace direction. Positing that no additional symmetry
  breaking occurs implies the reduction to five order parameters corresponding
  to elements of a superspace operator algebra generated by the spontaneously
  \textsc{susy}-breaking operator. We relate the order parameters to the
  geometry and spectra of stationary points, showing that the
  \textsc{susy}-breaking order parameter corresponds to the spectral density of
  the Hessian at zero eigenvalue. We give examples of this formalism applied to
  calculate the annealed complexity of several models, including the perceptron
  and the Sherrington--Kirkpatrick model. The framework is naturally extended
  to quenched complexity, where each order parameter corresponds to a replica
  matrix.
\end{abstract}

\tableofcontents

\section{Introduction}

Counting metastable minima and saddle points is a canonical way of
understanding the complex landscapes generated by the graphs of many-parameter
functions $H:\Omega\to\mathbb R$. The Kac--Rice formula
\begin{equation}
  \mathcal N[H]
  =\sum_{\substack{\boldsymbol x^*\in\Omega\\\boldsymbol\nabla H(\boldsymbol x^*)=0}}1
  =\int_\Omega d\boldsymbol x\,
  \delta\big(\boldsymbol\nabla H(\boldsymbol x)\big)
  \,|\det\operatorname{Hess}H(\boldsymbol x)|
\end{equation}
provides a convenient starting point for making this count, connecting it to an
integral over configuration space \cite{Rice_1939_The, Kac_1943_On}. In
practice one rarely performs this count for a specific function $H$, but
averages the count over an ensemble of random functions. The calculation is
facilitated by the presence of exponentially many stationary points in the
dimension $N$ of the configuration space $\Omega$, so that useful information
can be gleaned from the \emph{annealed complexity}
\begin{equation}
  \Sigma=\lim_{N\to\infty}\frac1N\log\overline{\mathcal N[H]}
\end{equation}
where the bar denotes the average over functions $H$. Calculation of the
complexity $\Sigma$ has been sought by researchers in diverse contexts for
insight into equilibrium and out-of-equilibrium dynamics in disordered systems
as diverse as spin-glasses, neural networks, and ecosystems
\cite{Bray_1980_Metastable, Ros_2023_Quenched, Maillard_2020_Landscape}.

Making this average is complicated by the absolute value function
wrapping the determinant of the Hessian, a nonanalyticity that stymies many
prospective approaches. Removing the absolute value is a disaster, because the
resulting integral produces a topological characteristic of the configuration
space $\Omega$ independent of $H$,\footnote{
  Strictly speaking, this is only true for compact, boundaryless $\Omega$. For
  noncompact configuration space or one with a boundary, the result is depends
  on the value of $H$ along the boundaries of the space, including those at
  infinity.
}
\begin{equation}
  \chi
  =\sum_{\substack{\boldsymbol x^*\in\Omega\\\boldsymbol\nabla H(\boldsymbol x^*)=0}}(-1)^{\operatorname{ind}(\boldsymbol x^*)}
  =\int_\Omega d\boldsymbol x\,
  \delta\big(\boldsymbol\nabla H(\boldsymbol x)\big)
  \,\det\operatorname{Hess}H(\boldsymbol x)
\end{equation}
useless if the goal is to study properties of $H$. If removing the absolute
value \emph{did} work, the expression could be written as a convenient field
theory. Representing the $\delta$ function using its inverse Fourier transform
and the determinant using a pair of Grassmann integrals, we could write
\begin{equation}
  \chi
  =\int\frac{d\boldsymbol x\,d\hat{\boldsymbol x}}{(2\pi)^N}
  \,d\bar{\boldsymbol\eta}\,d\boldsymbol\eta\,
  e^{i\hat{\boldsymbol x}\cdot\boldsymbol\nabla H(\boldsymbol x)
    -\bar{\boldsymbol\eta}^T\operatorname{Hess}H(\boldsymbol x)\boldsymbol\eta
  }
\end{equation}
which is an exponential integral susceptible to standard field-theoretic methods.
This action is especially appealing because it can be naturally represented in
2-index superspace, which eases the burden of calculations and suggests Ward
identities based on the symmetries of this space, detailed in Section \ref{sec:2-index}.

How can we proceed? There are several options. If $H$ is a Gaussian random
function, the determinant and gradient are statistically independent when
evaluated at the same point in configuration space and the average of the
integrand over $H$ can be factorized into separate averages of its two factors
\cite{Bray_2007_Statistics}. This is a powerful tool for studying Gaussian
random landscapes using annealed averages, but may not work to correctly
describe quenched averages, where replicas of configuration space necessitate
averages over the gradient and determinant evaluated at different points.

Conditions can be added to the count so that rather than counting all
stationary points, only a subset are counted. If the conditioned subset has a
high probability of being overwhelmingly populated by stationary points with
the same index (usually all minima), then the average count produced by
removing the absolute value function from the determinant is a good
approximation of the true one \cite{Castellani_2005_Spin-glass}. It is most
common to condition on the energy density of the stationary point and then
focus on regions where the most common population can be self-consistently
shown to be minima, but other properties can also be conditioned
\cite{Kent-Dobias_2024_Conditioning}. Unfortunately, beyond minima, maxima, or
saddle points with a small index (much less than the dimension $N$), this
technique cannot be used to reliably count saddle points.

The final option is to use or produce results on the statistics of the Hessian
conditioned on the gradient being zero \cite{Fyodorov_2004_Complexity}. This
approach will always work and has been applied to diverse models, but each
application can require dramatically different preliminary results.
Moreover it is difficult to generalize to the quenched case, where conditioning
must be made simultaneously on $n$ different stationary points. The difficulty
of this is glimpsed in existing calculations involving the simultaneous
conditioning on three points \cite{Pacco_2025_Triplets}.

Here, we introduce a general and concise framework for treating the absolute
value. Our approach is inspired by that of Fyodorov in Ref~\cite{Fyodorov_2004_Complexity}, who noted that the
absolute value can be preserved in a field-theoretic calculation by introducing
four additional fields, two real-valued and two Grassmann-valued, and using the identity
\begin{align} \label{eq:det.formula}
  |\det A|
  &=\lim_{\epsilon\to0}\frac{(\det A)^2}{\sqrt{\det(A+i\epsilon I)}\sqrt{\det(A-i\epsilon I)}}
  \\ \notag
  &=\lim_{\epsilon\to0}
  \int \frac{d\boldsymbol a\,d\boldsymbol b}{(2\pi)^N}\,d\bar{\boldsymbol\eta}\,d\boldsymbol\eta\,d\bar{\boldsymbol\gamma}\,d\boldsymbol\gamma\,
  e^{
    -\bar{\boldsymbol\eta}^TA\boldsymbol\eta
    -\bar{\boldsymbol\gamma}^TA\boldsymbol\gamma
    -\frac12\boldsymbol a^T(A+i\epsilon I)\boldsymbol a
    -\frac12\boldsymbol b^T(A-i\epsilon I)\boldsymbol b
  }
\end{align}
Along with $\boldsymbol x$ and $\hat{\boldsymbol x}$, this results in an
integral representation of the number of stationary points $\mathcal N$ depending on 4 real-valued fields and 4
Grassmann-valued ones. Even considering that Grassmann-valued order parameters
can be neglected, this results in $\frac124\times(4+1)+\frac124(4-1)=16$ order
parameters due to scalar products among these fields alone. Understanding which of these order parameters are equal to each other, which are zero, and which make nontrivial contributions can be a formidable task.

We show that an identity like \eqref{eq:det.formula} can be
naturally represented in a 4-index superspace. Besides easing the burden of
calculations, this representation and its symmetries suggest a systematic way
of understanding the effect of the perturbation made by $\epsilon$ and its
limit to zero. The perturbation creates the possibility of spontaneous breaking
of superspace symmetry, but only partially: its effect amounts to
condensation in only five independent order parameters out of a potential 256. We call this the \emph{minimally supersymmetry-breaking} subspace.
This quantitative reduction in the potential parameters of the problem provides
qualitative benefits to making and understanding calculations using the
technique. With only an algebra of five superoperators and a small dictionary
of identities, a wide variety of mean-field complexity problems can be directly
solved.

This paper begins by introducing the superspace representation of complexity,
first in Section \ref{sec:2-index} with its flawed 2-index form. Section
\ref{sec:4-index} introduces the 4-index form, shows how it correctly
reproduces the absolute value of the determinant, and discusses how the
minimally \textsc{susy}-breaking subspace of order parameters arises from
spontaneous symmetry breaking along a specific superspace axis. The later parts
of that section detail how the formalism works in various specific settings,
from non-Euclidean spaces to conditioning on energy and index. Several examples
of the formalism are given in Section \ref{sec:examples}, where we apply it to
random Gaussian fields in Euclidean space, the spherical spin glasses, canyon
landscapes, the perceptron, and the Sherrington--Kirkpatrick model. Finally,
Section \ref{sec:conclusion} summarises our results and forecasts the next
steps.

\section{Supersymmetry and its breaking in stationary point counts}
\label{sec:susy}

A pedagogical introduction to superspace can be found in Appendix
\ref{sec:superspace}.

\subsection{2-index superspace and the Euler characteristic}
\label{sec:2-index}

We want to understand properties of the stationary points of
functions $H$ with argument $\boldsymbol x\in\mathbb R^N$ (non-Euclidean spaces are treated in Section \ref{sec:lagrange}). One route is to use a two-index superspace representation. If $\bar\theta$ and $\theta$
are Grassmann superspace indices, $\boldsymbol x$ and $\hat{\boldsymbol x}$ are
$N$-dimensional real vectors, and $\bar{\boldsymbol\eta}$ and $\boldsymbol\eta$
are $N$-dimensional Grassmann vectors, we can define the supervector
$\boldsymbol\varphi\in\mathbb R^{N|2}$ by
\begin{equation}
  \boldsymbol\varphi(1)
  =\boldsymbol x+\bar\theta_1\boldsymbol\eta+\bar{\boldsymbol\eta}\theta_1+i\hat{\boldsymbol x}\bar\theta_1\theta_1
\end{equation}
Because all Grassmann elements come in even products, supervectors can be
treated in many expressions as if they were $N$-dimensional real vectors.
Integrating the action
\begin{equation}
  S_\textsc{susy}(\boldsymbol\varphi)
  =\int d\theta_1\,d\bar\theta_1\,H(\boldsymbol\varphi(1))
  =\int d1\,H(\boldsymbol\varphi(1))
\end{equation}
over
supervectors provides a compact way to formally represent the alternating sum
over stationary points of the function $H$ of increasing index. To see this,
first understand that because products of the same Grassmann variable vanish,
any series expansion as a function of the superspace indices $\bar\theta$ and
$\theta$ exactly terminates at second order. Therefore,
\begin{equation}
  H(\boldsymbol\varphi(1))
  =H(\boldsymbol x)
  +\bar{\boldsymbol\eta}\cdot\boldsymbol\nabla H(\boldsymbol x)\theta_1
  +\bar\theta_1\boldsymbol\eta\cdot\boldsymbol\nabla H(\boldsymbol x)
  +i\hat{\boldsymbol x}\cdot\boldsymbol\nabla H(\boldsymbol x)\bar\theta_1\theta_1
  -\bar{\boldsymbol\eta}^T\operatorname{Hess}H(\boldsymbol x)\boldsymbol\eta\bar\theta_1\theta_1
\end{equation}
It therefore follows that we can write the Euler characteristic as
\begin{align}
  \chi[H]
  &=\int_{\mathbb R^{N\mid2}} d\boldsymbol\varphi\,
    e^{\int d1\,H(\boldsymbol\varphi(1))}
  =\int \frac{d\boldsymbol x\,d\hat{\boldsymbol x}}{(2\pi)^N}
  d\bar{\boldsymbol\eta}\,d\boldsymbol\eta\,
  e^{
    i\hat{\boldsymbol x}\cdot\boldsymbol\nabla H(\boldsymbol x)
    -\bar{\boldsymbol\eta}^T\operatorname{Hess}H(\boldsymbol x)\boldsymbol\eta
  }
  \\ \notag
  &=\int_{\mathbb R^N} d\boldsymbol x\,\delta\big(\boldsymbol\nabla H(\boldsymbol x)\big)
  \det\operatorname{Hess}H(\boldsymbol x)
  =\sum_{\substack{\boldsymbol x^*\in\Omega\\\boldsymbol\nabla H(\boldsymbol x^*)=0}}(-1)^{\operatorname{ind}(\boldsymbol x^*)}
\end{align}
The Euler characteristic $\chi[H]$ is only weakly dependent on the function $H$. When the integration is
restricted to a compact manifold with the use of Lagrange multipliers (Section~\ref{sec:lagrange}),
$\chi[H]$ is a topological invariant of the configuration space for almost all functions $H$. For
integration over $\mathbb R^N$, $\chi$ depends only on the topology of the
behavior of $H$ at $\|\boldsymbol x\|\to\infty$.

\subsubsection{Symmetries of the action}

\paragraph{Supersymmetry.}

The resulting exponential action has several symmetries. It is supersymmetric,
meaning that it is invariant under translation of the superfield along the
superindicies. If $T_\theta(\varepsilon)\theta_1=\theta_1+\varepsilon$ is the
translation operator along $\theta$, then
\begin{equation}
  S_\textsc{susy}(T_\theta(\varepsilon)\boldsymbol\varphi)
  =\int d1\,H(T_\theta(\varepsilon)\boldsymbol\varphi(1))
  =\int d1\,H(\boldsymbol\varphi(1))
  =S_\textsc{susy}(\boldsymbol\varphi)
\end{equation}
The same is true for translation along $\bar\theta$. These two translations are
generated by $\frac\partial{\partial\theta_1}$ and
$\frac\partial{\partial\bar\theta_1}$.\footnote{
  Strictly speaking the generators of a supersymmetric algebra have a
  nontrivial anticommutator. This role is often played by \emph{time}. For
  instance, the generators $Q=\frac\partial{\partial\theta}$ and $\bar
  Q=\frac\partial{\partial\bar\theta}+\theta\frac\partial{\partial t}$ also
  satisfy $Q^2=\bar Q^2=0$ but have $Q\bar Q+\bar QQ=\frac\partial{\partial
  t}$. These are the generators of symmetries for the path integral
  representation of the generating functional of the Langevin equation, which
  has \emph{bona fide} supersymmetry \cite{Kurchan_1992_Supersymmetry}. Our
  action corresponds to the Langevin action in the limit of zero friction,
  where time translation becomes trivial. Despite the trivial anticommutator,
  we will continue to call these symmetries under translations of superindicies
  in superspace `supersymmetries' because it is a compact and relatively clear
  way to refer to them.
}
If applied to an operator depending on more than one set of superindicies, then
the generator is the sum over partial derivatives of $\theta$ from each set,
e.g., $\frac\partial{\partial\theta_1}+\frac\partial{\partial\theta_2}$.
Supersymmetry is a Grassmann symmetry, meaning that its generators are nilpotent.

\paragraph{Ghost number symmetry.}

The action also has a symmetry called \emph{ghost number symmetry}, a nonlinear symmetry generated by
\begin{equation}
  \theta\frac\partial{\partial\theta}-\bar\theta\frac\partial{\partial\bar\theta}
\end{equation}
When linearized about small transformations, the effect of this symmetry is a bosonic translation of the superindices $\theta\mapsto\theta+\varepsilon\theta$ and $\bar\theta\mapsto\bar\theta-\varepsilon\bar\theta$. Applied to $\boldsymbol\varphi$, the action is invariant up to $O(\varepsilon^2)$ because the integral $d1=d\theta\,d\bar\theta$ ensures that only terms with balanced pairs $\bar\theta\theta$ contribute. Ghost number symmetry is a real symmetry.

\subsubsection{The saddle point approximation}

When evaluating $\chi[H]$ averaged over an ensemble of functions $H$, one typically reduces the problem to an integral over an order parameter superoperator $\mathbb Q(1,2)=\frac1N\boldsymbol\varphi(1)\cdot\boldsymbol\varphi(2)$ of the form
\begin{equation}
  \overline{\chi[H]}=\int_{\mathbb R^{1|2\times 2}} d\mathbb Q\,e^{N\mathcal S(\mathbb Q)}
\end{equation}
for some \emph{effective action} $\mathcal S$. The form of the integral allows it
to be evaluated by a saddle point approximation for large dimension $N$, with
\begin{equation}
  \overline{\chi[H]}
  \simeq\sum_{\substack{\mathbb Q^*\in\mathbb R^{1|2\times 2}\\\nabla\mathcal S(\mathbb Q^*)=0}}
    [\operatorname{sdet}\operatorname{Hess}\mathcal S(\mathbb Q^*)]^{-\frac12}e^{N\mathcal S(\mathbb Q^*)}
\end{equation}
which results in the complexity being given by
\begin{equation}
  \Sigma=\lim_{N\to\infty}\frac1N\log\overline{\chi[H]}
  =\max_{\substack{\mathbb Q^*\in\mathbb R^{1|2\times 2}\\\nabla\mathcal S(\mathbb Q^*)=0}}\mathcal S(\mathbb Q^*)
\end{equation}
The supersymmetry of the action
is also reflected in the effective action.
Because $\mathcal S$ is supersymmetric, its saddle points $\nabla\mathcal
S(\mathbb Q^*)=0$ are at values of $\mathbb Q^*$ that are supersymmetric, or
else spontaneously break supersymmetry but come in a set $\{\mathbb Q^*\}$
which is itself invariant to supersymmetry transformations.  The
most general form of $\mathbb Q^*$ compatible with invariance under translation
of $\theta$ and $\bar\theta$ is
\begin{equation}
  \mathbb Q^*(1,2)=Q_1+Q_\delta\delta(1,2)
\end{equation}
where $\delta$ is the identity superoperator defined by
\begin{equation}
  \delta(1,2)=(\bar\theta_1-\bar\theta_2)(\theta_1-\theta_2)
\end{equation}
and has the property
\begin{equation}
  \mathbb A(1,2)=\int d3\,\delta(1,3)\mathbb A(3,2)
  =\int d3\,\mathbb A(1,3)\delta(3,2)
\end{equation}
for any superoperator $A\in\mathbb R^{1|2\times 2}$.

\subsubsection{Vanishing of supersymmetry-breaking saddle points}
\label{sec:2-index.vanishing}

The contribution of saddle points that break supersymmetry and only
supersymmetry is identically zero \cite{Kurchan_1991_Replica, Parisi_2004_On}.
One can see this at leading order by a simple argument. If the effective action
is supersymmetric, then its invariance to translations of $\theta$ implies
\begin{align}
  0
  &=\lim_{\varepsilon\to0}\frac{\mathcal S(T_\theta(\varepsilon)\mathbb Q)-\mathcal S(\mathbb Q)}{\varepsilon}
  =\int d1\,d2\,\frac{\partial\mathcal S(\mathbb Q)}{\partial\mathbb Q(1,2)}\frac{\partial T_\theta(\varepsilon)\mathbb Q(1,2)}{\partial\varepsilon}
  \\ \notag
  &=\int d1\,d2\,\frac{\partial\mathcal S(\mathbb Q)}{\partial\mathbb Q(1,2)}\left(\frac{\partial}{\partial\theta_1}+\frac{\partial}{\partial\theta_2}\right)\mathbb Q(1,2)
\end{align}
Differentiating both sides of this equation with respect to $\mathbb Q$ and applying the product rule yields
\begin{align}
  0
  =\int d3\,d4\,\frac{\partial^2\mathcal S(\mathbb Q)}{\partial\mathbb Q(1,2)\partial\mathbb Q(3,4)}\left(\frac{\partial}{\partial\theta_3}+\frac{\partial}{\partial\theta_4}\right)\mathbb Q(3,4)
  \\ \notag
  +
  \int d3\,d4\,\frac{\partial\mathcal S(\mathbb Q)}{\partial\mathbb Q(3,4)}\left(\frac{\partial}{\partial\theta_3}+\frac{\partial}{\partial\theta_4}\right)\delta(1,3)\delta(2,4)
\end{align}
Evaluating this expression at a saddle point $\mathbb Q^*$ of the effective action gives
\begin{align}
  0
  =\int d3\,d4\,\frac{\partial^2\mathcal S(\mathbb Q^*)}{\partial\mathbb Q^*(1,2)\partial\mathbb Q^*(3,4)}\left(\frac{\partial}{\partial\theta_3}+\frac{\partial}{\partial\theta_4}\right)\mathbb Q^*(3,4)
\end{align}
since by definition the derivative of the effective action with respect to $\mathbb Q$ evaluated at the saddle point vanishes. If $\mathbb Q^*$ is supersymmetric, then $(\frac\partial{\partial\theta_1}+\frac\partial{\partial\theta_2})\mathbb
Q^*(1,2)=0$ and the relation is trivially satisfied.
If instead supersymmetry is broken and
$(\frac\partial{\partial\theta_1}+\frac\partial{\partial\theta_2})\mathbb
Q^*(1,2)$ is nonzero, then the relation implies that it is an eigensupervector of
$\operatorname{Hess}\mathcal S(\mathbb Q^*)$ with eigenvalue zero. Since the Grassmann components of $\mathbb
Q^*$ are zero for any saddle point obeying fermion number symmetry, $\mathbb Q^*$ contains only components with
an even number of superindices, and therefore
$(\frac\partial{\partial\theta_1}+\frac\partial{\partial\theta_2})\mathbb
Q^*(1,2)$ contains only ones with an odd number of superindices. This means that
the zero eigenvalue is present in the sector of $\operatorname{Hess}\mathcal
S(\mathbb Q^*)$ that maps between odd subspaces. Since the superdeterminant is
inversely proportional to the determinant of the linear mapping in the odd
subspace, it follows that $[\operatorname{sdet}\operatorname{Hess}\mathcal
S(\mathbb Q^*)]^{-\frac12}=0$.

\subsection{4-index superspace and the number of stationary points}
\label{sec:4-index}

We have seen that the standard supersymmetric approach to counting stationary
points results in an alternating count whose total equals a topological invariant
of the configuration space, with restricted possibility of spontaneous symmetry
breaking at the saddle-point evaluation for large $N$. In order to correctly
count, we need to break supersymmetry in a precise way, and to do this we
require a larger supervector space. We consider now supervectors with 4
Grassmann indices: $\bar\theta$, $\theta$, $\bar\vartheta$, $\vartheta$. A
generic element of the resulting supervector space $\boldsymbol\phi\in\mathbb
R^{N|4}$ takes the form
\begin{align}
  \boldsymbol\phi(1)
  =\boldsymbol x+i\hat{\boldsymbol x}\bar\theta_1\theta_1\bar\vartheta_1\vartheta_1
  +\boldsymbol a\bar\theta_1\vartheta_1-\bar{\boldsymbol a}\theta_1\bar\vartheta_1
  +\boldsymbol b\bar\theta_1\theta_1-\bar{\boldsymbol b}\bar\vartheta_1\vartheta_1
  +\boldsymbol c\bar\theta_1\bar\vartheta_1+\bar{\boldsymbol c}\theta_1\vartheta_1
  \hspace{4em}
  \\ \notag
  +\bar\theta_1\boldsymbol\eta+\bar{\boldsymbol\eta}\theta_1\bar\vartheta_1\vartheta_1
  +\bar\vartheta_1\boldsymbol\xi+\bar{\boldsymbol\xi}\bar\theta_1\theta_1\vartheta_1
  +\bar{\boldsymbol\gamma}\theta_1+\bar\theta_1\bar\vartheta_1\vartheta_1\boldsymbol\gamma
  +\bar{\boldsymbol\chi}\vartheta_1+\bar\theta_1\theta_1\bar\vartheta_1\boldsymbol\chi
\end{align}
where $\boldsymbol x,\hat{\boldsymbol x}\in\mathbb R^N$, $\boldsymbol
a,\boldsymbol b,\boldsymbol c\in\mathbb C^N$, $\bar{\boldsymbol
a},\bar{\boldsymbol b},\bar{\boldsymbol c}$ are their complex conjugates, and
$\bar{\boldsymbol\eta}, \boldsymbol\eta, \bar{\boldsymbol\xi},
\boldsymbol\xi, \bar{\boldsymbol\gamma}, \boldsymbol\gamma,
\bar{\boldsymbol\chi}, \boldsymbol\chi$ are $N$-dimensional Grassmann
vectors. The result of writing the analogous action
\begin{equation}
  S_\textsc{susy}(\boldsymbol\phi)=\int d1\,H(\boldsymbol\phi(1))
\end{equation}
with $d1=d\vartheta_1\,d\bar\vartheta_1\,d\theta_1\,d\bar\theta_1$
is the same as that in the 2-index superspace, since
\begin{align}
  &\int_{\mathbb R^{N\mid 4}}d\boldsymbol\phi\,e^{S_\textsc{susy}(\boldsymbol\phi)}
  =
  \int_{\mathbb R^{N\mid 4}}d\boldsymbol\phi\,e^{\int d1\,H(\boldsymbol\phi(1))}
  \\ \notag
  &=\int\frac{d\boldsymbol x\,d\hat{\boldsymbol x}}{(2\pi)^N}\frac{d\bar{\boldsymbol a}\,d\boldsymbol a}{(2\pi)^N}
  \frac{d\bar{\boldsymbol b}\,d\boldsymbol b}{(2\pi)^N}\frac{d\bar{\boldsymbol c}\,d\boldsymbol c}{(2\pi)^N}
  d\bar{\boldsymbol\eta}\,d\boldsymbol\eta\,
  d\bar{\boldsymbol\xi}\,d\boldsymbol\xi\,
  d\bar{\boldsymbol\gamma}\,d\boldsymbol\gamma\,
  d\bar{\boldsymbol\chi}\,d\boldsymbol\chi
    \\ \notag
  &  \hspace{2em}
  \times\exp\bigg(
    i\hat{\boldsymbol x}\cdot\boldsymbol\nabla H(\boldsymbol x)
  -\bar{\boldsymbol a}^T\operatorname{Hess}H(\boldsymbol x)\boldsymbol a
    -\bar{\boldsymbol b}^T\operatorname{Hess}H(\boldsymbol x)\boldsymbol b
    -\bar{\boldsymbol c}^T\operatorname{Hess}H(\boldsymbol x)\boldsymbol c
    \\ \notag
  &  \hspace{2em}-\bar{\boldsymbol\eta}^T\operatorname{Hess}H(\boldsymbol x)\boldsymbol\eta
    -\bar{\boldsymbol\xi}^T\operatorname{Hess}H(\boldsymbol x)\boldsymbol\xi
    -\bar{\boldsymbol\gamma}^T\operatorname{Hess}H(\boldsymbol x)\boldsymbol\gamma
    -\bar{\boldsymbol\chi}^T\operatorname{Hess}H(\boldsymbol x)\boldsymbol\chi
    \\ \notag
  &\hspace{2em}+\big(
      a_i\bar\gamma_j\xi_k-b_i\bar\chi_j\xi_k+c_i\bar\gamma_j\bar\chi_k
      +\bar a_i\bar\chi_j\eta_k+\bar b_i\bar\gamma_j\eta_k+\bar c_i\eta_j\xi_k
    \big)\partial_i\partial_j\partial_kH(\boldsymbol x)
  -\bar\gamma_i\bar\chi_j\eta_k\xi_l\partial_i\partial_j\partial_k\partial_lH(\boldsymbol x)
  \bigg)
\end{align}
All terms with more than two derivatives of $H$ do not contribute to the result
because they do not have the necessary combinations of the Grassmann fields. The
expression therefore reduces to
\begin{align}
  \int_{\mathbb R^{N\mid 4}}d\boldsymbol\phi\,e^{S_\textsc{susy}(\boldsymbol\phi)}
  &=\int\frac{d\boldsymbol x\,d\hat{\boldsymbol x}}{(2\pi)^N}\frac{d\bar{\boldsymbol a}\,d\boldsymbol a}{(2\pi)^N}
  \frac{d\bar{\boldsymbol b}\,d\boldsymbol b}{(2\pi)^N}\frac{d\bar{\boldsymbol c}\,d\boldsymbol c}{(2\pi)^N}
  d\bar{\boldsymbol\eta}\,d\boldsymbol\eta\,
  d\bar{\boldsymbol\xi}\,d\boldsymbol\xi\,
  d\bar{\boldsymbol\gamma}\,d\boldsymbol\gamma\,
  d\bar{\boldsymbol\chi}\,d\boldsymbol\chi
    \\ \notag
  &  \hspace{2em}
  \times\exp\bigg(
    i\hat{\boldsymbol x}\cdot\boldsymbol\nabla H(\boldsymbol x)
  -\bar{\boldsymbol a}^T\operatorname{Hess}H(\boldsymbol x)\boldsymbol a
    -\bar{\boldsymbol b}^T\operatorname{Hess}H(\boldsymbol x)\boldsymbol b
    -\bar{\boldsymbol c}^T\operatorname{Hess}H(\boldsymbol x)\boldsymbol c
    \\ \notag
  &  \hspace{4em}-\bar{\boldsymbol\eta}^T\operatorname{Hess}H(\boldsymbol x)\boldsymbol\eta
    -\bar{\boldsymbol\xi}^T\operatorname{Hess}H(\boldsymbol x)\boldsymbol\xi
    -\bar{\boldsymbol\gamma}^T\operatorname{Hess}H(\boldsymbol x)\boldsymbol\gamma
    -\bar{\boldsymbol\chi}^T\operatorname{Hess}H(\boldsymbol x)\boldsymbol\chi
  \bigg)
  \\ \notag
  &=\int d\boldsymbol x\,\delta\big(\boldsymbol\nabla H(\boldsymbol x)\big)
  \frac{(\det\operatorname{Hess}H(\boldsymbol x))^4}{(\det\operatorname{Hess}H(\boldsymbol x))^3}
  =\int d\boldsymbol x\,\delta\big(\boldsymbol\nabla H(\boldsymbol x)\big)
  \det\operatorname{Hess}H(\boldsymbol x)
  =\chi[H]
\end{align}
Changing the dimension of superspace does not change the result of
integrating the simple supersymmetric action over it. However, the result can
be changed by spontaneous breaking of the supersymmetry.
We can write the count of stationary points $\mathcal N[H]$ as the limit
\begin{equation}
  \mathcal N[H]=\int_{\mathbb R^N}d\boldsymbol x\,\delta\big(\boldsymbol\nabla H(\boldsymbol x)\big)\,|\det\operatorname{Hess}H(\boldsymbol x)|
  =\lim_{\epsilon\to0}\mathcal N_\epsilon[H]
\end{equation}
where we have defined the perturbed count
\begin{equation}
  \mathcal N_\epsilon[H]
  =\int_{\mathbb R^N} d\boldsymbol x
  \,\delta\big(\nabla H(\boldsymbol x)\big)
  \,\sqrt{\det(\operatorname{Hess}H(\boldsymbol x)+i\epsilon I)}
  \,\sqrt{\det(\operatorname{Hess}H(\boldsymbol x)-i\epsilon I)}
\end{equation}
This is a small modification of the identity \eqref{eq:det.formula} of Ref
\cite{Fyodorov_2004_Complexity}. The perturbed count of stationary points can
be expressed compactly using a superspace integral over $\mathbb R^{N|4}$, with
\begin{equation}
  \mathcal N_\epsilon[H]
  =\int_{\mathbb R^{N|4}}d\boldsymbol\phi\,e^{
    S_\textsc{susy}(\boldsymbol\phi)+\epsilon S_\text{pert}(\boldsymbol\phi)
  }
\end{equation}
where the \textsc{susy} and perturbative actions are given by
\begin{align}
  S_\textsc{susy}(\boldsymbol\phi)=\int d1\,H\big(\boldsymbol\phi(1)\big)
  &&
  S_\text{pert}(\boldsymbol\phi)=-\frac i2\int d1\,d2\,\Psi(1,2)\boldsymbol\phi(1)\cdot\boldsymbol\phi(2)
\end{align}
We have introduced a superoperator $\Psi$ in the perturbative action which is
defined in terms of the superindices by
\begin{align}
  \Psi(1,2)
  &=
  \bar\theta_1\vartheta_1\bar\theta_2\vartheta_2
  +\theta_1\bar\vartheta_1\theta_2\bar\vartheta_2
  +
  \bar\theta_1\theta_1\bar\theta_2\theta_2
  +\bar\vartheta_1\vartheta_1\bar\vartheta_2\vartheta_2
  +
  \bar\theta_1\bar\vartheta_1\bar\theta_2\bar\vartheta_2
  +\theta_1\vartheta_1\theta_2\vartheta_2
  +\bar\theta_1\theta_1\vartheta_1\bar\theta_2
  \\ \notag
  &-\bar\vartheta_1\theta_2\bar\vartheta_2\vartheta_2
  +\theta_1\bar\vartheta_1\vartheta_1\bar\vartheta_2-\bar\theta_1\bar\theta_2\theta_2\vartheta_2
  +\theta_1\bar\theta_2\theta_2\bar\vartheta_2-\bar\theta_1\bar\vartheta_1\vartheta_1\vartheta_2
  +\vartheta_1\bar\theta_2\bar\vartheta_2\vartheta_2-\bar\theta_1\theta_1\bar\vartheta_1\theta_2
\end{align}
To see why this perturbation to the action produces the perturbed count
$\mathcal N_\epsilon[H]$, we expand the component pieces of $S_\text{pert}$ and
perform the superindex integrals, yielding
\begin{align}
  &S_\text{pert}(\boldsymbol\phi)
  =-\frac i2\left(
    \boldsymbol a^T\boldsymbol a
    +\bar{\boldsymbol a}^T\bar{\boldsymbol a}
    +\boldsymbol b^T\boldsymbol b
    +\bar{\boldsymbol b}^T\bar{\boldsymbol b}
    +\boldsymbol c^T\boldsymbol c
    +\bar{\boldsymbol c}^T\bar{\boldsymbol c}
  \right)
  +i(\bar{\boldsymbol\eta}^T\boldsymbol\xi
  +\bar{\boldsymbol\xi}^T\boldsymbol\eta
  +\bar{\boldsymbol\gamma}^T\boldsymbol\chi
  +\bar{\boldsymbol\chi}^T\boldsymbol\gamma)
  \\ \notag
  &=-i\left(
    \|\operatorname{Re}\boldsymbol a\|^2
    -\|\operatorname{Im}\boldsymbol a\|^2
    +\|\operatorname{Re}\boldsymbol b\|^2
    -\|\operatorname{Im}\boldsymbol b\|^2
    +\|\operatorname{Re}\boldsymbol c\|^2
    -\|\operatorname{Im}\boldsymbol c\|^2
  \right)
  +i\left(\bar{\boldsymbol\eta}^T\boldsymbol\xi
  +\bar{\boldsymbol\xi}^T\boldsymbol\eta
  +\bar{\boldsymbol\gamma}^T\boldsymbol\chi
  +\bar{\boldsymbol\chi}^T\boldsymbol\gamma
  \right)
\end{align}
Combined with the contribution from $S_\textsc{susy}$, the integral over each
of the three complex fields produces
\begin{align}
  \int\frac{d\bar{\boldsymbol a}\,d\boldsymbol a}{(2\pi)^N}\exp\left(
    -\begin{bmatrix}\operatorname{Re}\boldsymbol a\\\operatorname{Im}\boldsymbol a\end{bmatrix}^T
  \begin{bmatrix}
    \operatorname{Hess}H(\boldsymbol x)+i\epsilon I & 0 \\
    0 & \operatorname{Hess}H(\boldsymbol x)-i\epsilon I
  \end{bmatrix}
  \begin{bmatrix}\operatorname{Re}\boldsymbol a\\\operatorname{Im}\boldsymbol a\end{bmatrix}
  \right)
  \hspace{8em}
  \\ \notag
  =\frac1{\sqrt{\det(\operatorname{Hess}H(\boldsymbol x)+i\epsilon I)}
  \sqrt{\det(\operatorname{Hess}H(\boldsymbol x)-i\epsilon I)}}
\end{align}
while the integral over each set of two pairs of Grassmann fields produces
\begin{align}
  \int d\bar{\boldsymbol\eta}\,d\bar{\boldsymbol\xi}\,d\boldsymbol\eta\,d\boldsymbol\xi\,\exp\left(
    -
  \begin{bmatrix}
    \bar{\boldsymbol\eta} \\
    \bar{\boldsymbol\xi}
  \end{bmatrix}^T
  \begin{bmatrix}
    \operatorname{Hess}H(\boldsymbol x) & -i\epsilon I \\
    -i\epsilon I & \operatorname{Hess}H(\boldsymbol x)
  \end{bmatrix}
  \begin{bmatrix}
    \boldsymbol\eta \\
    \boldsymbol\xi
  \end{bmatrix}
  \right)
  \hspace{14em}
  \\ \notag
  =\det(\operatorname{Hess}H(\boldsymbol x)+i\epsilon I)
  \det(\operatorname{Hess}H(\boldsymbol x)-i\epsilon I)
\end{align}
The result of all complex and Grassmann integrals is precisely $\mathcal
N_\epsilon[H]$. It was not necessary to invoke 4-index superspace in order to
express $S_\text{pert}$; fewer fields could accomplish the same result.
However, we will see that because $S_\text{pert}$ has the form of a linear
perturbation to the superoverlap along a specific direction $\Psi$ in
superspace, the form of symmetry breaking it generates is constrained to a
small subspace of possible \textsc{susy}-broken superoperators. We will call
this the subspace of \emph{minimally \textsc{susy}-broken superoperators}.

\subsubsection{Symmetries of the action}

\paragraph{Supersymmetry.}

As in the 2-index case, the $\epsilon=0$ action is invariant under translation of the superfield along each of the superindicies. This results in four Grassmann symmetries generated by
\begin{align}
  \frac\partial{\partial\bar\theta}
  &&
  \frac\partial{\partial\theta}
  &&
  \frac\partial{\partial\bar\vartheta}
  &&
  \frac\partial{\partial\vartheta}
\end{align}
The perturbation $S_\text{pert}$ breaks all four supersymmetries.

\paragraph{Ghost number symmetry.}

As is in the 2-index case, the $\epsilon=0$ action is invariant under the ghost number symmetry corresponding to any pair of superindicies. This gives $\binom 42=6$ such real symmetries generated by
\begin{align}
  \theta\frac\partial{\partial\theta}-\bar\theta\frac\partial{\partial\bar\theta}
  &&
  \vartheta\frac\partial{\partial\vartheta}-\bar\vartheta\frac\partial{\partial\bar\vartheta}
  &&
  \theta\frac\partial{\partial\theta}-\bar\vartheta\frac\partial{\partial\bar\vartheta}
  &&
  \vartheta\frac\partial{\partial\vartheta}-\bar\theta\frac\partial{\partial\bar\theta}
  &&
  \theta\frac\partial{\partial\theta}-\vartheta\frac\partial{\partial\vartheta}
  &&
  \bar\vartheta\frac\partial{\partial\bar\vartheta}-\bar\theta\frac\partial{\partial\bar\theta}
\end{align}
The perturbation $S_\text{pert}$ breaks all six ghost number symmetries.

\paragraph{Permutation symmetry.}

Besides the continuous symmetries noted above, the action also has a set of
discrete symmetries. Because the $\epsilon=0$ action depends only on terms
including all four superindicies $\bar\theta\theta\bar\vartheta\vartheta$, any
permutation of the indices that leave this combination unchanged also leaves
the \textsc{susy} action unchanged. However, many fewer symmetries leave
$S_\text{pert}$ unchanged. This is because our choice to pair $\boldsymbol\eta$
with $\boldsymbol\xi$ and $\boldsymbol\gamma$ with $\boldsymbol\chi$ was
arbitrary; we could have paired the eight fermionic fields together in any
other way not involving a field and its own bar. Any such choice would lead to
a different form for $\Psi$ but equivalent once the fields are integrated away.
Before our choice there was no formal difference between the four superindices,
but $S_\text{pert}$ breaks their full permutation symmetry into the reduced set
\begin{align}
  (\bar\theta,\theta,\bar\vartheta,\vartheta)
  \mapsto
  (\bar\vartheta,\vartheta,\bar\theta,\theta)
  &&
  (\bar\theta,\theta,\bar\vartheta,\vartheta)
  \mapsto
  (\theta,\bar\theta,\vartheta,\bar\vartheta)
  &&
  (\bar\theta,\theta,\bar\vartheta,\vartheta)
  \mapsto
  (\vartheta,\bar\vartheta,\theta,\bar\theta)
\end{align}
which are all even permutations which preserve the division into barred and unbarred and $\theta$ and $\vartheta$.

\subsubsection{The saddle point approximation}
\label{sec:saddle.point.4}

Calculation of the average number of stationary points in a mean-field system
typically leads to an integral over the superoverlap order parameter $\mathbb
Q(1,2)=\frac1N\boldsymbol\phi(1)\cdot\boldsymbol\phi(2)$ of the form
\begin{equation}
  \overline{\mathcal N_\epsilon[H]}
  =\int_{\mathbb R^{1|4\times 4}}d\mathbb Q\,e^{N\mathcal S(\mathbb Q)-N\epsilon\frac i2\int d1\,d2\,\Psi(1,2)\mathbb Q(1,2)}
\end{equation}
where the effective action $\mathcal S$ is supersymmetric and the second term
explicitly breaks supersymmetry along the direction $\Psi$.
If supersymmetry is preserved, then like in the 2-index case the most general form for the saddle point value $\mathbb Q^*$ compatible with the symmetry is
\begin{equation}
  \mathbb Q^*(1,2)=Q_1+Q_\delta\delta(1,2)
\end{equation}
where the identity superoperator for 4-index superspace is defined by
\begin{equation}
  \delta(1,2)=(\bar\theta_1-\bar\theta_2)(\theta_1-\theta_2)(\bar\vartheta_1-\bar\vartheta_2)(\vartheta_1-\vartheta_2)
\end{equation}
If instead supersymmetry is
spontaneously broken, the perturbation
favors one of the otherwise equivalent \textsc{susy}-broken solutions $\mathbb
Q^*$ based on the value of the component
\begin{equation}
  \frac i2\int d1\,d2\,\Psi(1,2)\mathbb Q(1,2)=Q_\Psi
\end{equation}
The most general form of $\mathbb Q^*$ consistent with supersymmetry breaking
condensed along the $\Psi$ axis involves all superaxes generated by the
superspace algebra of addition, multiplication, and convolution of $1$,
$\delta$, and $\Psi$. The set of orthogonal operators that can be generated
this way are 1, $\Psi$ and
\begin{align}
  \pi(1,2)=\bar\theta_1\theta_1\bar\vartheta_1\vartheta_1
  &&
  \Phi(1,2)=\delta(1,2)-\bar\pi(1,2)-\pi(1,2)
  \\ \notag
  \bar\pi(1,2)=\bar\theta_2\theta_2\bar\vartheta_2\vartheta_2
  &&
  \Theta(1,2)=\bar\pi(1,2)\pi(1,2)\hspace{4.9em}
\end{align}
We additionally expect that the saddle points $\mathbb Q^*$ are symmetric superoperators, and therefore $\mathbb Q^*$ will not depend on $\pi(1,2)$ and $\bar\pi(1,2)$ independently but instead only on the symmetric combination
\begin{equation}
  \Pi(1,2)=\bar\pi(1,2)+\pi(1,2)
\end{equation}
Thus, the most general \textsc{susy}-broken operator $\mathbb Q^*$ resulting
from condensation along the direction $\Psi$ has the form
\begin{equation}
  \mathbb Q^*(1,2)=Q_1+Q_\Phi\Phi(1,2)+iQ_\Psi\Psi(1,2)+Q_\Pi\Pi(1,2)+Q_\Theta\Theta(1,2)
\end{equation}
This defines the minimally \textsc{susy}-broken subspace, whose algebra will be
specified in Section \ref{sec:algebra}. An operator in this subspace preserves supersymmetry if $Q_\Phi=Q_\Pi$ and $Q_\Theta=Q_\Psi=0$.

\subsubsection{Nonvanishing of supersymmetry-breaking saddle points}
\label{sec:4-index.vanishing}

What if $\mathcal S$ is not supersymmetric, but contains a supersymmetry
breaking term proportional to $\epsilon$? Following the same procedure as Section \ref{sec:2-index.vanishing} for the 2-index superspace results in the relation
\begin{equation}
  \int d3\,d4\,\frac{\partial^2\mathcal S(\mathbb Q^*)}{\partial\mathbb Q^*(1,2)\mathbb Q^*(3,4)}\left(\frac{\partial}{\partial\theta_3}+\frac{\partial}{\partial\theta_4}\right)\mathbb Q^*(1,2)
  =
  \epsilon\frac i2\left(\frac{\partial}{\partial\theta_1}+\frac{\partial}{\partial\theta_2}\right)\Psi(1,2)
\end{equation}
for any $\mathbb Q^*$ that extremizes the action.
This implies that
$(\frac\partial{\partial\theta_1}+\frac\partial{\partial\theta_2})\mathbb
Q^*(1,2)$ is mapped to
$\epsilon\frac i2(\frac\partial{\partial\theta_1}+\frac\partial{\partial\theta_2})\Psi(1,2)$
by the Hessian. Each of the four supersymmetries will have a corresponding soft
mode in the odd-to-odd part of the Hessian proportional to $\epsilon$. This
seems to imply that, just as in the 2-index superspace, the inverse
superdeterminant of the Hessian is identically zero in the limit of zero $\epsilon$. However, the action $S_\text{pert}$ also breaks all six
ghost number symmetries, each of which lead to a relation of the form
\begin{align}
  \int d3\,d4\,\frac{\partial^2\mathcal S(\mathbb Q^*)}{\partial\mathbb Q^*(1,2)\mathbb Q^*(3,4)}\left(\theta_3\frac{\partial}{\partial\theta_3}-\bar\theta_3\frac{\partial}{\partial\theta_3}+\theta_4\frac{\partial}{\partial\theta_4}-\bar\theta_4\frac{\partial}{\partial\bar\theta_4}\right)\mathbb Q^*(1,2)
  \hspace{4em}
  \\ \notag
  =
  \epsilon\frac i2\left(\theta_3\frac{\partial}{\partial\theta_3}-\bar\theta_3\frac{\partial}{\partial\theta_3}+\theta_4\frac{\partial}{\partial\theta_4}-\bar\theta_4\frac{\partial}{\partial\bar\theta_4}\right)\Psi(1,2)
\end{align}
and a soft mode proportional to $\epsilon$ in the \emph{even-to-even} part of
the Hessian. If $Q_\Psi$ is nonzero in the limit of $\epsilon$ to zero, then
having six broken real ghost-number symmetries and four broken Grassmann
supersymmetries means that determinant of the even-to-even sector of the
Hessian is proportional to $\epsilon^6$ and the determinant of the odd-to-odd
sector of the Hessian is proportional to $\epsilon^4$. Since the
superdeterminant is the ratio of the determinants of the even sector and the
odd sector, the end result is
\begin{equation}
  [\operatorname{sdet}\operatorname{Hess}\mathcal S(\mathbb Q^*)]^{-\frac12}\propto\epsilon^{-1}
\end{equation}
which diverges as $\epsilon$ goes to zero rather than vanishing. A divergent
prefactor is typical when continuous real symmetries are spontaneously broken,
and the resulting zero modes in the Hessian require that higher-order
corrections be used to evaluate the prefactor. We don't concern ourselves with
the prefactor, but only with the fact that by simultaneously breaking
\emph{both} supersymmetries and ghost number symmetries, these
\textsc{susy}-breaking saddle points have prefactors that do not vanish.

\subsubsection{The algebra of minimally \textsc{susy}-breaking operators}
\label{sec:algebra}

In Section \ref{sec:saddle.point.4} we argued that when supersymmetry is
minimally broken along the necessary axis to produce the absolute value of the
determinant, superoperators will only condense inside the six-dimensional subspace
\begin{align}
  1(1,2)&=1
  \\
  \pi(1,2)&=\bar\theta_1\theta_1\bar\vartheta_1\vartheta_1
  \\
  \bar\pi(1,2)&=\bar\theta_2\theta_2\bar\vartheta_2\vartheta_2
  \\
  \Theta(1,2)&=\bar\theta_1\theta_1\bar\vartheta_1\vartheta_1\bar\theta_2\theta_2\bar\vartheta_2\vartheta_2
  \\
  \Phi(1,2)
  &=
  \bar\theta_1\theta_1\bar\vartheta_2\vartheta_2+\bar\vartheta_1\vartheta_1\bar\theta_2\theta_2
  -\bar\theta_1\bar\vartheta_1\theta_2\vartheta_2-\theta_1\vartheta_1\bar\theta_2\bar\vartheta_2
  +\bar\theta_1\vartheta_1\theta_2\bar\vartheta_2+\theta_1\bar\vartheta_1\bar\theta_2\vartheta_2
  -\bar\theta_1\theta_1\bar\vartheta_1\vartheta_2
  \\ \notag
  &
  +\vartheta_1\bar\theta_2\theta_2\bar\vartheta_2
  +\bar\theta_1\theta_1\vartheta_1\bar\vartheta_2-\bar\vartheta_1\bar\theta_2\theta_2\vartheta_2
  -\bar\theta_1\bar\vartheta_1\vartheta_1\theta_2+\theta_1\bar\theta_2\bar\vartheta_2\vartheta_2
  +\theta_1\bar\vartheta_1\vartheta_2\bar\theta_2-\bar\theta_1\theta_2\bar\vartheta_2\vartheta_2
  \\
  \Psi(1,2)
  &=
  \bar\theta_1\vartheta_1\bar\theta_2\vartheta_2
  +\theta_1\bar\vartheta_1\theta_2\bar\vartheta_2
  +
  \bar\theta_1\theta_1\bar\theta_2\theta_2
  +\bar\vartheta_1\vartheta_1\bar\vartheta_2\vartheta_2
  +
  \bar\theta_1\bar\vartheta_1\bar\theta_2\bar\vartheta_2
  +\theta_1\vartheta_1\theta_2\vartheta_2
  +\bar\theta_1\theta_1\vartheta_1\bar\theta_2
  \\ \notag
  &
  -\bar\vartheta_1\theta_2\bar\vartheta_2\vartheta_2
  +\theta_1\bar\vartheta_1\vartheta_1\bar\vartheta_2-\bar\theta_1\bar\theta_2\theta_2\vartheta_2
  +\theta_1\bar\theta_2\theta_2\bar\vartheta_2-\bar\theta_1\bar\vartheta_1\vartheta_1\vartheta_2
  +\vartheta_1\bar\theta_2\bar\vartheta_2\vartheta_2-\bar\theta_1\theta_1\bar\vartheta_1\theta_2
\end{align}
The five-dimensional subspace of symmetric minimally \textsc{susy}-breaking
superoperators only involves $\pi$ and $\bar\pi$ in the symmetric combination
\begin{equation}
  \Pi(1,2)=\pi(1,2)+\bar\pi(1,2)=
  \bar\theta_1\theta_1\bar\vartheta_1\vartheta_1
  +\bar\theta_2\theta_2\bar\vartheta_2\vartheta_2
\end{equation}
Understanding the algebra of this six-dimensional subspace of $\mathbb
R^{1|4\times 4}$ allows us to dispense with reference to the 256-dimensional
embedding space and work directly with the operators above. Besides addition,
whose algebra is trivial, the relevant operations between superoperators are
convolution and multiplication, respectively defined by
\begin{align}
  (A\ast B)(1,2)=\int d3\,A(1,3)B(3,2)
  &&
  (A\times B)(1,2)=A(1,2)B(1,2)
\end{align}
The behavior of the
symmetric operators under convolution and multiplication is given in Table~\ref{tab:algebra}.
Recall that $\delta=\Phi+\Pi$ is the identity operator under convolution, which
the algebra verifies.

\begin{table}
  \centering
  \hspace{1em}
  \begin{tabular}{c|ccccccc}
    $\ast$   &1    &$\Phi$&$\Psi$&$\Theta$ &$\Pi$    &$\pi$&$\bar\pi$\\
    \hline
    1        &0    &0     &0     &$\bar\pi$&1        &1    &0        \\
    $\Phi$   &0    &$\Phi$&$\Psi$&0        &0        &0    &0        \\
    $\Psi$   &0    &$\Psi$&$\Phi$&0        &0        &0    &0        \\
    $\Theta$ &$\pi$&0     &0     &0        &$\Theta$ &0    &$\Theta$ \\
    $\Pi$    &1    &0     &0     &$\Theta$ &$\Pi$    &$\pi$&$\bar\pi$\\
    $\bar\pi$&1    &0     &0     &0        &$\bar\pi$&0    &$\bar\pi$\\
    $\pi$    &0    &0     &0     &$\Theta$ &$\pi$    &$\pi$&0        \\
  \end{tabular}
  \hfill
  \begin{tabular}{c|ccccccc}
    $\times$ &1        &$\Phi$    &$\Psi$    &$\Theta$&$\Pi$    &$\pi$   &$\bar\pi$\\
    \hline
    1        &1        &$\Phi$    &$\Psi$    &$\Theta$&$\Pi$    &$\pi$   &$\bar\pi$\\
    $\Phi$   &$\Phi$   &$-2\Theta$&0         &0       &0        &0       &0        \\
    $\Psi$   &$\Psi$   &0         &$-2\Theta$&0       &0        &0       &0        \\
    $\Theta$ &$\Theta$ &0         &0         &0       &0        &0       &0        \\
    $\Pi$    &$\Pi$    &0         &0         &0       &$2\Theta$&$\Theta$&$\Theta$ \\
    $\bar\pi$&$\bar\pi$&0         &0         &0       &$\Theta$ &$\Theta$&0        \\
    $\pi$    &$\pi$    &0         &0         &0       &$\Theta$ &0       &$\Theta$ \\
  \end{tabular}
  \hspace{1em}
  \caption{
    The algebra of minimally \textsc{susy}-breaking operators under (left)
    convolution and (right) multiplication. An entry whose row is labeled by
    $A$ and whose column is labeled by $B$ is given by $A\ast B$ or $A\times
    B$, respectively.
  } \label{tab:algebra}
\end{table}

Order parameters like Lagrange multipliers (Section \ref{sec:lagrange}) and overlaps with a signal or spike (Section \ref{sec:signal}) often take the form of superspace vectors in $\mathbb
R^{1|4}$. The space of minimally \textsc{susy}-breaking supervectors is
two-dimensional, involving 1 and
\begin{equation}
  \varpi(1)=\bar\theta_1\theta_1\bar\vartheta_1\vartheta_1
\end{equation}
The algebraic relation of minimally \textsc{susy}-breaking supervectors with
each other and with minimally \textsc{susy}-breaking superoperators can be
inferred from the relations above, since any supervector $B(1)=b+\hat
b\varpi(1)$ can be written as a superoperator $B(1,2)=b1(1,2)+\hat
b\pi(1,2)=b1(2,1)+\hat b\bar\pi(2,1)$ that does not depend on the 2
superindices.

\subsubsection{Identities in the minimally \textsc{susy}-breaking subspace}
\label{sec:identities}

Because all elements of the reduced subspace are nilpotent under multiplication
with index at most 3, series expansions in the soul of any superoperator in
this subspace terminate at second order. If
\begin{equation}
  \mathbb A(1,2)=A_1+A_\Phi\Phi(1,2)+iA_\Psi\Psi(1,2)+A_{\pi}\pi(1,2)+A_{\bar\pi}\bar\pi(1,2)+A_\Theta\Theta(1,2)
\end{equation}
is a generic minimally \textsc{susy}-breaking operator, then the following
identities hold. An arbitrary function of $\mathbb A$ can be written
\begin{align}
  f(\mathbb A(1,2))
  &=f(A_1)+f'(A_1)A_\Phi\Phi(1,2)
  +f'(A_1)A_{\pi}\pi(1,2)+f'(A_1)A_{\bar\pi}\bar\pi(1,2)
  \\ \notag
  &\hspace{4em}
  +if'(A_1)A_\Psi\Psi(1,2)
  +\big(f'(A_1)A_\Theta+f''(A_1)(A_{\pi}A_{\bar\pi}-A_\Phi^2+A_\Psi^2)\big)\Theta(1,2)
\end{align}
The inverse of $\mathbb A$ under convolution, defined by
\begin{equation}
  \int d3\,\mathbb A^{-1}(1,3)\mathbb A(3,2)=\delta(1,2)
\end{equation}
is also a minimally \textsc{susy}-breaking superoperator with coefficients given by
\begin{align}
  A^{-1}_1=\frac{-A_1}{A_{\pi}A_{\bar\pi}-A_1A_\Theta}
  &&
  A^{-1}_{\pi}=\frac{A_{\bar\pi}}{A_{\pi}A_{\bar\pi}-A_1A_\Theta}
  &&
  A^{-1}_\Phi=\frac{A_\Phi}{A_\Phi^2+A_\Psi^2}
  \\
  A^{-1}_\Theta=\frac{-A_\Theta}{A_{\pi}A_{\bar\pi}-A_1A_\Theta}
  &&
  A^{-1}_{\bar\pi}=\frac{A_{\pi}}{A_{\pi}A_{\bar\pi}-A_1A_\Theta}
  &&
  A^{-1}_\Psi=\frac{-A_\Psi}{A_\Phi^2+A_\Psi^2}
\end{align}
The superdeterminant, supertrace, and double integral over its elements are given by
\begin{align}
  \operatorname{sdet}\mathbb A=\frac{A_{\pi}A_{\bar\pi}-A_1A_\Theta}{A_\Phi^2+A_\Psi^2}
  &&
  \operatorname{sTr}\mathbb A=A_\pi+A_{\bar\pi}-2A_\Phi
  &&
  \int d1\,d2\,\mathbb A(1,2)=A_\Theta
\end{align}
The above formulae are immediately generalized to the case of a symmetric superoperator by setting $A_\pi=A_{\bar\pi}=A_\Pi$.

Finally, we will frequently encounter super-Gaussian integrals with covariance
given by a symmetric superoperator $\mathbb A$. These are given by
\begin{align}
  &\int_{\mathbb R^{1|4}} d\phi\,(\operatorname{sdet}\mathbb A)^{-\frac12}e^{-\frac12\int d1\,d2\,\phi(1)\mathbb A^{-1}(1,2)\phi(2)+\int d1\,(1-\beta\varpi(1))U(\phi(1))}
  \\ \notag
  &\hspace{10em}
  =\int dx\,
  \sqrt{\frac{(1-A_\Phi U''(x))^2+A_\Psi^2U''(x)^2}{2\pi A_1}}
  e^{
    -\frac1{2A_1}(x-A_\Pi U'(x))^2+\frac12A_\Theta U'(x)^2
    -\beta U(x)
  }
\end{align}
or alternatively
\begin{align}
  &\int_{\mathbb R^{1|4}} d\phi\,e^{\frac12\int d1\,d2\,\phi(1)\mathbb A(1,2)\phi(2)+\int d1\,(1-\beta\varpi(1))U(\phi(1))}
  \\ \notag
  &\hspace{10em}
  =\int dx\,
  \sqrt{\frac{( A_\Phi+U''(x))^2+ A_\Psi^2}{2\pi A_1}}
  e^{
    -\frac1{2 A_1}( A_\Pi x+U'(x))^2+\frac12 A_\Theta x^2
    -\beta U(x)
  }
\end{align}
In the absence of supersymmetry breaking and with $\beta=0$, both integrals
evaluate exactly to $1$ if $U$ is a monotonic function. This can be seen in the first expression by changing
integration variables to $z=y-A_\delta U'(y)$. Then
$dz=dy\,\frac{dz}{dy}=dy\,(1-A_\delta U''(y))$ and the resulting integral is
that of the \textsc{pdf} of the normal distribution in $z$ with mean zero and
variance $A_1$.

\subsubsection{Interpretation of the order parameters}

The order parameters $Q_1$, $Q_\Phi$, $Q_\Pi$, $Q_\Theta$, and $Q_\Psi$ making
up the minimally \textsc{susy}-breaking operator $\mathbb Q$ can be interpreted
in terms of properties averaged over the stationary points, or variations of
their count with respect to modifications of the function $H$.
Define the average over stationary points of any quantity $A(\boldsymbol x)$ depending on position $\boldsymbol x$ as
\begin{equation}
  \langle A\rangle
  =\frac1{\mathcal N}\sum_{\boldsymbol x^*\in\mathrm{Crit}(H)}A(\boldsymbol x^*)
  =\frac1{\mathcal N}\int d\boldsymbol\phi\,e^{S(\boldsymbol\phi)}A\left(\int d1\,\varpi(1)\boldsymbol\phi(1)\right)
\end{equation}

\paragraph{\boldmath{$Q_1$}: the average norm of stationary points.}

The parameter $Q_1$ is proportional to the average norm of $\boldsymbol x$ over the
stationary points. Since $\boldsymbol x=\int d1\,\varpi(1)\boldsymbol\phi(1)$,
it follows that
\begin{equation}
  \frac1N\langle\|\boldsymbol x\|^2\rangle
  =\frac1{\mathcal N}\int d\boldsymbol\phi\,e^{S(\boldsymbol\phi)}\int d1\,d2\,\Theta(1,2)\frac{\boldsymbol\phi(1)\cdot\boldsymbol\phi(2)}N
  =Q_1
\end{equation}

\paragraph{\boldmath{$Q_\Pi$}: the average response of stationary points to a linear potential.}

Consider varying the function $H$ by a linear field, defining a new function $H(\boldsymbol x)+\boldsymbol h\cdot\boldsymbol x$.
The order parameter $Q_\Pi$ is proportional to the average change of stationary point position with $\boldsymbol h$, or
\begin{align}
  \frac1N\sum_{i=1}^N\frac{\partial\langle x_i\rangle}{\partial h_i}\bigg|_{\boldsymbol h=0}
  &=\frac1{\mathcal N}\frac1N\sum_{i=1}^N\frac\partial{\partial h_i}\int d\boldsymbol\phi\,e^{S(\boldsymbol\phi)+\int d1\,\boldsymbol h\cdot\boldsymbol\phi(1)}\int d1\,\varpi(1)\phi_i(1)
  \\ \notag
  &\qquad -\frac1N\frac1{\mathcal N^2}\sum_{i=1}^N\frac{\partial\mathcal N}{\partial h_i}\int d\boldsymbol\phi\,e^{S(\boldsymbol\phi)+\int d1\,\boldsymbol h\cdot\boldsymbol\phi(1)}\int d1\,\varpi(1)\phi_i(1)\bigg|_{\boldsymbol h=0}
  \\ \notag
  &=\frac1{\mathcal N}\int d\boldsymbol\phi\,e^{S(\boldsymbol\phi)}\int d1\,d2\,\varpi(1)\frac{\boldsymbol\phi(1)\cdot\boldsymbol\phi(2)}N
  \\ \notag
  &\qquad-\frac1N\sum_{i=1}^N\langle x_i\rangle\frac1{\mathcal N}\int d\boldsymbol\phi\,e^{S(\boldsymbol\phi)+\int d1\,\boldsymbol h\cdot\boldsymbol\phi(1)}\int d1\,\phi_i(1)\bigg|_{\boldsymbol h=0}
  \\ \notag
  &=Q_\Pi-\frac1N\langle\boldsymbol x\rangle\cdot\langle\hat{\boldsymbol x}\rangle
\end{align}
In isotropic systems $\langle\boldsymbol x\rangle=\langle\hat{\boldsymbol x}\rangle=0$, but they are generically nonzero in systems with a symmetry-breaking field, spike, or signal (see Section \ref{sec:signal}).

\paragraph{\boldmath{$Q_\Theta$}: the curvature of the complexity with respect to a linear potential.}

The order parameter $Q_\Theta$ is proportional to the second variation of of the complexity $\Sigma=\frac1N\log\mathcal N$ with respect to $\boldsymbol h$, or
\begin{align}
  \sum_{i=1}^N\frac{\partial^2\Sigma}{\partial h_i^2}\bigg|_{\boldsymbol h=0}
  &=\frac1N\sum_{i=1}^N\frac\partial{\partial h_i}\frac1{\mathcal N}\frac{\partial\mathcal N}{\partial h_i}\bigg|_{\boldsymbol h=0}
  =\frac1N\sum_{i=1}^N\left(
    \frac1{\mathcal N}\frac{\partial^2\mathcal N}{\partial h_i^2}
    -\frac1{\mathcal N^2}\left(\frac{\partial\mathcal N}{\partial h_i}\right)^2
  \right)\bigg|_{\boldsymbol h=0}
  \\ \notag
  &=\frac1{\mathcal N}\int d\boldsymbol\phi\,e^{S(\boldsymbol\phi)}\int d1\,d2\,\frac{\boldsymbol\phi(1)\cdot\boldsymbol\phi(2)}{N}
  -\frac1N\langle\hat{\boldsymbol x}\rangle\cdot\langle\hat{\boldsymbol x}\rangle
  \\ \notag
  &=Q_\Theta
  -\frac1N\langle\hat{\boldsymbol x}\rangle\cdot\langle\hat{\boldsymbol x}\rangle
\end{align}

\paragraph{\boldmath{$Q_\Phi$}: the average inverse curvature at a stationary point.}

To interpret $Q_\Phi$ and $Q_\Psi$, we must turn to average properties of the
Hessian matrix $\operatorname{Hess}H(\boldsymbol x)$. The order parameter $Q_\Phi$ is proportional to the average of the trace of the inverse of the Hessian. This is based on the general identities for symmetric matrices $A$ of
\begin{align}
  A^{-1}_{ij}\int d\bar{\boldsymbol\eta}\,d\boldsymbol\eta\,
  e^{-\bar{\boldsymbol\eta}^TA\boldsymbol\eta}
  &=\int d\bar{\boldsymbol\eta}\,d\boldsymbol\eta\,
  e^{-\bar{\boldsymbol\eta}^TA\boldsymbol\eta}(-\bar\eta_i\eta_j)
  \\
  A^{-1}_{ij}
  \int d\bar{\boldsymbol a}\,d\boldsymbol a\,
  e^{-\bar{\boldsymbol a}^TA\boldsymbol a}
  &=\int d\bar{\boldsymbol a}\,d\boldsymbol a\,
  e^{-\bar{\boldsymbol a}^TA\boldsymbol a}\bar a_ia_j
\end{align}
Since $\int d1\,d2\,\Phi(1,2)\phi_i(1)\phi_j(2)$ is composed of such all combinations, it follows that
\begin{align}
  \frac1{N}\langle\operatorname{Tr}\operatorname{Hess}H(\boldsymbol x)^{-1}\rangle
  &=-\frac1{\mathcal N}\int d\boldsymbol\phi\,e^{S(\boldsymbol\phi)}\frac1{N}\big(
    \bar{\boldsymbol\eta}\cdot\boldsymbol\eta
    +\bar{\boldsymbol\xi}\cdot\boldsymbol\xi
    +\bar{\boldsymbol\gamma}\cdot\boldsymbol\gamma
    +\bar{\boldsymbol\chi}\cdot\boldsymbol\chi
    +\bar{\boldsymbol a}\cdot\boldsymbol a
    +\bar{\boldsymbol b}\cdot\boldsymbol b
    +\bar{\boldsymbol c}\cdot\boldsymbol c
  \big)
  \\ \notag
  &=\frac1{\mathcal N}\int d\boldsymbol\phi\,e^{S(\boldsymbol\phi)}\int d1\,d2\,\Phi(1,2)\frac{\boldsymbol\phi(1)\cdot\boldsymbol\phi(2)}{2N}
  =-Q_\Phi
\end{align}
Note that since supersymmetry implies $Q_\Pi=Q_\Phi$, preserved supersymmetry also implies that two alternative definitions of the susceptibility of stationary points coincide: with $Q_\Pi$ corresponding to their response to an external field and $Q_\Phi$ corresponding to their inverse curvature.

\paragraph{\boldmath{$Q_\Psi$}: the spectral density of the Hessian at zero eigenvalue.}

Finally, the supersymmetry breaking order parameter $Q_\Psi$ is proportional to
the spectral density of the Hessian at zero eigenvalue:
\begin{align}
  Q_\Psi
  &=\frac1{\mathcal N}\int d\boldsymbol\phi\,e^{S(\boldsymbol\phi)}
  \frac i{2N}\int d1\,d2\,\Psi(1,2)\boldsymbol\phi(1)\cdot\boldsymbol\phi(2)
  =\frac1N\lim_{\epsilon\to0}\frac{\partial}{\partial\epsilon}\mathcal N_\epsilon
  \\ \notag
  &=\frac1N\lim_{\epsilon\to0}\int d\boldsymbol x\,\delta\big(\boldsymbol\nabla H(\boldsymbol x)\big)
  \frac\partial{\partial\epsilon}\sqrt{\det(\operatorname{Hess}H(\boldsymbol x)-i\epsilon I)\det(\operatorname{Hess}H(\boldsymbol x)+i\epsilon I)}
  \\ \notag
  &=
  \frac i{2N}\lim_{\epsilon\to0}\big\langle
    \operatorname{Tr}(\operatorname{Hess}H(\boldsymbol x)+i\epsilon I)^{-1}-\operatorname{Tr}(\operatorname{Hess}H(\boldsymbol x)-i\epsilon I)^{-1}
    \big\rangle
  \\ \notag
  &=
  \frac i2\lim_{\epsilon\to0}\big\langle G_{\operatorname{Hess}H(\boldsymbol x)}(i\epsilon)-G_{\operatorname{Hess}H(\boldsymbol x)}(-i\epsilon)\big\rangle
  =\pi\big\langle\rho_{\operatorname{Hess}H(\boldsymbol x)}(0)\big\rangle
\end{align}
where $G_{\operatorname{Hess}H(\boldsymbol x)}$ is the resolvent of the Hessian
matrix at the point $\boldsymbol x$ and $\rho_{\operatorname{Hess}(\boldsymbol
x)}$ is its spectral density.

\subsubsection{Models on non-Euclidean configuration space}
\label{sec:lagrange}

Many models of interest are defined on a nontrivial configuration space, like
the hypersphere. In these cases, the gradient and Hessian in the Kac--Rice
formula must be interpreted as the Riemannian gradient and Hessian \cite{Ros_2019_Complex}. Expressing these directly can make expressions dramatically more complicated.
When the configuration space can be expressed as a series of smooth
constraints, the method of Lagrange multipliers is a convenient
solution. Consider a configuration space $\Omega\subset\mathbb R^N$ defined by
$m$ constraints $g_i(\boldsymbol x)=0$. Define the Lagrangian
\begin{equation}
  L(\boldsymbol x,\boldsymbol\lambda)
  =H(\boldsymbol x)+\sum_{i=1}^m\lambda_ig_i(\boldsymbol x)
\end{equation}
The number of stationary points on the configuration space manifold is then
given by the Kac--Rice formula for the Lagrangian, or
\begin{align}
  \mathcal N[H]
  &=\int_{\mathbb R^{N|4}}d\boldsymbol\phi\int_{\mathbb R^{m|4}}d\boldsymbol\Lambda\,
  e^{\int d1\,L(\boldsymbol\phi(1),\boldsymbol\Lambda(1))}
  \\ \notag
  &=\int_{\mathbb R^N}d\boldsymbol x\int_{\mathbb R^m}d\boldsymbol\lambda\,
  \bigg[\prod_{i=1}^m\delta\big(g_i(\boldsymbol x)\big)\bigg]
  \delta\bigg(\boldsymbol\nabla H(\boldsymbol x)+\sum_{i=1}^m\lambda_i\boldsymbol\nabla g_i(\boldsymbol x)\bigg)
  \\ \notag
  &\hspace{12em}\times\left|\det\begin{bmatrix}
    \operatorname{Hess}H(\boldsymbol x)+\sum_{i=1}^m\lambda_i\operatorname{Hess}g_i(\boldsymbol x) & \boldsymbol\nabla g_1(\boldsymbol x) & \cdots & \boldsymbol\nabla g_m(\boldsymbol x) \\
    \boldsymbol\nabla g_1(\boldsymbol x)^T & 0 & \cdots & 0 \\
    \vdots & \vdots & \ddots & \vdots \\
    \boldsymbol\nabla g_m(\boldsymbol x)^T & 0 & \cdots & 0
  \end{bmatrix}\right|
\end{align}
The result naturally produces the Riemannian gradient and Hessian: for a given
point $\boldsymbol x$, the integral over the Lagrange multipliers
$\boldsymbol\lambda$ attempts every possible subtraction of the normal vectors
to the constraint surface from the Euclidean gradient. Meanwhile, the determinant is
\begin{equation}
  \det\bigg[\operatorname{Hess}H(\boldsymbol x)+\sum_{i=1}^m\lambda_i\operatorname{Hess}g_i(\boldsymbol x)\bigg]
  \det\bigg[-\sum_{j=1}^m\boldsymbol\nabla g_j(\boldsymbol x)^T\bigg(\operatorname{Hess}H(\boldsymbol x)+\sum_{i=1}^m\lambda_i\operatorname{Hess}g_i(\boldsymbol x)\bigg)^{-1}\boldsymbol\nabla g_j(\boldsymbol x)\bigg]
\end{equation}
The first factor produces the product over the eigenvalues of the Euclidean
Hessian. The second produces the product over eigenvalues of the inverse
of the Euclidean Hessian projected into the directions normal to the
configuration space manifold. The eigenvalues of the inverse Hessian cancel precisely those
eigenvalues from the Euclidean Hessian that correspond to excitations in
directions that would leave the configuration space manifold.

\subsubsection{Energy dependence}
\label{sec:energy}

The total number of stationary points is interesting, but it is often more
interesting to know their count conditioned on properties like their energy. We
do this by first biasing the count over stationary stationary points by the value
$H(\boldsymbol x)$ at each stationary point, or
\begin{equation}
  \mathcal N_\beta[H]
  =\int_{\mathbb R^N} d\boldsymbol x\,\delta\big(\boldsymbol\nabla H(\boldsymbol x\big)\,
  \big|\det\operatorname{Hess}H(\boldsymbol x)\big|\,
  e^{-\beta H(\boldsymbol x)}
  =\int_{\mathbb R^{N|4}} d\boldsymbol\phi\,e^{\int d1\,\beta(1)H(\boldsymbol\phi(1))}
\end{equation}
where we have introduced the compact notation $\beta(1)=1-\beta\varpi(1)$. The complexity as a function of energy can be derived from the complexity as a function of $\beta$ by a Legendre transform, or
\begin{equation}
  \Sigma(E)=\max_\beta(\Sigma(\beta)+\beta E)
\end{equation}
If the complexity is not maximized at the absolute minimum or maximum of the energy, this gives a condition on the effective action in addition to the regular extremal ones of
\begin{equation}
  0=\frac{\partial\mathcal S_\beta}{\partial\beta}+E
\end{equation}
Biasing the count of stationary points by their energy manifestly breaks
supersymmetry, since the action depends explicitly on the superindices.
However, a set of four fermionic Becchi--Rouet--Stora--Tyutin (BRST) symmetries
persist, corresponding to the translations
\begin{equation}
  T_\theta(\varepsilon)+T_{\hat{\boldsymbol x}}(i\beta\bar{\boldsymbol\gamma}\varepsilon)
  \qquad
  T_{\bar\theta}(\bar\varepsilon)+T_{\hat{\boldsymbol x}}(i\beta\bar\varepsilon{\boldsymbol\eta})
  \qquad
  T_\vartheta(\varepsilon)+T_{\hat{\boldsymbol x}}(i\beta\bar{\boldsymbol\chi}\varepsilon)
  \qquad
  T_{\bar\vartheta}(\bar\varepsilon)+T_{\hat{\boldsymbol x}}(i\beta\bar\varepsilon{\boldsymbol\xi})
\end{equation}
simultaneously applied to a superindex and $\hat{\boldsymbol x}$.
The most generic form of $\mathbb Q$ associated with this symmetry has nonzero $Q_1$, $Q_\Phi$, $Q_\Pi$, and $Q_\Theta$, with $Q_\Phi=Q_\Pi$ and $Q_\Theta=\beta Q_\Pi$.
Translation in $\hat{\boldsymbol x}$ cannot be expressed as translation in the superindices, which is why this is not a supersymmetry, despite being often referred to as such in the literature (including by this author) \cite{Annibale_2003_Supersymmetric, Annibale_2003_The, Annibale_2004_Coexistence, Kent-Dobias_2023_How}.
Ghost number symmetries are preserved, since $\varpi(1)=\bar\theta_1\theta_1\bar\vartheta_1\vartheta_1$ is balanced in the superindices.

\subsubsection{Index dependence}
\label{sec:index}

Besides energy, another property of interest is the index of stationary points.
Rather than fix this directly, we fix another property that whose tuning
indirectly tunes the index density. The property we fix depends on the nature
of the problem at hand.

\paragraph{In models with a spherical constraint.}

For models with a spherical constraint, the gradient on the sphere
$\boldsymbol\nabla$ and the derivative $\boldsymbol\partial$ with respect to
$\boldsymbol x$ are different. Though the gradient must be zero at any
stationary point, the derivative can be nonzero in the radial direction,
perpendicular to the constraint surface. The normalized magnitude of the
derivative in the radial direction $\mu=\frac1N\boldsymbol
x\cdot\boldsymbol\partial H(\boldsymbol x)$ is called the \emph{radial
reaction}, since it is the magnitude of the radial force that would be necessary to
enforce the constraint at that point \cite{Folena_2020_Rethinking}. It is this radial reaction that is fixed
in order to indirectly tune the index density.

We will always enforce the spherical constraint using the method of Lagrange
multipliers. For a function defined on the sphere, the total number of stationary points is
\begin{align}
  \mathcal N[H]
  &=\int_{\mathbb R^N}d\boldsymbol x\int_\mathbb Rd\lambda\,\delta\big(\tfrac12(N-\|\boldsymbol x\|^2)\big)\,
  \delta\big(\boldsymbol\partial H(\boldsymbol x)-\lambda\boldsymbol x\big)\,
  \bigg|\det\begin{bmatrix}\boldsymbol\partial\boldsymbol\partial^TH(\boldsymbol x)-\lambda I&-\boldsymbol x\\-\boldsymbol x^T&0\end{bmatrix}\bigg|
  \\ \notag
  &=\int_{\mathbb R^{N|4}}d\boldsymbol\phi\int_{\mathbb R^{1|4}}\,d\Lambda\,
  e^{\int d1\,[H(\boldsymbol\phi(1))+\frac12\Lambda(1)(N-\|\boldsymbol\phi(1)\|^2)]}
\end{align}
The count of stationary points biased by radial reaction is instead given by
\begin{align}
  \mathcal N_\omega[H]
  &=\int_{\mathbb R^N}d\boldsymbol x\int_\mathbb Rd\lambda\,
  e^{-\omega\boldsymbol x\cdot\boldsymbol\partial H(\boldsymbol x)}
  \delta\big(\tfrac12(N-\|\boldsymbol x\|^2)\big)\,
  \delta\big(\boldsymbol\partial H(\boldsymbol x)-\lambda\boldsymbol x\big)\,
  \bigg|\det\begin{bmatrix}\boldsymbol\partial\boldsymbol\partial^TH(\boldsymbol x)-\lambda I&-\boldsymbol x\\-\boldsymbol x^T&0\end{bmatrix}\bigg|
  \\ \notag
  &=\int_{\mathbb R^{N|4}}d\boldsymbol\phi\int_{\mathbb R^{1|4}}\,d\Lambda\,
  e^{\int d1\,[H(\omega(1)\boldsymbol\phi(1))+\frac12\Lambda(1)(N-\|\boldsymbol\phi(1)\|^2)]}
\end{align}
where we have introduced the compact notation $\omega=1-\omega\varpi(1)$. The
complexity of stationary points with particular radial reaction $\mu$ can be derived from the complexity as a function of $\omega$ by a Legendre transform, or
\begin{equation}
  \Sigma(\mu)=\max_\omega(\Sigma(\omega)+\omega\mu)
\end{equation}
This gives a condition on the effective action in addition to the regular extremal ones of
\begin{equation}
  0=\frac{\partial\mathcal S_\omega}{\partial\omega}+\mu
\end{equation}
Unlike the case of biasing the energy, the resulting superspace action
explicitly breaks supersymmetry in a way that cannot be accommodated by an
alternative set of fermionic symmetries.

When the superspace Lagrange multiplier $\Lambda$ is evaluated by saddle-point
under the minimal \textsc{susy}-breaking ansatz
$\Lambda(1)=\lambda+\hat\lambda\varpi(1)$, fixing the radial reaction to $\mu$
will likewise fix $\lambda=\mu$. This is true because the scalar product
of the gradient with $\boldsymbol x$ must vanish because $\boldsymbol x$ is the normal vector to the sphere. It follows that
\begin{equation}
  0=\boldsymbol x\cdot\boldsymbol\nabla H
  =\boldsymbol x\cdot\boldsymbol\partial H(\boldsymbol x)-\lambda\|\boldsymbol x\|^2
  =\boldsymbol x\cdot\boldsymbol\partial H(\boldsymbol x)-N\lambda
\end{equation}
This coincidence leads many to fix the radial reaction by simply setting
$\lambda$ to a constant \cite{Folena_2020_Rethinking, Kent-Dobias_2023_How}.
However, this leads to problems with interpreting the other order
parameters, as we will see in Section \ref{sec:ssg}.

\paragraph{In models with single-spin confinement.}

Other models pair random coupling between spins with a single-spin confinement,
with energy of the form
\begin{equation}
  H(\boldsymbol x)=V(\boldsymbol x)+\sum_iU(x_i)
\end{equation}
with $V$ random and $U$ deterministic. In such cases, the index can be
indirectly fixed by biasing the mean value of some function $g(x_i)$ over the
single spins, or
\begin{align}
  \mathcal N_\omega
  &=\int_{\mathbb R^N}d\boldsymbol x\,e^{-\omega\sum_{i=1}^Ng(x_i)}
  \delta\big(\boldsymbol\partial H(\boldsymbol x)\big)\,
  \big|\det\boldsymbol\partial\boldsymbol\partial^TH(\boldsymbol x)\big|
  \\ \notag
  &=\int_{\mathbb R^{N|4}}d\boldsymbol\phi\,e^{
    \int d1\,H(\boldsymbol\phi(1))
    -\omega\sum_{i=1}^N\int d1\,\varpi(1)g(\phi_i(1))
  }
\end{align}
The complexity of stationary points biased by the mean value of $g$ over the
spins can then be translated into the complexity of stationary points with mean value of $g$ fixed to $\mu$ using the Legendre transform
\begin{equation}
  \Sigma(\mu)=\max_{\sigma}[\Sigma(\omega)+\omega\mu]
\end{equation}
In Section~\ref{sec:sk.model} we use this method to bias the TAP complexity of the Sherrington--Kirkpatrick model using $g(x)=1/U''(x)$, so that conditioning on $\mu$ corresponds with conditioning on the trace of the inverse Hessian of the deterministic part of the energy.

\subsubsection{Thresholds and marginal complexity}
\label{sec:threshold}

A common goal of counting stationary points conditioned on their energy is the
determination of the energy density at which typical stationary points
transition from being minima to being saddle points. This energy density is
often called the threshold energy $E_\text{th}$, and sometimes has significance
for out-of-equilibrium dynamics. The methods presented in this article cannot
distinguish between minima and saddle points with a finite number of unstable
directions, but eliding this possibility we can determine the threshold as defined by the
dominance or not of saddle points with extensive index.

On the side of the threshold with typical minima or maxima, the spectral
density of the Hessian at stationary points has zero density at zero
eigenvalue. Therefore, the \textsc{susy}-breaking order parameter $Q_\Psi$ is
zero. At the threshold, unless the
spectral density of the Hessian has a discontinuous drop, new non-zero
solutions for the value of $Q_\Psi$ emerge continuously from zero. This point
can be found by extremizing the effective action under the assumption that
$Q_\Psi=0$ under the additional condition
\begin{align} \label{eq:marginal.condition}
  0=\frac{\partial^2\mathcal S}{\partial Q_\Psi^2}\bigg|_{Q_\Psi=0}
\end{align}
All such points correspond to situations where a continuous piece of the spectral
density touches zero eigenvalue, and are therefore \emph{marginal}. Marginal
complexity can be measured by varying the index until this condition is
satisfied.

\subsubsection{Models with a signal or spike}
\label{sec:signal}

Many models of interest in inference and learning involve a disordered loss
function depending on one or more \emph{ground truth} vectors $\boldsymbol
x^*\in\mathbb R^N$ representing the solution that should be found by minimizing
the loss. This generically results in the presence of another supervector order
parameter $\mathbb M(1)=\frac1N\boldsymbol\phi(1)\cdot\boldsymbol x^*$ in the
saddle-point approximation resulting from the overlap of the superfield with
the ground truth vector. The minimally \textsc{susy} breaking form of such a
supervector contains only two nonzero terms, with
\begin{equation}
  \mathbb M(1)=m+\hat m\varpi(1)
\end{equation}
Each of these can be interpreted in terms of the anisotropy of
stationary points along the direction corresponding to the ground truth.
The order parameter
\begin{equation}
  m=\frac1N\langle\boldsymbol x\rangle\cdot\boldsymbol x^*
\end{equation}
is the \emph{magnetization}, the overlap between the position and the ground
truth averaged over stationary points. The order parameter
\begin{equation}
  \hat m=\frac1N\langle\hat{\boldsymbol x}\rangle\cdot\boldsymbol x^*
  =\boldsymbol x^*\cdot\frac{\partial\Sigma}{\partial\boldsymbol h}\bigg|_{\boldsymbol h=0}
\end{equation}
is the rate change of the complexity with respect to a linear perturbation of
the potential along the direction of the ground truth.

\section{Application to models}
\label{sec:examples}

\subsection{Gaussian field in Euclidean space}

Consider the number of stationary points of a Gaussian field in a quadratic
potential in $\mathbb R^N$. The Hamiltonian is
\begin{equation}
  H(\boldsymbol x)=\frac\mu2\|\boldsymbol x\|^2+V(\boldsymbol x)
\end{equation}
where $\mu$ is the stiffness of the potential and $V$ is a Gaussian random field with zero mean and covariance
\begin{align} \label{eq:gaussian.covariance}
  \overline{V(\boldsymbol x)V(\boldsymbol y)}
  &=N\Gamma\left(\frac1{2N}\|\boldsymbol x-\boldsymbol y\|^2\right)
  =N\Gamma\left(\frac1{2N}(\|\boldsymbol x\|^2+\|\boldsymbol y\|^2)-\frac1N\boldsymbol x\cdot\boldsymbol y\right)
\end{align}
depending only on the distance between two points \cite{Bray_2007_Statistics}.
The number of stationary points averaged over the disordered field is
\begin{equation}
  \overline{\mathcal N[H]}
  =\overline{\int d\boldsymbol\phi\,e^{\frac\mu2\int d1\,\boldsymbol\phi(1)\cdot\boldsymbol\phi(1)+\int d1\,V(\boldsymbol\phi(1))}}
  =\int d\boldsymbol\phi\,e^{\frac\mu2\int d1\,\boldsymbol\phi(1)\cdot\boldsymbol\phi(1)+\frac12\int d1\,d2\,\overline{V(\boldsymbol\phi(1))V(\boldsymbol\phi(2))}}
\end{equation}
Because of the form of the covariance \eqref{eq:gaussian.covariance}, the
integrand only depends on $\boldsymbol\phi$ through the superoverlap $\mathbb
Q(1,2)\equiv\frac1N\boldsymbol\phi(1)\cdot\boldsymbol\phi(2)$. We are therefore
inspired to make the change of variables in the integral from $\boldsymbol\phi$
to $\mathbb Q$. This translation can be make with the change of measure $d\boldsymbol\phi\mapsto d\mathbb Q\,(\operatorname{sdet}\mathbb Q)^\frac N2$, and yields
\begin{equation}
  \overline{\mathcal N[H]}=\int_{\mathbb R^{1\mid4\times 4}} d\mathbb Q\,e^{N\mathcal S(\mathbb Q)}
\end{equation}
where we have defined an effective action $\mathcal S$ by
\begin{align}
  \mathcal S(\mathbb Q)
  &=\frac\mu2\int d1\,\mathbb Q(1,1)
  +\frac12\int d1\,d2\,\Gamma\left(
    \tfrac12\mathbb Q(1,1)+\tfrac12\mathbb Q(2,2)-\mathbb Q(1,2)
  \right)
  +\frac12\log\operatorname{sdet}\mathbb Q
\end{align}
Based on the considerations of Section \ref{sec:saddle.point.4}, we expect $\mathbb Q$ to condense only in a restricted subspace of the space of all superoperators. Representing $\mathbb Q$ in this subspace as
\begin{equation}
  \mathbb Q(1,2)=Q_1+Q_\Phi\Phi(1,2)+iQ_\Psi\Psi(1,2)+Q_{\Pi}\Pi(1,2)+Q_\Theta\Theta(1,2)
\end{equation}
and applying the properties of these operators presented in Sections
\ref{sec:algebra} and \ref{sec:identities} results in
\begin{align}
  \mathcal S(\mathbb Q)
  &=\mu(Q_\Pi-Q_\Phi)
  -\frac12(\Gamma'(0)Q_\Theta+\Gamma''(0)Q_\Psi^2)
  +\frac12\log\frac{Q_\Pi^2-Q_1Q_\Theta}{Q_\Phi^2+Q_\Psi^2}
\end{align}
We can evaluate the integral over $\mathbb Q$ as a saddle point in these
coefficients when $N$ is very large. There are two solutions to the extremal
conditions on the effective action. One solution is
\begin{align}
  Q_1=-\frac{\Gamma'(0)}{\mu^2}
  &&
  Q_\Phi=-\frac1\mu
  &&
  Q_\Psi=0
  &&
  Q_\Pi=-\frac1\mu
  &&
  Q_\Theta=0
\end{align}
which is supersymmetric because $Q_\Phi=Q_\Pi$ and $Q_\Psi=Q_\Theta=0$. The
supersymmetric solution has $\Sigma=\mathcal S(\mathbb Q^*)=0$, and corresponds to the
regime where $\mu$ is so large that only the one minimum of the confining
potential is present. The other solution is well-defined for
$\mu^2<\Gamma''(0)$ and is
\begin{align}
  Q_1=-\frac{\Gamma'(0)}{\mu^2}
  &&
  Q_\Phi=-\frac\mu{\Gamma''(0)}
  &&
  Q_\Psi=\pm\frac{\sqrt{\Gamma''(0)-\mu^2}}{\Gamma''(0)}
  &&
  Q_\Pi=-\frac1\mu
  &&
  Q_\Theta=0
\end{align}
which spontaneously breaks supersymmetry in our restricted subspace because
$Q_\Phi\neq Q_\Pi$ and $Q_\Psi\neq0$. The supersymmetry breaking solution has $\Sigma=\mathcal S(\mathbb Q^*)>0$, and corresponds to the regime where many stationary points exist. The topological
trivialization transition thus corresponds to spontaneous supersymmetry breaking
via condensation along $Q_\Psi$
below the critical mass $\mu^2=\Gamma''(0)$.

\subsection{Spherical spin glasses}
\label{sec:ssg}

The spherical spin glasses are Gaussian random fields on the surface of the
sphere defined for $\boldsymbol x\in\mathbb R^N$ by $\|\boldsymbol x\|^2=N$ and
are a canonical toy model for disordered mean-field theory
\cite{Kirkpatrick_1987_p-spin-interaction, Crisanti_1992_The,
Cavagna_1997_Structure, Cavagna_1997_An, Castellani_2005_Spin-glass,
Crisanti_2004_Spherical, Crisanti_2006_Spherical, Folena_2020_Rethinking}. The
Hamiltonian is a Gaussian random field $H$ with zero mean and covariance
\begin{equation}
  \overline{H(\boldsymbol x)H(\boldsymbol y)}=Nf\left(\frac{\boldsymbol x\cdot\boldsymbol y}N\right)
\end{equation}
depending only on the scalar product between configurations. The spherical
constraint can be enforced by Lagrange multipliers as described in Section
\ref{sec:lagrange}, allowing us to write the number of stationary points biased
by inverse temperature $\beta$ as
\begin{align}
  \mathcal N_\beta[H]
  =\int_{\mathbb R^{N\mid4}}d\boldsymbol\phi\int_{\mathbb R^{1\mid4}} d\Lambda\,e^{\int d1\,[\frac12\Lambda(1)(N-\boldsymbol\phi(1)\cdot\boldsymbol\phi(1))+\beta(1)H(\boldsymbol\phi(1))]}
\end{align}
Averaging over the noise and making the change of variables from $\boldsymbol\phi$ to $\mathbb Q$, this becomes
\begin{equation}
  \overline{\mathcal N_\beta[H]}
  =\int_{\mathbb R^{1\mid4\times4}}d\mathbb Q\int_{\mathbb R^{1\mid4}}
  d\Lambda\,e^{N\mathcal S_{\beta}(\mathbb Q,\Lambda)}
\end{equation}
where we have defined the effective action
\begin{equation}
  \mathcal S_\beta(\mathbb Q,\Lambda)
  =\frac12\int d1\,\Lambda(1)(1-\mathbb Q(1,1))
  +\frac12\int d1\,d2\,\beta(1)\beta(2)f(\mathbb Q(1,2))
  +\frac12\operatorname{sdet}\mathbb Q
\end{equation}
The extremal conditions can be enforced at the level of the superoperators. The resulting extremal conditions are
\begin{align}
  0
  &=2\frac{\partial\mathcal S_\beta(\mathbb Q,\Lambda)}{\partial\mathbb Q(1,2)}
  =\frac12(\Lambda(1)+\Lambda(2))\delta(1,2)+\beta(1)\beta(2)f'(\mathbb Q(1,2))
  +\mathbb Q^{-1}(1,2)
  \\
  0
  &=2\frac{\partial\mathcal S_\beta(\mathbb Q,\Lambda)}{\partial\Lambda(1)}
  =\mathbb Q(1,1)-1
\end{align}
When the minimally \textsc{susy} breaking forms
\begin{align}
  \mathbb Q(1,2)&=Q_1+Q_\Phi\Phi(1,2)+iQ_\Psi\Psi(1,2)+Q_{\Pi}\Pi(1,2)+Q_\Theta\Theta(1,2)
  \\
  \Lambda(1)&=\lambda+\hat\lambda\varpi(1)
\end{align}
are substituted for the order parameters, the equations become
\begin{align}
  0&=-\big(\lambda+\tfrac12\hat\lambda\Pi(1,2)\big)
  \big(\Phi(1,2)+\Pi(1,2)\big)
  +\big(1-\beta\Pi(1,2)+\beta^2\Theta(1,2)\big)
  \\ \notag
   &\hspace{8em}\times
   \Big[f'(Q_1)+f''(Q_1)Q_\Phi\Phi(1,2)+f''(Q_1)Q_\Psi\Psi(1,2)+f''(Q_1)Q_\Pi\Pi(1,2)
   \\ \notag
   &\hspace{12em}
   +(f''(Q_1)Q_\Theta+f'''(Q_1)(Q_\Pi^2-Q_\Phi^2-Q_\Psi^2))\Theta(1,2)\Big]
   +\mathbb Q^{-1}(1,2)
   \\
  0&=Q_1-1+2(Q_\Pi-Q_\Phi)\varpi(1)
\end{align}
The resulting equations must be satisfied independently along each direction in superspace. Expanding products and using the algebra defined in Section \ref{sec:algebra} results in the set of equations
\begin{align}
  0&=f'(Q_1)-\frac{Q_1}{Q_\Pi^2-Q_1Q_\Theta}
  \hspace{3em}
  0=-\lambda+f''(Q_1)Q_\Phi+\frac{Q_\Phi}{Q_\Phi^2+Q_\Psi^2}
  \\ \notag
  0&=f''(Q_1)Q_\Psi-\frac{Q_\Psi}{Q_\Phi^2+Q_\Psi^2}
  \hspace{2.5em}
  0=-\lambda-\beta f'(Q_1)+f''(Q_1)Q_\Pi+\frac{Q_\Pi}{Q_\Pi^2-Q_1Q_\Theta}
  \\ \notag
  0&=-\hat\lambda+\beta^2f'(Q_1)+f''(Q_1)Q_\Theta+f'''(Q_1)(Q_\Pi^2-Q_\Phi^2-Q_\Psi^2)+2\beta f''(Q_1)Q_\Pi-\frac{Q_\Theta}{Q_\Pi^2-Q_1Q_\Theta}
  \\ \notag
  0&=Q_1-1
  \hspace{9.75em}
  0=2(Q_\Pi-Q_\Phi)
\end{align}
The final equations imply $Q_1=1$ and $Q_\Pi=Q_\Phi$, conditions that are
generically found under spherical constraint. Eliminating $Q_1$, $Q_\Pi$,
and $\hat\lambda$, we then write
\begin{align}
  0&=f'(1)-\frac1{Q_\Phi^2-Q_\Theta}
  \hspace{5.75em}
  0=-\lambda+f''(1)Q_\Phi+\frac{Q_\Phi}{Q_\Phi^2+Q_\Psi^2}
  \\ \notag
  0&=f''(1)Q_\Psi-\frac{Q_\Psi}{Q_\Phi^2+Q_\Psi^2}
  \hspace{4em}
  0=-\lambda-\beta f'(1)+f''(1)Q_\Phi+\frac{Q_\Phi}{Q_\Phi^2-Q_\Theta}
\end{align}
We add to these conditions one further to fix the energy density $E$ via a Legendre transform from $\beta$, which requires $\beta$ to verify
\begin{equation}
  0=E+\frac{\partial\mathcal S_\beta(\mathbb Q,\Lambda)}{\partial\beta}
  =E+\beta f(Q_1)-Q_\Pi f'(Q_1)
  =E+\beta f(1)-Q_\Phi f'(1)
\end{equation}
Solving these equations produces the annealed complexity of the spherical spin
glasses. Because we have set the energy by hand, there is no regime where the
result is supersymmetric, but there is a regime where the result maintains the
BRST symmetry described in Section \ref{sec:energy} and a regime where
this symmetry is broken. In the former regime the \textsc{susy}-breaking order parameter $Q_\Psi$ is zero, while in the latter it condenses with value
\begin{equation}
  Q_\Psi=\sqrt{\frac1{f''(1)}-\left(\frac{Ef'(1)}{f'(1)^2+f(1)(f''(1)-f'(1))}\right)^2}
\end{equation}
The boundary between these regimes occurs at $E=E_\text{th}$, the threshold
energy, above which typical stationary points are saddles with extensive index
and below which typical stationary points are minima. As expected from the
interpretation of $Q_\Psi$, the symmetry breaking occurs when the spectrum of
typical stationary points develops support over zero.

The index density can be fixed as described in Section~\ref{sec:index}.
Inserting the operator $\omega(1)=1-\omega\varpi(1)$ to bias the count by radial reaction and repeating the
steps above results in the revised action
\begin{equation}
  \mathcal S_{\beta,\omega}(\mathbb Q,\Lambda)
  =\frac12\int d1\,\Lambda(1)(1-\mathbb Q(1,1))
  +\frac12\int d1\,d2\,\beta(1)\beta(2)f(\omega(1)\omega(2)\mathbb Q(1,2))
  +\frac12\operatorname{sdet}\mathbb Q
\end{equation}
Following the same steps as above results in revised equations for $Q_\Theta$, $Q_\Phi$, $Q_\Psi$, $\lambda$, and $\beta$ of the form
\begin{align}
  0&=f'(1)-\frac1{Q_\Phi^2-Q_\Theta}
  \hspace{5.75em}
  0=-\lambda+f''(1)Q_\Phi+\frac{Q_\Phi}{Q_\Phi^2+Q_\Psi^2}
  \\ \notag
  0&=f''(1)Q_\Psi-\frac{Q_\Psi}{Q_\Phi^2+Q_\Psi^2}
  \hspace{4em}
  0=-\lambda-(\beta+\omega)f'(1)+(Q_\Phi-\omega)f''(1)+\frac{Q_\Phi}{Q_\Phi^2-Q_\Theta}
  \\ \notag
  0
  &=E+\beta f(1)+(\omega-Q_\Phi)f'(1)
\end{align}
to which must be added the equation for the Legendre transform from $\omega$ to the radial reaction $\mu$,
\begin{align}
  0
  =\mu+\frac{\partial\mathcal S_{\beta,\omega}(\mathbb Q,\Lambda)}{\partial\omega}
  =\mu+\beta f'(1)+(\omega-Q_\Phi)[f''(1)+f'(1)]
\end{align}
These equations can be solved to give the complexity as a function of energy density $E$ and radial reaction $\mu$.
Marginal stationary points have the added constraint
\begin{equation}
  0=\frac{\partial^2\mathcal S}{\partial Q_\Psi^2}\bigg|_{Q_\Psi=0}
  =f''(1)-\frac1{Q_\Phi^2}
\end{equation}
due to their position at the \textsc{susy}-breaking phase transition line.
Solving this whole system gives
\begin{align}
  &Q_\Theta=\frac1{f''(1)}-\frac1{f'(1)}
  \hspace{5em}
  Q_\Phi=Q_\Pi=\mp\frac1{\sqrt{f''(1)}}
  \hspace{5em}
  \lambda=\mu=\mp\sqrt{4f''(1)}
  \\
  &\beta=\frac{-(f'(1)+f''(1))E\mp f'(1)\sqrt{4f''(1)}}{f(1)(f'(1)+f''(1))-f'(1)^2}
  \qquad
  \omega=\frac{-f'(1)E\mp f(1)\sqrt{4f''(1)}}{f(1)(f'(1)+f''(1))-f'(1)^2}
  \mp\frac1{\sqrt{f''(1)}}
\end{align}
The result reproduces the expected value of the radial reaction $\mp\sqrt{4f''(1)}$ for marginal stationary points
in the spherical spin glasses, with the minus branch describing minima and the
plus one maxima. Note that this approach gives the appropriate susceptibility
$-Q_\Pi=\frac1{\sqrt{f''(1)}}$ for the marginal minima, something not done when
$\lambda$ is fixed by hand in the calculation.

\subsection{Least squares \& canyon landscapes}

The method developed here works equally well in non-Gaussian random landscapes.
Perhaps the simplest such landscape is that resulting from taking the sum of
squared Gaussian fields restricted to the hypersphere. This model has been used
to study confluent tissues, least squares optimization, continuous
constraint satisfaction, and two-layer neural networks \cite{Fyodorov_2019_A,
  Fyodorov_2020_Counting, Fyodorov_2022_Optimization, Tublin_2022_A,
  Kamali_2023_Dynamical, Kamali_2023_Stochastic, Urbani_2023_A,
  Urbani_2024_Statistical, Urbani_2024_Quantum, Montanari_2023_Solving,
  Montanari_2024_On, Montanari_2025_Dynamical, Kent-Dobias_2024_Conditioning,
Kent-Dobias_2025_On}.
The Hamiltonian is given for $\boldsymbol x\in\mathbb R^N$ satisfying the
spherical constraint $\|\boldsymbol x\|^2=N$ by
\begin{equation}
  H(\boldsymbol x)=\frac12\sum_{\mu=1}^M(V^\mu(\boldsymbol x)-V_0)^2
\end{equation}
where the target $V_0$ is a fixed parameter and the $M$ functions $V^\mu$ are
Gaussian random fields with zero mean and covariance
\begin{equation}
  \overline{V^\mu(\boldsymbol x)V^\nu(\boldsymbol y)}
  =N\delta^{\mu\nu}f\left(\frac{\boldsymbol x\cdot\boldsymbol y}N\right)
\end{equation}
This model has novel structure: when the load $\alpha=M/N$ is less than one,
there is the possibility of finding configurations where $H(\boldsymbol x)=0$,
and moreover when they exist these configurations form a smooth manifold. The
number of stationary points biased by their energy and by their radial reaction
is given by
\begin{equation}
  \mathcal N_{\beta,\omega}[H]
  =\int_{\mathbb R^{N|4}}d\boldsymbol\phi
  \int_{\mathbb R^{1|4}}d\Lambda\,e^{\int d1\,[\frac12\Lambda(1)(N-\boldsymbol\phi(1)\cdot\boldsymbol\phi(1))+\frac12\beta(1)\sum_{\mu=1}^M(V^\mu(\omega(1)\boldsymbol\phi(1))-V_0)^2]}
\end{equation}
We insert $M$ $\delta$ functions fixing $v^\mu(1)=V^\mu(\omega(1)\boldsymbol\phi(1))-V_0$ for
each $\mu$, and represent them in their Fourier form. This gives
\begin{align}
  \mathcal N_{\beta,\omega}[H]
  &=\int_{\mathbb R^{N|4}}d\boldsymbol\phi
  \int_{\mathbb R^{1|4}}d\Lambda\,e^{\int d1\,\frac12\Lambda(1)(N-\boldsymbol\phi(1)\cdot\boldsymbol\phi(1))}
  \\ \notag
  &\hspace{6em}
  \times\prod_{\mu=1}^M\int_{\mathbb R^{1|4}}dv^\mu\,d\hat{v}^\mu\,e^{\int d1\,[\frac12\beta(1)v^\mu(1)^2+i\hat v^\mu(1)(V^\mu(\omega(1)\boldsymbol\phi(1))-V_0-v^\mu(1))]}
\end{align}
The random Gaussian fields can be averaged over, which makes each integral over
$\mu$ identical. This yields
\begin{align}
  \overline{\mathcal N_{\beta,\omega}[H]}
  &=\int_{\mathbb R^{N|4}}d\boldsymbol\phi
  \int_{\mathbb R^{1|4}}d\Lambda\,e^{\int d1\,\frac12\Lambda(1)(N-\boldsymbol\phi(1)\cdot\boldsymbol\phi(1))}
  \\
  &\hspace{6em}
  \times\left(\int_{\mathbb R^{1|4}}dv\,d\hat{v}\,e^{\int d1\,[\frac12\beta(1)v(1)^2-i\hat v(1)(V_0+v(1))]-\frac12\,d1\,d2\,\hat v(1)\hat v(2)f\big(\omega(1)\omega(2)\frac{\boldsymbol\phi(1)\cdot\boldsymbol\phi(2)}N\big)}\right)^M
  \notag
\end{align}
Introducing the superoverlap order parameter $\mathbb Q(1,2)=\frac1N\boldsymbol\phi(1)\cdot\boldsymbol\phi(2)$, this gives
\begin{align}
  \overline{\mathcal N_{\beta,\omega}[H]}
  &=\int_{\mathbb R^{1|4\times 4}}d\mathbb Q
  \int_{\mathbb R^{1|4}}d\Lambda\,e^{N\int d1\,\frac12\Lambda(1)(1-\mathbb Q(1,1))+N\frac12\operatorname{sdet}\mathbb Q}
  \\
  &\hspace{2pc}
  \times\left(\int_{\mathbb R^{1|4}}dv\,d\hat{v}\,e^{\int d1\,[\frac12\beta(1)v(1)^2-i\hat v(1)(V_0+v(1))]-\frac12\,d1\,d2\,\hat v(1)\hat v(2)f(\omega(1)\omega(2)\mathbb Q(1,2))}\right)^M
  \notag
\end{align}
The integrals over $v$ and $\hat v$ are Gaussian and can be evaluated explicitly, which gives
\begin{equation}
  \overline{\mathcal N_{\beta,\omega}[H]}
  =\int_{\mathbb R^{1|4\times 4}}d\mathbb Q\int_{\mathbb R^{1|4}}d\Lambda\,
  e^{N\mathcal S_{\beta,\omega}(\mathbb Q,\Lambda)}
\end{equation}
with effective action
\begin{align}
  \mathcal S_{\beta,\omega}(\mathbb Q,\Lambda)
  =\frac12\int d1\,\Lambda(1)(1-\mathbb Q(1,1))
  +\frac12\operatorname{sdet}\mathbb Q
  -\frac12\alpha\operatorname{sdet}\mathbb A
  +\frac12\alpha V_0^2\int d1\,d2\,\beta(1)\mathbb A^{-1}(1,2)
\end{align}
where we have defined the load $\alpha=M/N$ and the superoperator
\begin{equation}
  \mathbb A(1,2)=\delta(1,2)-\beta(1)f(\omega(1)\omega(2)\mathbb Q(1,2))
\end{equation}
The extremal conditions on the Lagrange multiplier $\Lambda$ dictate that $Q_1=1$ and
$Q_\Pi=Q_\Phi$, just as in the spherical spin glasses. The remaining action can
be evaluated explicitly for the minimally supersymmetry breaking $\mathbb Q$ using the rules of Sections \ref{sec:algebra} and \ref{sec:identities},
which give
\begin{align}
  &\mathcal S_{\beta,\omega}(Q_\Phi,Q_\Theta,Q_\Psi)
  =\frac12\log\frac{Q_\Phi^2-Q_\Theta}{Q_\Phi^2+Q_\Psi^2}
  \\
  &\qquad-\frac12\alpha\log\left[
    \frac{
      1+(\beta-Q_\Psi^2f'')f+(\omega-Q_\Theta f)f'+(\omega-Q_\Phi)^2f'^2
      -(\omega-2Q_\Phi)(\omega(f''+f')f-f')
    }{
      (1-Q_\Phi f')^2+(Q_\Psi f')^2
    }
  \right]
  \notag \\
  &\qquad-\frac12\alpha V_0^2\frac{
    \beta^2f+\beta-Q_\Psi^2f''-\omega(\omega-2Q_\Phi)(f'+f'')-Q_\Theta f'
  }{
    1+(\beta-Q_\Psi^2f'')f+(\omega-Q_\Theta f)f'+(\omega-Q_\Phi)^2f'^2
    -(\omega-2Q_\Phi)(\omega(f''+f')f-f')
  }
  \notag
\end{align}
where $f=f(1)$, $f'=f'(1)$, and $f''=f''(1)$.
This expression extremized over the remaining order parameters and with the
Legendre transform from $\beta$ to energy density $E$ and from $\omega$ to
radial reaction $\mu$ gives the complexity as a function of energy and radial reaction.

\begin{figure}
  \centering
  \includegraphics{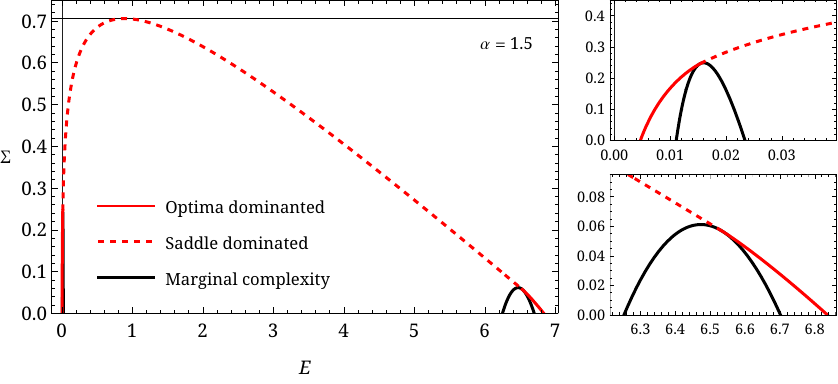}
  \caption{
    \textbf{Total complexity and marginal complexity of underparameterized nonlinear least squares.}
    Complexities as a function of energy for the nonlinear least squares model
    with $f(q)=q^2+q^3$, $\alpha=1.5$, and $V_0=0$. Total complexity of
    stationary points is shown in red, with the line format indicating whether
    most stationary points are optima ($Q_\Psi=0$) or saddles ($Q_\Psi>0)$.
    Complexity of marginal optima is shown in black. Plots on the right show
    detail around the minimum and maximum energies.
  } \label{fig:ls.complexity}
\end{figure}

When $V_0=0$, the maximum complexity is either zero or
\begin{equation} \label{eq:total.complexity.ls}
  \Sigma=
  \begin{cases}
    \frac12\log\left(\frac{f'(1)}{f(1)}+\frac{f''(1)}{f'(1)}\right)
    &
    \alpha\geq\left(1+\frac{f(1)f''(1)}{f'(1)^2}\right)^{-1}
    \\
    \frac12\left(\log\frac{f''(1)}{(1-\alpha)f'(1)}-\alpha\log\frac{\alpha f(1)f''(1)}{(1-\alpha)f'(1)^2}\right)
    &
    \alpha<\left(1+\frac{f(1)f''(1)}{f'(1)^2}\right)^{-1}
  \end{cases}
\end{equation}
where the two cases depend on whether the maximum complexity occurs at $E>0$ or $E=0$.
The zero complexity solution is supersymmetric with $Q_\Theta=Q_\Psi=0$ and
\begin{equation}
  Q_\Phi=\left(f'(1)\pm\sqrt{\alpha f(1)f'(1)}\right)^{-1}
\end{equation}
while the \textsc{susy} breaking solution for $E>0$ has $Q_\Phi=0$ and
\begin{align}
  Q_\Theta=-\frac1{f(1)f'(1)}\frac{f'(1)^2+f(1)f''(1)}{(\alpha-1)f'(1)^2+\alpha f(1)f''(1)}
  &&
  Q_\Psi=\left((\alpha-1)f'(1)^2+\alpha f(1)f''(1)\right)^{-\frac12}
\end{align}
As $E$ approaches zero both $Q_\Theta$ and $Q_\Psi$ diverge. Writing $Q_\Theta=\delta_\Theta E^{-1}$ and $Q_\Psi=\delta_\Psi E^{-\frac12}$, the equations can be solved in the zero-energy limit to give
\begin{align}
  \delta_\Theta=-\frac1{2f'(1)}
  &&
  \delta_\Psi=\sqrt{\frac{1-\alpha}{2f''(1)}}
  &&
  Q_\Phi=\frac{(1-\alpha)f'(1)^2-\alpha f(1)f''(1)}{(1-\alpha)f'(1)[f'(1)^2+f(1)(f''(1)-f'(1)]}
\end{align}

\begin{figure}
  \centering
  \includegraphics{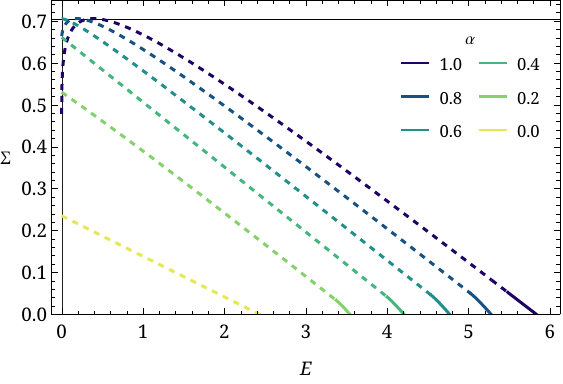}
  \caption{
    \textbf{Total complexity of overparameterized nonlinear least squares.}
    Complexities as a function of energy for the nonlinear least squares model
    with $f(q)=q^2+q^3$, $V_0=0$, and a variety of loads $\alpha$. The
    horizontal line shows the maximum complexity at saturation in $\alpha$ which
    matches the maximum complexity when it occurs at positive energy $E$. Solid lines depict regimes dominated by optima ($Q_\Psi=0$) and dashed lines depict regimes with a nonzero spectral density at zero eigenvalue ($Q_\Psi>0$).
    Nonzero complexity in the ground state $E=0$ is characteristic of canyon
    landscapes.
  } \label{fig:ls.complexity.under}
\end{figure}

It is straightforward to use this action to evaluate the complexity as a
function of energy density. An example is shown in
Fig~\ref{fig:ls.complexity}, where the total complexity and marginal
complexity are plotted for a specific ensemble of canyon landscapes. A thin
horizontal line also depicts the maximum complexity predicted by
\eqref{eq:total.complexity.ls}, which coincides with the peak of the complexity
curve. The total and marginal complexity of minima in the same model was
analyzed in Ref~\cite{Kent-Dobias_2024_Conditioning}.

As the load $\alpha$ is varied, the maximum complexity shifts from occurring at
nonzero to zero energy density. This trend is seen in
Fig~\ref{fig:ls.complexity.under}, which shows total complexity versus energy
density curves for the same model as Fig~\ref{fig:ls.complexity} as the load
$\alpha$ is brought towards zero. In all examples there is a residual
complexity at zero energy density, but the complexity is only maximized there
for load $\alpha$ under $\alpha\simeq0.6098$. Note that the annealed complexity
always predicts that zero energy density stationary points have a nonzero
spectral density at zero eigenvalue, meaning that $Q_\Psi>0$.

\begin{figure}
  \centering
  \includegraphics{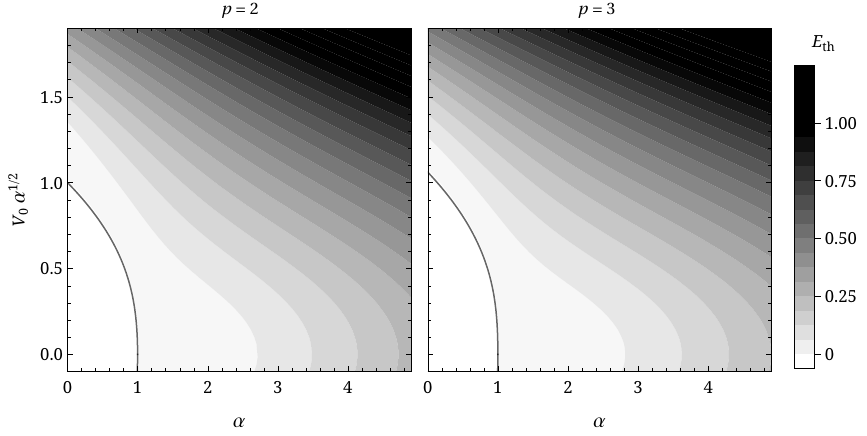}
  \caption{
    \textbf{Threshold energy of nonlinear least squares as a function of load and target.}
    Threshold energy $E_\text{th}$ in nonlinear least squares as a function of
    load $\alpha$ and target $V_0$ for two different disorder covariances
    $f(q)=\frac12q^p$. When the threshold energy is predicted to fall below
    zero, stationary points with spectral weight on zero eigenvalue exist all
    the way to the ground state.
  } \label{fig:ls.eth}
\end{figure}

Of particular interest is the value of the threshold energy $E_\text{th}$,
defined as the energy density at which the total complexity transitions from
being optima dominated to being saddle dominated. As explained in Section \ref{sec:threshold}, the threshold can be found by solving the extremal problem for the effective action under the conditions that $Q_\Psi=0$ and
\begin{equation}
  0=\frac{\partial^2\mathcal S}{\partial Q_\Psi^2}\bigg|_{Q_\Psi=0}
\end{equation}
Plots of the threshold energy in two different least squares models with
covariance $f(q)=\frac12q^p$ are shown in Fig~\ref{fig:ls.eth} as a function of
load $\alpha$ and target $V_0$. For certain parameter combinations the
threshold is predicted at positive energy density, while for others it is
predicted to be negative, which implies that the threshold is not present in
the annealed complexity and saddles (or stationary points with a spectral
density with nonzero mass on zero eigenvalue) are predicted all the way to zero
energy density. The transition between these two regimes is of particular
interest. It can be found by looking for the value of the load $\alpha$ at
which $E_{\text{th}}=0$, which is given by
\begin{equation}
  \alpha(E_\text{th}=0)
  =\frac{f'(1)[f'(1)-f(1)]+f(1)f''(1)\frac{2V_0^2+f(1)}{V_0^2+f(1)}}{
    f'(1)[f'(1)-f(1)]+f''(1)(V_0^2\frac{f(1)}{f'(1)}+[V_0^2+f(1)])
  }
\end{equation}
This value is plotted as the thin black line in Fig~\ref{fig:ls.eth}.

\subsection{The perceptron}

The perceptron is a model of a single neuron and is the basis for mean-field studies of neural networks \cite{Rosenblatt_1958_The}. Given $M$ patterns
$\boldsymbol\xi^\mu\in\mathbb R^N$, the perceptron with weights $\boldsymbol x\in\mathbb R^N$
has energy
\begin{equation}
  H(\boldsymbol x)
  =\sum_{\mu=1}^M\sigma(\boldsymbol x\cdot\boldsymbol\xi^\mu)
\end{equation}
where $\sigma:\mathbb R\to\mathbb R$ is the activation function. A canonical
example of an activation is the rectified linear unit with margin $\kappa$,
which is $\sigma(y)=(\kappa-y)\Theta(\kappa-y)$, and gives zero energy for all
patterns where the scalar product between the weights and the pattern is
sufficiently positive. In the simplest case one takes the patterns to be
independent and mean-zero Gaussian random variables, and the perceptron's
properties correspond to the storage capacity of a neuron to random data
\cite{Gardner_1988_The, Gardner_1988_Optimal, Gardner_1989_Three}. We take the
weights to be constrained to lie on the sphere $\|\boldsymbol x\|^2=N$. In the
spherical version of the perceptron with nonnegative margin, the energy is a
convex function and the landscape is trivial. Rich structure emerges when the margin is negative \cite{Stojnic_2013_Negative, Franz_2017_Universality,
Annesi_2023_Star-shaped, Annesi_2025_Exact, Kent-Dobias_2026_Structure}. The
complexity of the perceptron has been studied previously with uncorrelated
\cite{Maillard_2020_Landscape} and correlated \cite{Tsironis_2025_Landscape}
patterns, but the resulting expressions are very difficult to explicitly
evaluate.

The energy- and radial reaction-weighted number of stationary points of the perceptron is given by
\begin{equation}
  \mathcal N_{\beta,\omega}[H]
  =\int_{\mathbb R^{N|4}} d\boldsymbol\phi\int_{\mathbb R^{1|4}}d\Lambda\,
  e^{\int d1\,\frac12\Lambda(1)(\boldsymbol\phi(1)\cdot\boldsymbol\phi(1)-N)}
  \prod_{\mu=1}^M
  e^{
    \int d1\,\beta(1)\sigma(\omega(1)\boldsymbol\phi(1)\cdot\boldsymbol\xi^\mu)
  }
\end{equation}
Defining the super-preactivations
$\nu^\mu(1)=\boldsymbol\phi(1)\cdot\boldsymbol\xi^\mu$ using $\delta$ functions
and then representing the $\delta$ functions in their Fourier form with auxiliary fields $\hat\nu^\mu(1)$, we have
\begin{align}
  \mathcal N_{\beta,\omega}[H]
  &=\int_{\mathbb R^{N|4}} d\boldsymbol\phi\int_{\mathbb R^{1|4}}d\Lambda\,
  e^{\int d1\,\frac12\Lambda(1)(\boldsymbol\phi(1)\cdot\boldsymbol\phi(1)-N)}
  \\ \notag
  &\hspace{4pc}
  \times\prod_{\mu=1}^M\int_{\mathbb R^{1|4}}dv^\mu\,d\hat v^\mu\,
  e^{
    \int d1\,[\beta(1)\sigma(\omega(1)v^\mu(1))+i\hat v^\mu(1)(\boldsymbol\phi(1)\cdot\boldsymbol\xi^\mu-v^\mu(1))]
  }
\end{align}
In this form the average over random patterns can be made. Afterwards, each factor in the product over patterns becomes identical, and we find
\begin{align}
  \overline{\mathcal N_{\beta,\omega}[H]}
  &=\int_{\mathbb R^{N|4}} d\boldsymbol\phi\int_{\mathbb R^{1|4}}d\Lambda\,
  e^{\int d1\,\frac12\Lambda(1)(\boldsymbol\phi(1)\cdot\boldsymbol\phi(1)-N)}
  \\
  &\qquad\times\left(
    \int_{\mathbb R^{1|4}}dv\,d\hat v\,
  e^{
    \int d1\,[\beta(1)\sigma(\omega(1)v(1))+i\hat v(1)v(1)]
    -\frac12\int d1\,d2\,\hat v(1)\hat v(2)\frac{\boldsymbol\phi(1)\cdot\boldsymbol\phi(2)}N
  }
  \right)^M
  \notag
\end{align}
Making the Gaussian integral over the auxiliary fields $\hat v$ and introducing the superoverlap $\mathbb
Q=\frac1N\boldsymbol\phi(1)\cdot\boldsymbol\phi(2)$, the energy- and radial reaction-biased count of
stationary points is given by the exponential of the extremal value of the
effective action
\begin{align}
  \mathcal S_{\beta,\omega}(\mathbb Q,\Lambda)
  &=\frac12\int d1\,\Lambda(1)(\mathbb Q(1,1)-1)+\frac12\log\operatorname{sdet}\mathbb Q
  \\
  &\qquad+\alpha\log\int_{\mathbb R^{1|4}} dv\,
    (\operatorname{sdet}\mathbb Q)^{-\frac12}e^{-\frac12\int d1\,d2\,v(1)\mathbb Q^{-1}(1,2)v(2)+\int d1\,\beta(1)\sigma(\omega(1)v(1))}
  \notag
\end{align}
The extremal conditions on the Lagrange multiplier $\Lambda$ give $Q_1=1$ and $Q_\Pi=Q_\Phi$, as usual
for spherically constrained problems. Inserting these conditions and using the
formulae of Section \ref{sec:identities}, we arrive at an effective action in
three order parameters
\begin{align} \label{eq:action.perceptron.general}
  &\mathcal S_{\beta,\omega}(Q_\Phi,Q_\Theta,Q_\Psi)
  =\frac12\log\frac{Q_\Phi^2-Q_\Theta}{Q_\Phi^2+Q_\Psi^2}
  \\
  &\hspace{4pc}+\alpha\log\int_{\mathbb R} dy\,
  \sqrt{\frac{(1-Q_\Phi\sigma''(y))^2+Q_\Psi^2\sigma''(y)^2}{2\pi}}
  e^{
    -\frac12(y-Q_\Phi\sigma'(y))^2+\frac12Q_\Theta\sigma'(y)^2
    -\beta\sigma(y)
    -\omega y\sigma'(y)
  }
  \notag
\end{align}
We can simplify this action considerably by inserting a specific activation
$\sigma$. The rectified linear activation referenced above cannot be used
because ti does not have a well-defined second derivative everywhere. Taking
instead the rectified quadratic activation
\begin{equation} \label{eq:quad.activation}
  \sigma(y)=\frac12(y-\kappa)^2\Theta(\kappa-y)
\end{equation}
with margin $\kappa$, the action becomes
\begin{align} \label{eq:action.perceptron}
  &\mathcal S_{\beta,\omega}(Q_\Phi,Q_\Theta,Q_\Psi)
  =\frac12\log\frac{Q_\Phi^2-Q_\Theta}{Q_\Phi^2+Q_\Psi^2}
  \\
  &\hspace{2pc}+\alpha\log\left[
    \int_{\kappa}^{\infty}\frac{dy}{\sqrt{2\pi}}\,e^{-\frac12y^2}
    +
    \int_{-\infty}^{\kappa} dy\,
    \sqrt{\frac{(1-Q_\Phi)^2+Q_\Psi^2}{2\pi}}
    e^{
      -\frac12((1-Q_\Phi)y+Q_\Phi\kappa)^2-\frac12(\beta-Q_\Theta)(y-\kappa)^2
      -\omega y(y-\kappa)
    }
  \right]
  \notag
  \\ \notag
  &
  =\frac12\log\frac{Q_\Phi^2-Q_\Theta}{Q_\Phi^2+Q_\Psi^2}
  +\alpha\log\frac12\Bigg[
    \operatorname{erfc}\frac\kappa{\sqrt2}
    \\ \notag
  &\hspace{4em}+
    \sqrt{
      \frac{(1-Q_\Phi)^2+Q_\Psi^2}{\beta+2\omega-Q_\Theta+(1-Q_\Phi)^2}
    }
    e^{-\frac12\frac{\beta-Q_\Theta+\omega(2Q_\Phi-\omega)}{\beta+2\omega-Q_\Theta+(1-Q_\Phi)^2}\kappa^2}
    \operatorname{erfc}\left(
      \frac\kappa{\sqrt2}
      \frac{1-Q_\Phi+\omega}{\sqrt{\beta+2\omega-Q_\Theta+(1-Q_\Phi)^2}}
    \right)
  \Bigg]
  \notag
\end{align}
When $\kappa>0$ the Hamiltonian is convex and in the \textsc{unsat} phase there
is exactly one minimum. This gives a \textsc{susy} solution, with
$\beta=\omega=Q_\Psi=Q_\Theta=0$ and
\begin{equation}
  Q_\Phi=-\left(
    \sqrt{\alpha\left[
      \frac{\kappa}{\sqrt{2\pi}}e^{-\frac12\kappa^2}
      +\frac12(1+\kappa^2)\operatorname{erfc}\frac{-\kappa}{\sqrt 2}
    \right]}-1
  \right)^{-1}
  =-\left(\sqrt{\frac{\alpha}{\alpha_\textsc{sat}^\textsc{rs}(\kappa)}}-1\right)^{-1}
\end{equation}
where
\begin{equation}
  \alpha_\textsc{sat}^\textsc{rs}(\kappa)
  =\left(\frac{\kappa}{\sqrt{2\pi}}e^{-\frac12\kappa^2}
  +\frac12(1+\kappa^2)\operatorname{erfc}\frac{-\kappa}{\sqrt 2}\right)^{-1}
\end{equation}
is the load corresponding to the \textsc{sat}--\textsc{unsat} transition predicted
by a replica symmetric analysis of the zero-temperature limit of the Gibbs
measure \cite{Franz_2017_Universality}.
Writing
\begin{equation}
  E=-\frac{\partial\mathcal S_\beta}{\partial\beta}\bigg|_{\beta=0}
  =\frac12\frac{\alpha/\alpha_\textsc{sat}^\textsc{rs}(\kappa)}{(1-Q_\Phi)^2}
  =\frac12\left(\sqrt{\frac{\alpha}{\alpha_\textsc{sat}^\textsc{rs}(\kappa)}}-1\right)^2
\end{equation}
for this solution gives the energy density of the ground state as predicted by
the zero-temperature limit of the Gibbs measure \cite{Franz_2017_Universality}.
The \textsc{susy} solution exists for all $\kappa$ and
$\alpha>\alpha_\textsc{sat}^\textsc{rs}(\kappa)$ but is not correct when
$\kappa<0$.

\begin{figure}
  \centering
  \includegraphics{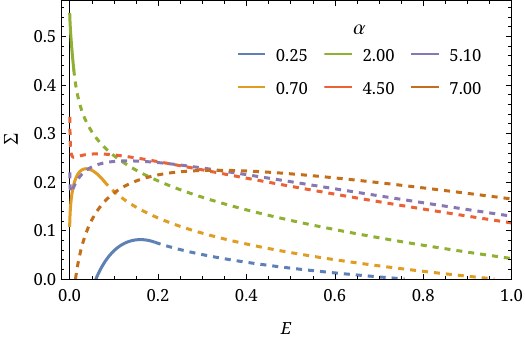}
  \caption{
    \textbf{Complexity curves for the negative margin spherical perceptron.}
    Complexity as a function of energy density $E$ for the spherical perceptron
    with quadratic activation and margin $\kappa=-\frac12$ for several values
    of load $\alpha$. Solid lines depict energies dominated by optima
    ($Q_\Psi=0$) and dashed lines depict energies dominated by saddles
    ($Q_\Psi>0$). Several distinct regimes are exhibited. When $\alpha=0.5$,
    the complexity is maximized at positive energy density by a
    \textsc{susy}-preserving solution. When $\alpha=0.70$, the complexity is
    still maximized for a \textsc{susy}-preserving $E>0$ solution, but the
    complexity at $E=0$ is positive. When $\alpha=2.00$, the complexity is
    maximized at $E=0$. When $\alpha=4.50$, the complexity is still maximized
    at $E=0$ but a second local maximum whose solution breaks \textsc{susy} has
    formed at positive energy density. When $\alpha=5.10$, the second maximum
    has become the global maximum. When $\alpha=7.00$, the local maximum at
    $E=0$ has vanished and the complexity is negative at $E=0$.
  }
  \label{fig:perceptron.examples}
\end{figure}

For negative margin $\kappa$ there is an alternative \textsc{susy}-preserving solution with nonzero
$\Sigma$. Naïvely examining the effective action, such a solution should not
exist, since with $Q_\Theta=Q_\Psi=\beta=\omega=0$ the change of variables to
$u=y-Q_\Phi\sigma'(y)$ in \eqref{eq:action.perceptron.general} seems to produce
a normalized Gaussian integral in $u$. However, this is complicated when $u$ is
not a monotonic function of $y$, which is realized for the quadratic activation
when $Q_\Phi>1$. When this happens, the resulting Gaussian integral in $u$ has double limits from $\kappa$ to $\infty$, which yields
\begin{equation}
  \Sigma=\alpha\log\left(\operatorname{erfc}\frac\kappa{\sqrt2}\right)
\end{equation}
These conditions are realized for
\begin{equation}
  \alpha<\alpha_\textsc{susy}(\kappa)
  =\left(\frac12(1+\kappa^2)-\frac{\kappa e^{-\frac12\kappa^2}}{\sqrt{2\pi}\operatorname{erfc}\frac\kappa{\sqrt2}}\right)^{-1}
\end{equation}
where
\begin{equation}
  Q_\Phi=\left(
    1-\sqrt{\frac\alpha{\alpha_\textsc{susy}(\kappa)}}
  \right)^{-1}
\end{equation}
The energy density associated with this solution is
\begin{equation}
  E=\frac12\frac{\alpha/\alpha_\textsc{susy}(\kappa)}{(Q_\Phi-1)^2}
  =\frac12\left(1-\sqrt{\frac{\alpha}{\alpha_\textsc{susy}(\kappa)}}\right)^2
\end{equation}
which vanishes as $\alpha$ goes to $\alpha_\textsc{susy}(\kappa)$. This
\textsc{susy}-preserving maximum complexity is present in the $\alpha=0.25$ and
$\alpha=0.70$ examples of Fig~\ref{fig:perceptron.examples}.

There is also a \textsc{susy}-breaking solution that typically predicts a
nonzero complexity. Like in the nonlinear least squares problem, the maximum
complexity occurs at either positive or zero energy density. When the maximum occurs at positive energy density $E>0$, the complexity must be evaluated numerically
by extremizing the effective action. When it occurs at zero energy density $E=0$, certain manipulations must be
made to extract a prediction. We find in general that solutions to the extremal conditions in the limit of $E\to0$ in the portion of the nonconvex phase where zero-energy solutions are the majority sees $Q_\Psi=0$ and the other order parameters diverge like
\begin{align}
  Q_\Theta=\delta_\Theta E^{-1}
  &&
  Q_\Phi=\delta_\Phi E^{-\frac12}
  &&
  \beta=\delta_\beta E^{-\frac12}
\end{align}
The saddle point conditions are asymptotically satisfied by
$\delta_\Theta=\delta_\Phi^2-\frac12$ and result in the limiting effective
action
\begin{equation}
  \mathcal S_{E=0}(\delta_\Phi)=-\log(\sqrt2\delta_\Phi)
  +\alpha\log\frac\alpha2\frac{\sqrt{\frac8\pi}e^{-\frac12\kappa^2}\kappa\delta_\Phi^2+(1+2\kappa^2\delta_\Phi^2)\operatorname{erfc}\frac\kappa{\sqrt2}}{
    \alpha(1+2\kappa^2\delta_\Phi^2)-1
  }
\end{equation}
depending only on $\delta_\Phi$. The transition from the \textsc{susy} solution
with nonzero complexity to the zero-energy \textsc{susy}-breaking solution is
continuous at $\alpha=\alpha_\textsc{susy}(\kappa)$, where
$\delta_\Phi=\frac1{\sqrt2}$. This transition occurs between $\alpha=0.70$ and
$\alpha=2.00$ in Fig~\ref{fig:perceptron.examples}.
Below $\alpha_\textsc{susy}(\kappa)$ the $E=0$ \textsc{susy}-breaking solution is
still valid for understanding the complexity at zero energy density.

The transition between the \textsc{susy}-breaking maximal complexity at zero
and nonzero energy is discontinuous for all negative margins $\kappa<0$. In the regime where the global maximum of the complexity occurs at zero energy
density, a secondary local
maximum forms in the complexity at nonzero energy density. This is exemplified
by the $\alpha=4.50$ curve in Fig~\ref{fig:perceptron.examples}. Raising
$\alpha$ increases the height of this second maximum relative to the one at
zero energy density until it becomes the global maximum, exemplified by the
$\alpha=5.10$ curve in Fig~\ref{fig:perceptron.examples}. For sufficiently
large $\alpha$, the complexity at zero energy density vanishes
entirely.

\begin{figure}
  \centering
  \includegraphics{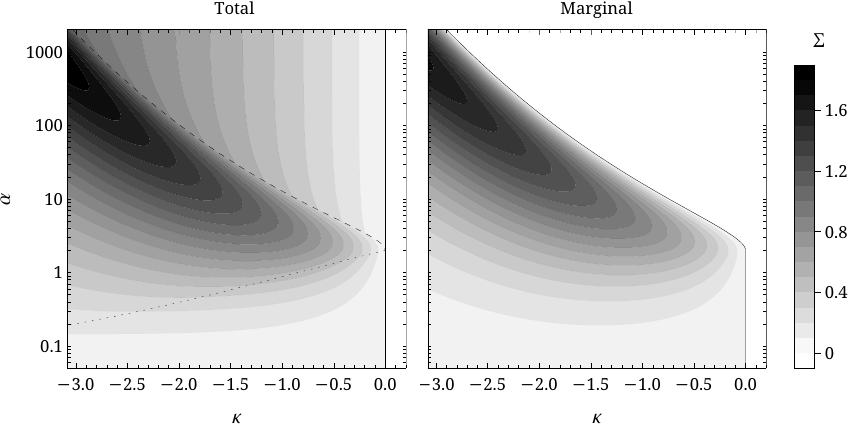}
  \caption{
    \textbf{Maximum total and marginal complexity of the perceptron.}
    Complexities for the spherical perceptron with activation
    $\sigma(y)=\frac12(\kappa-y)^2\Theta(\kappa-y)$ and load $\alpha=M/N$.
    Total complexity is divided into three phases. Below the dotted line the
    total complexity occurs at positive energy density with a
    \textsc{susy}-preserving solution. Between the dotted line and the dashed
    line the total complexity occurs at zero energy density and \textsc{susy}
    is broken. Above the dashed line the total complexity occurs at positive
    energy density with a \textsc{susy}-breaking solution. The marginal
    complexity is always maximized at positive energy density.
  }
  \label{fig:perceptron.complexity}
\end{figure}

The maximum total complexity and maximum marginal complexity of the spherical
perceptron as a function of margin $\kappa$ and load $\alpha$ is shown in
Fig~\ref{fig:perceptron.complexity}. The lines of transition between different
phases of the total complexity characterized by the presence or not of
\textsc{susy} and by whether the maximum complexity is at zero energy density
are shown. The marginal complexity is always maximized at nonzero energy
density. The marginal complexity is found by starting from a solution for total
complexity and varying $\mu$ until the marginal condition
\eqref{eq:marginal.condition} is satisfied.

There is a regime of fixed $\mu$ and $E$ which drives the action \eqref{eq:action.perceptron} to a point
where the integral over $y$ becomes ill-defined. This occurs when the Gaussian integral
from $-\infty$ to $\kappa$ has a negative covariance, or when
$\beta+2\omega-Q_\Theta+(1-Q_\Phi)^2<0$. This occurring does not appear to
correspond with a singularity in either the complexity or in any of the order
parameters; everything remains smooth up to the transition. However, directly
solving the extremal conditions for the effective action in this regime becomes
impossible. It seems unlikely this breakdown is related to an inherent feature
of the underlying properties of the perceptron, but instead is related to a
breakdown of the mean-field theoretical approach to the problem. These
conditions are encountered when attempting to evaluate the marginal complexity
far from its maximum value. It is possible that exact results can be extracted
in this regime through analytic continuation of the effective action.

\subsection{The Sherrington--Kirkpatrick model}
\label{sec:sk.model}

The examples we have given heretofore have either been Gaussian, spherically
constrained, or both. In many problems of interest, the configurations are
constrained not by a global constraint but on one that applies individually to
each coordinate. Examples include soft-spin Ising models, which approximate configurations of binary variables by a potential on each spin with minima at $\pm1$ like
\begin{equation}
  U(x)=U_0(1-2x^2+x^4)
\end{equation}
As $U_0$ is increased to infinity, the configuration space is restricted to the
desired binary variables. Besides recent interest in the complexity of this
kind of system \cite{Aspelmeier_2022_Free-energy, Yamamura_2024_Geometric, Ghimenti_2026_Geometry}, they
also represent one of the first systems whose complexity was ever
studied, in the context of the Sherrington--Kirkpatrick model
\cite{Bray_1980_Metastable}.

The Sherrington--Kirkpatrick (SK) model consists of $N$ binary spins $s_i=\pm1$ with
Hamiltonian
\begin{equation}
  V(\boldsymbol s)=\frac1{\sqrt{2N}}\sum_{ij=1}^NJ_{ij}s_is_j
\end{equation}
for independent Gaussian couplings $J$ \cite{Sherrington_1975_Solvable}. The complexity of this model cannot be
studied directly, since $V$ is not differentiable. However, Thouless,
Anderson and Palmer (TAP) introduced a free energy functional for the model \cite{Thouless_1977_Solution}
parameterized by $N$ magnetizations $x_i=\langle s_i\rangle$ taking values
continuously between $\pm1$ given by
\begin{equation} \label{eq:energy.SK}
  H(\boldsymbol x)=
  V(\boldsymbol x)
  -\frac12NT^{-1}W\left(\frac{\|\boldsymbol x\|^2}N\right)
  +T\sum_{i=1}^NU(x_i)
\end{equation}
where $V$ is naturally continued to real-valued argument,
$W(q)=\frac12(1-q)^2$, and the single-site potential $U$ is defined by
\begin{equation}
  U(x)=-\log 2+\frac12\log(1-x^2)+x\tanh^{-1}x
\end{equation}
The minima $\boldsymbol x^*$ of $H$ give metastable thermodynamic states of the
SK model at temperature $T$ whose free energy is $H(\boldsymbol x^*)$. Because
$H$ is a differentiable function of continuous parameters, the complexity of
these states can be found by the Kac--Rice method.

Applying Kac--Rice to the SK model has been a contentious affair, due to the
model's novel structure and symmetry breaking. At all temperatures and (free)
energy densities, typical stationary points have no spectral density at zero
eigenvalue, yet \textsc{susy} or \textsc{brst} symmetry is broken. In the
language of this paper, there is \textsc{susy} breaking but no condensation
along $\Psi$. This is due to the fact that while there is no \emph{density} of
modes with zero eigenvalue at these stationary points, a \emph{single, exactly
zero} eigenvalue is present at each one in the thermodynamic limit
\cite{Aspelmeier_2004_Complexity}.
Though typical stationary points at all temperatures and (free) energy
densities have this zero eigenvalue, stationary points conditioned on other
properties do not. Refs \cite{Rizzo_2005_TAP, Muller_2006_Marginal} showed that
stationary points conditioned on atypical values of the overlap
$q=\frac1N\|\boldsymbol x\|^2$ do not have the zero eigenvalue. By biasing the count of stationary points with respect to the overlap and taking the bias to zero, like
\begin{equation}
  \mathcal N[H]
  =\lim_{\epsilon\to0}\int_{(-1,1)^N} d\boldsymbol x\,
  e^{-\frac12\epsilon\|\boldsymbol x\|^2}
  \delta(\boldsymbol\nabla H(\boldsymbol x))\,
  |\det\operatorname{Hess}H(\boldsymbol x)|\,
\end{equation}
the presence of a broken symmetry at zero bias is understood to be the result of spontaneous symmetry breaking with condensation along the bias direction. Written in the superspace formalism, the count resulting from biasing with respect to the overlap and then taking the bias to zero is given by
\begin{align}
  \mathcal N[H]
  &=\lim_{\epsilon\to0}\int_{\mathbb R^{N|4}}d\boldsymbol\phi\,
  e^{
    \int d1\,H(\boldsymbol\phi(1))
    -\frac12\epsilon\int d1\,d2\,\Theta(1,2)\boldsymbol\phi(1)\cdot\boldsymbol\phi(2)
  }
  \\ \notag
  &=\lim_{\epsilon\to0}\int_{\mathbb R^{1|4\times 4}}d\mathbb Q\,
  e^{
    N\mathcal S(\mathbb Q)
    -\frac N2\epsilon\int d1\,d2\,\Theta(1,2)\mathbb Q(1,2)
  }
\end{align}
The $\epsilon>0$ action explicitly breaks supersymmetry along the direction of
the operator $\Theta$. Perturbation along this direction results in generation
of spontaneous \textsc{susy}-breaking along the directions $\Theta$, $\Phi$,
and $\Pi$ but not $\Psi$. Unlike the case of perturbation along $\Psi$,
breaking along $\Theta$ does not explicitly break ghost number symmetries,
which was a crucial part of our argument in Section \ref{sec:4-index.vanishing} for why the prefactor of
$\Psi$-broken \textsc{susy} is nonzero.
Resolving the tension of the zero prefactor in the SK model appears to be an open problem.

To treat the complexity of the TAP free energy \eqref{eq:energy.SK}, first note
that the random part $V$ of the energy is a Gaussian random function with zero
mean and covariance
\begin{align}
  \overline{V(\boldsymbol x)V(\boldsymbol y)}
  =Nf\left(\frac{\boldsymbol x\cdot\boldsymbol y}N\right)
  =\frac N2\left(\frac{\boldsymbol x\cdot\boldsymbol y}N\right)^2
\end{align}
where $f(q)=\frac12q^2$ for the traditional SK model and takes different forms
for generalized $p$-spin versions. Note that for generic $f$, the function $W$
in \eqref{eq:energy.SK} takes the form \cite{Rieger_1992_The}
\begin{equation}
  W(q)=f(1)-f(q)-(1-q)f'(q)
\end{equation}
We count the number of stationary points biased by (free) energy density as
described in Section~\ref{sec:energy} and by another property as a proxy for
index as described in Section~\ref{sec:index}. Here, instead of index we will
bias the count by value of
\begin{equation}
  \sum_{i=1}^N\frac1{U''(x_i)}
\end{equation}
or the trace of the inverse Hessian of the single-spin energy. The number of stationary points biased by energy and inverse trace is
\begin{equation}
  \mathcal N_{\beta,\omega}[H]
  =\int_{\mathbb R^{N|4}}d\boldsymbol\phi\,e^{
    \int d1\,\beta(1)[
    V(\boldsymbol\phi(1))
    -\frac N{2T}W\big(\frac{\|\boldsymbol\phi(1)\|^2}N\big)
    +T\sum_{i=1}^NU(\phi_i(1))
    ]
    -\omega\sum_{i=1}^N\int d1\,\varpi(1)U''(\phi_i(1))^{-1}
  }
\end{equation}
The average over random functions $V$ gives
\begin{equation}
  \overline{\mathcal N_{\beta,\omega}[H]}
  =\int_{\mathbb R^{N|4}}d\boldsymbol\phi\,e^{
    \frac N2\int d1\,d2\,\beta(1)\beta(2)f\big(\frac{\boldsymbol\phi(1)\cdot\boldsymbol\phi(2)}{N}\big)
    -\frac N{2T}\int d1\,\beta(1)
      W\big(\frac{\|\boldsymbol\phi(1)\|^2}N\big)
    +\sum_{i=1}^N\int d1\,[
      T\beta(1)U(\phi_i(1))
      -\omega\varpi(1)U''(\phi_i(1))^{-1}
    ]
  }
\end{equation}
Define the superoverlap order parameter $\mathbb Q(1,2)=\frac1N\boldsymbol\phi(1)\cdot\boldsymbol\phi(2)$ in the integral by inserting a $\delta$ function fixing its value. This gives
\begin{align}
  \overline{\mathcal N_{\beta,\omega}[H]}
  &=\int_{\mathbb R^{1|4\times 4}}d\mathbb Q\,
  e^{
    \frac N2\int d1\,d2\,\beta(1)\beta(2)f(\mathbb Q(1,2))
    -\frac N{2T}\int d1\,\beta(1)W(\mathbb Q(1,1))
  }
  \\
  &\qquad\times\int_{\mathbb R^{N|4}}d\boldsymbol\phi\,
  \delta\big(\boldsymbol\phi(1)\cdot\boldsymbol\phi(2)-N\mathbb Q(1,2)\big)
  e^{
    \sum_{i=1}^N\int d1\,[
      T\beta(1)U(\phi_i(1))
      -\omega\varpi(1)U''(\phi_i(1))^{-1}
    ]
  }
  \notag
\end{align}
Representing the $\delta$ function by its Fourier transform with the
introduction of a conjugate field $\tilde{\mathbb Q}$, the integral over
$\boldsymbol\phi$ can be factorized over its $N$ components, which gives
\begin{align}
  \overline{\mathcal N_{\beta,\omega}[H]}
  &=\int_{\mathbb R^{1|4\times 4}}d\mathbb Q\,d\tilde{\mathbb Q}\,
  e^{
    \frac N2\int d1\,d2\,\beta(1)\beta(2)f(\mathbb Q(1,2))
    -\frac N{2T}\int d1\,\beta(1)W(\mathbb Q(1,1))
    -\frac N2\int d1\,d2\,\tilde{\mathbb Q}(1,2)\mathbb Q(1,2)
  }
  \\
  &\hspace{6em}\times\left(\int_{\mathbb R^{1|4}}d\phi\,
  e^{\frac12\int d1\,d2\,\phi(1)\tilde{\mathbb Q}(1,2)\phi(2)
  +\int d1\,[T\beta(1)U(\phi(1))-\omega\varpi(1)U''(\phi(1))^{-1}]}\right)^N
  \notag
\end{align}
We have therefore reduced the count to an extremal problem over two
superoperator order parameters $\mathbb Q$ and $\tilde{\mathbb Q}$ of the
effective action
\begin{align} \notag
  \mathcal S_{\beta,\omega}(\mathbb Q,\tilde{\mathbb Q})
  &=
  \frac12\int d1\,d2\,\beta(1)\beta(2)f(\mathbb Q(1,2))
  -\frac1{2T}\int d1\,\beta(1)W(\mathbb Q(1,1))
  -\frac12\int d1\,d2\,\tilde{\mathbb Q}(1,2)\mathbb Q(1,2)
  \\
  & \qquad
  +\log\int_{\mathbb R^{1|4}} d\phi\,
  e^{\frac12\int d1\,d2\,\phi(1)\tilde{\mathbb Q}(1,2)\phi(2)+\int d1\,[T\beta(1)U(\phi(1))-\omega\varpi(1)U''(\phi(1))^{-1}]}
\end{align}
Expanding the order parameters into their minimally \textsc{susy}-breaking forms yields
\begin{align}
  \mathcal S_{\beta,\omega}(\mathbb Q,\tilde{\mathbb Q})
  &=\frac12\left[
    \beta^2f(Q_1)+(Q_\Theta-2\beta Q_\Pi)f'(Q_1)+(Q_\Pi^2-Q_\Phi^2+Q_\Psi^2)f''(Q_1)
  \right]
  \\ \notag
  &\hspace{4em}
  +\frac1{2T}\left[
    \beta[f(1)-f(Q_1)-(1-Q_1)f'(Q_1)]
    +2(1-Q_1)(Q_\Pi-Q_\Phi)f''(Q_1)
  \right]
  \\ \notag
  &\hspace{4em}
  -\frac12(\tilde Q_1Q_\Theta+\tilde Q_\Theta Q_1)-\tilde Q_\Pi Q_\Pi+\tilde Q_\Phi Q_\Phi-\tilde Q_\Psi Q_\Psi
  \\ \notag
  &\hspace{4em}
  +\log\int_{-1}^1 dx\,
  \sqrt{\frac{(\tilde Q_\Phi+TU''(x))^2+\tilde Q_\Psi^2}{2\pi\tilde Q_1}}
  e^{
    -\frac1{2\tilde Q_1}(\tilde Q_\Pi x+TU'(x))^2+\frac12\tilde Q_\Theta x^2
    -\beta TU(x)
    -\omega U''(x)^{-1}
  }
  \notag
\end{align}
For the specific case of the SK model where $f(q)=\frac12q^2$, the saddle point conditions for $\mathbb Q$ can be exactly solved and give
\begin{align}
  Q_1=\tilde Q_1
  \hspace{4em}
  Q_\Pi=\tilde Q_\Pi+\beta\tilde Q_1+T^{-1}(\tilde Q_1-1)
  \hspace{4em}
  Q_\Phi=\tilde Q_\Phi+T^{-1}(\tilde Q_1-1)
  \\ \notag
  Q_\Theta=\tilde Q_\Theta+\beta(2\tilde Q_\Pi+\beta\tilde Q_1)
  +T^{-1}(2\tilde Q_\Pi-2\tilde Q_\Phi+\beta(3\tilde Q_1-1))
  \hspace{6em}
  Q_\Psi=\tilde Q_\Psi
\end{align}
Substituting this solution into the action results in a five-parameter effective action depending on the minimally \textsc{susy}-breaking form of $\tilde{\mathbb Q}$ alone, or
\begin{align}
  \notag
  \mathcal S_{\beta,\omega}(\tilde{\mathbb Q})
  &=\frac{1-\tilde Q_1}T\left[
    \tilde Q_\Pi-\tilde Q_\Phi+\frac\beta4(1+3\tilde Q_1)
  \right]
  -\frac12\left[
    \frac12\beta^2\tilde Q_1^2+(\tilde Q_\Theta+2\beta\tilde Q_\Pi)\tilde Q_1
    +\tilde Q_\Pi^2-\tilde Q_\Phi^2+\tilde Q_\Psi^2
  \right]
  \\
  &\hspace{1em}
  +\log\int_{-1}^1\frac{dx}{\sqrt{2\pi\tilde Q_1}}\sqrt{\tilde Q_\Psi^2+\left(\tilde Q_\Phi^2+\frac T{1-x^2}\right)^2}e^{-\frac1{2\tilde Q_1}(\tilde Q_\Pi x+T\tanh^{-1}(x))^2+\frac12\tilde Q_\Theta x^2-\beta U(x)-\omega(1-x^2)}
\end{align}
When $\omega$ and $\tilde Q_\Psi$ are zero, this expression coincides with those in the literature \cite{Bray_1980_Metastable, Cavagna_2003_On} with the identification
\begin{align}
  \tilde Q_1=PT^2
  &&
  \tilde Q_\Pi=-T\Delta
  &&
  \tilde Q_\Phi=TB
  &&
  \tilde Q_\Theta=2\lambda
\end{align}
However, the presence of $\tilde Q_\Psi$ allows for the possibility of
describing atypical stationary points whose spectral densities have weight over
zero eigenvalue.
The complexity is found by locating extremal points of this action with respect
to its order parameters and making Legendre transformations from $\beta$ and
$\omega$ to (free) energy density $E$ and trace of the inverse
Hessian $\mu$.

\begin{figure}
  \centering
  \includegraphics{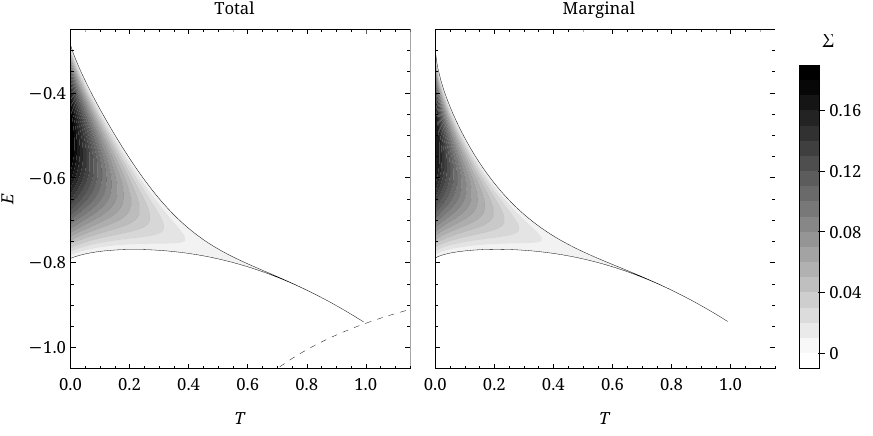}
  \caption{
    \textbf{Complexities of the Sherrington--Kirkpatrick model.} Total and
    marginal complexity of the SK model as a function of temperature $T$ and
    energy density $E$. The dashed line in the total complexity shows the
    $\tilde Q_1=0$ solution, corresponding to the paramagnetic state, where the
    complexity is always zero. The marginal complexity is defined by the point
    where the spectral density of the stationary points has a pseudogap and
    verifies the marginal condition \eqref{eq:marginal.condition}.
  } \label{fig:sk.complexity}
\end{figure}

When the total complexity is evaluated as a function of energy density but
without fixing the inverse Hessian, the relevant extremal point of the action
always has $Q_\Psi=Q_\Phi=0$. Because the resulting complexity can be evaluated
while neglecting the absolute value of the determinant in the Kac--Rice
formula, it recreates known results. However, when the trace of the inverse
Hessian $\mu$ is conditioned to an atypical value, the situation can change. For certain atypical $\mu$, a new branch appears characterized by zero $Q_\Phi$ but nonzero
$Q_\Psi$. The transition point gives precisely the marginal complexity, defined
by the complexity of stationary points whose spectral density has a pseudogap and which verify the marginal condition \eqref{eq:marginal.condition}. Both the total and marginal complexities as a function of temperature $T$ and free energy density $E$ are shown in Fig~\ref{fig:sk.complexity}.

\section{Conclusions}
\label{sec:conclusion}

We have introduced a systematic method for performing mean-field complexity
calculations. We account for the absolute value of the determinant present in
the Kac--Rice formula by introducing many auxiliary fields and looking for
their spontaneous symmetry breaking with respect to a specific perturbation.
Adopting a 4-index superspace notation, we organize the many fields into a compact
expression and show that the perturbation drives spontaneous supersymmetry
breaking along a specific axis in the space of superoperators. Condensation
along this axis engenders condensation along several others, leading us to
consider five order parameters. We interpret these order parameters in terms of
the geometry and spectral properties of the stationary points they
characterize. The formalism extends naturally to non-Euclidean geometries and
arbitrary forms of conditioning, and conditions of interest like marginality
are expressed naturally in terms of the order parameters. We apply the
formalism to several models, recovering old results and deriving new ones.

One strength of our approach is that it generalizes naturally to calculating
the quenched complexity, defined using the average of the logarithm of the number of stationary points rather than the logarithm of the average, or
\begin{equation}
  \Sigma'=\lim_{N\to\infty}\frac1N\overline{\log\mathcal N[H]}
\end{equation}
To generalize, treat the logarithm with replicas and introduce the 4-index
superfields to represent the number of stationary points. The superoverlap order parameter then becomes an $n\times n$ matrix of superoperators, with
\begin{equation}
  \mathbb Q_{ab}(1,2)=\frac1N\boldsymbol\phi_a(1)\cdot\boldsymbol\phi_b(2)
\end{equation}
Many of the formulas in this paper can be immediately applied to a quenched
calculation without modification by interpreting the order parameters as
replica matrices rather than scalars. An example is found in Appendix
\ref{sec:quenched}, where we derive the quenched effective action for the
spherical spin glasses found in Ref~\cite{Kent-Dobias_2023_How}, but without
assuming the independence of the gradient and Hessian in different
replicas. We will dedicate a future manuscript to quenched complexity.

Researchers are increasingly interested in the properties of stationary points
of flow fields that do not derive from the derivative of a potential, which
arise naturally in descriptions of nonreciprocal systems like ecologies,
evolution, and artificial and natural neural networks \cite{Lacroix-A-Chez-Toine_2022_Counting, Ros_2023_Quenched, Fournier_2026_Nonreciprocal}. The complexity of
stationary points in such flows can be calculated using the Kac--Rice formula,
but the superspace representation changes. Determining a minimally
\textsc{susy}-breaking prescription for such systems is a priority of future
research.

\appendix

\section{A short introduction to superspace}
\label{sec:superspace}

In this appendix we review the notation and linear algebra of superspace \cite{DeWitt_1992_Supermanifolds}.
Grassmann variables can be thought of as spin-zero fermions. Grassmann
variables anticommute: if $\bar\theta$ and $\theta$ are Grassmann variables,
then
\begin{equation}
  \bar\theta\theta=-\theta\bar\theta
\end{equation}
This immediately establishes that $\theta^2=\bar\theta^2=0$. As a result,
series expansions of functions with respect to Grassmann variables terminate
at finite order. For instance,
\begin{equation}
  f(a+\theta b)=f(a)+\theta bf'(a)+\frac12(\theta b)^2f''(a)+\cdots
  =f(a)+\theta bf'(a)
\end{equation}
A linear combination like $a+\theta b$ of a regular number and products of
Grassmann indices like $\theta$ is a supernumber. The regular part of the
supernumber (here $a$) is called its body, while the part involving products of
Grassmann indices (here $\theta b$) is called its soul. Given a supernumber
involving $d$ Grassmann indices, the series expansion of any function of that
supernumber around its body terminates exactly at $d$th order.
A classic result of this is that if $\bar{\boldsymbol\eta}$ and
$\boldsymbol\eta$ are $N$-dimensional vectors of Grassmann variables and $A$ is an $N\times N$ matrix, then
\begin{equation}
  \int d\bar{\boldsymbol\eta}\,d\boldsymbol\eta\,e^{-\bar{\boldsymbol\eta}^TA\boldsymbol\eta}
  =\int d\bar\eta_1\cdots d\bar\eta_N\,d\eta_1\cdots\,d\eta_N\,\sum_{n=0}^\infty\frac1{n!}\left(-\sum_{ij}A_{ij}\bar\eta_i\eta_j\right)^n
  =\det A
\end{equation}
Differentiation
and integration of Grassmann variables are identical operations and are defined
by
\begin{align}
  \frac\partial{\partial\theta}\theta=1
  &&
  \frac\partial{\partial\theta}1=0
  &&
  \int d\theta\,\theta=1
  &&
  \int d\theta\,1=0
\end{align}
For instance,
\begin{equation}
  \frac\partial{\partial\theta}f(a+\theta b)
  =\int d\theta\,f(a+\theta b)
  =bf'(a)
\end{equation}
The superspace $\mathbb R^{N|d}$ consists of $N$-dimensional supernumbers
involving linear combinations of $d$ superindicies $\theta_1,\ldots,\theta_d$. For instance, elements of the superspace $\mathbb R^{N|2}$ take the form
\begin{equation}
  \boldsymbol\varphi(1)=\boldsymbol x+\theta_1(1)\boldsymbol\eta+\bar{\boldsymbol\eta}\theta_2(1)+i\hat{\boldsymbol x}\theta_1(1)\theta_2(1)
\end{equation}
where $\theta_1$ and $\theta_2$ are the two Grassmann indices, $\boldsymbol x$
and $\hat{\boldsymbol x}$ are $N$-dimensional regular vectors, and
$\bar{\boldsymbol\eta}$ and $\boldsymbol\eta$ are $N$-dimensional Grassmann
vectors.\footnote{
  In this appendix we adopt a notation of numbered superindices in order to
  treat arbitrary superspaces on equal footing. In the manuscript, we denote
  the two superindices of $\mathbb R^{N|2}$ by $\theta_1=\bar\theta$ and
  $\theta_2=\theta$, and we denote the two additional superindices of $\mathbb
  R^{N|4}$ by $\theta_3=\bar\vartheta$ and $\theta_4=\vartheta$.
}
Because the Grassmann indices and Grassmann vectors always come in
even combinations, supervectors like $\boldsymbol\varphi(1)$ behave like regular
commuting vectors. We write the integral over all $d$ superindicies as
\begin{equation}
  d1=d\theta_d(1)\cdots d\theta_1(1)
\end{equation}
so that, for instance,
\begin{equation}
  \int d1\,\boldsymbol\varphi(1)=i\hat{\boldsymbol x}
\end{equation}
Supermatrices $\mathbb M\in\mathbb R^{N\times N|d\times d}$ are representations of
linear operations on superspace that act simultaneously as regular matrices on
the $N$ vector indices and by convolution on the $d$ Grassmann indices. For instance,
an element $\mathbb M\in\mathbb R^{N\times N|2\times 2}$ can be written
\begin{align} \label{eq:superop}
  \mathbb M(1,2)
  &=M^{(1)}
  +M^{(2)}\theta_1(1)\theta_2(1)\theta_1(2)\theta_2(2)
  +M^{(3)}\theta_1(1)\theta_2(1)+M^{(4)}\theta_1(2)\theta_2(2)
  \\ \notag
  &\qquad
  -M^{(5)}\theta_1(1)\theta_2(2)+M^{(6)}\theta_1(2)\theta_2(1)
  +M^{(7)}\theta_1(1)\theta_1(2)-M^{(8)}\theta_2(1)\theta_2(2)
  \\ \notag
  &\qquad
  -M^{(9)}\theta_1(1)-M^{(10)}\theta_1(2)
  -M^{(11)}\theta_2(1)\theta_1(2)\theta_2(2)-M^{(12)}\theta_1(1)\theta_2(1)\theta_2(2)
  \\ \notag
  &\qquad
  +M^{(13)}\theta_2(1)-M^{(14)}\theta_2(2)
  +M^{(15)}\theta_1(1)\theta_2(1)\theta_1(2)-M^{(16)}\theta_1(2)\theta_2(2)\theta_1(1)
\end{align}
where $M^{(1)}$ through $M^{(8)}$ are regular $N\times N$ matrices and
$M^{(9)}$ through $M^{(16)}$ are $N\times N$ Grassmann-valued matrices.
In general, the action of a superoperator on a supervector is defined by
\begin{equation}
  (\mathbb M\boldsymbol\varphi)_i(1)=\sum_{j=1}^N\int d2\,\mathbb M_{ij}(1,2)\varphi_j(2)
\end{equation}
The convolution of two superoperators is defined by
\begin{equation}
  (\mathbb M\ast\mathbb B)_{ij}(1,2)
  =\sum_{k=1}^N\int d3\,\mathbb M_{ik}(1,3)\mathbb B_{kj}(3,2)
\end{equation}
The identity superoperator is given by
\begin{equation}
  \delta(1,2)=I(\theta_1(1)-\theta_1(2))\times\cdots\times(\theta_d(1)-\theta_d(2))
\end{equation}
where $I$ is the regular $N\times N$ identity matrix. Superlinear algebra operators can be defined similarly to the traditional ones. The supertrace of an operator is defined by
\begin{equation}
  \operatorname{sTr}\mathbb M=\sum_{i=1}^N\int d1\,\mathbb M_{ii}(1,1)
\end{equation}
To define the superdeterminant of an operator, we first define canonical bases
for superspace. Let $\boldsymbol e(1)$ and $\boldsymbol f(1)$ be
$\frac122^d$-dimensional vectors whose components give the basis elements of
the even and odd subspaces of superspace, respectively. For instance, in
$\mathbb R^{N|2}$,
\begin{align}
  \boldsymbol e(1)=\{1,\theta_1(1)\theta_2(1)\}
  &&
  \boldsymbol f(1)=\{\theta_1(1),\theta_2(1)\}
\end{align}
while for $\mathbb R^{N|4}$ these bases are
\begin{align}
  \boldsymbol e(1)&=\{1,
    \theta_1(1)\theta_2(1),\theta_3(1)\theta_4(1),
    \theta_1(1)\theta_4(1),\theta_3(1)\theta_2(1),
    \\ \notag
                  &\hspace{14em}
    \theta_1(1)\theta_3(1),\theta_2(1)\theta_4(1),
    \theta_1(1)\theta_2(1)\theta_3(1)\theta_4(1)
  \}
  \\ \notag
    \boldsymbol f(1)&=\{
    \theta_1(1),\theta_2(1),\theta_3(1),\theta_4(1),
    \theta_1(1)\theta_2(1)\theta_3(1),
    \\ \notag
                  &\hspace{14em}
    \theta_3(1)\theta_4(1)\theta_1(1),
    \theta_1(1)\theta_2(1)\theta_4(1),
    \theta_3(1)\theta_4(1)\theta_2(1)
  \}
\end{align}
A dual basis $\boldsymbol e^\dagger$ and $\boldsymbol f^\dagger$ is defined by the requirement
\begin{align}
  \int d1\,e^\dagger_i(1)e_j(1)=\delta_{ij}
  &&
  \int d1\,e^\dagger_i(1)f_j(1)=0_{\hphantom{ij}}
  \\ \notag
  \int d1\,f^\dagger_i(1)e_j(1)=0_{\hphantom{ij}}
  &&
  \int d1\,f^\dagger_i(1)f_j(1)=\delta_{ij}
\end{align}
For instance, the dual basis of $\mathbb R^{N|2}$ is given by
\begin{align}
  \boldsymbol e^\dagger(1)=\{\theta_1(1)\theta_2(1),1\}
  &&
  \boldsymbol f^\dagger(1)=\{-\theta_2(1),\theta_1(1)\}
\end{align}
Break the superoperator $\mathbb M$ into four $\frac122^dN\times\frac122^dN$ regular
matrices defined by
\begin{align}
  A_{injm}=\int d1\,d2\,e^\dagger_i(1)\mathbb M_{nm}(1,2)e_j(2)
  &&
  B_{injm}=\int d1\,d2\,e^\dagger_i(1)\mathbb M_{nm}(1,2)f_j(2)
  \\ \notag
  C_{injm}=\int d1\,d2\,f^\dagger_i(1)\mathbb M_{nm}(1,2)e_j(2)
  &&
  D_{injm}=\int d1\,d2\,f^\dagger_i(1)\mathbb M_{nm}(1,2)f_j(2)
\end{align}
For instance, the 2-index superoperator $\mathbb M$ defined by \eqref{eq:superop} has blocks
\begin{align}
  A=\begin{bmatrix}
    M^{(4)} & M^{(1)} \\
    M^{(2)} & M^{(3)}
  \end{bmatrix}
  &&
  B=\begin{bmatrix}
    M^{(14)} & M^{(13)} \\
    M^{(12)} & M^{(15)}
  \end{bmatrix}
  \\ \notag
  C=\begin{bmatrix}
    M^{(16)} & M^{(9)} \\
    M^{(11)} & M^{(10)}
  \end{bmatrix}
  &&
  D=\begin{bmatrix}
    M^{(5)} & M^{(7)} \\
    M^{(8)} & M^{(6)}
  \end{bmatrix}
\end{align}
The superdeterminant of $\mathbb M$ is written in terms of these blocks by
\begin{equation}
  \operatorname{sdet}\mathbb M=\frac{\det(A-BD^{-1}C)}{\det D}
\end{equation}
The superoperators resulting from saddle point evaluations never have Grassmann
elements. In that case, the superdeterminant is simply the ratio of the
determinants of the even-to-even map $A$ and the odd-to-odd map $D$.
The supertrace can also be written in terms of this block representation, with
\begin{equation}
  \operatorname{sTr}\mathbb M=\operatorname{Tr}A-\operatorname{Tr}D
\end{equation}
The superdeterminant appears in similar contexts to where the determinant appears in ordinary calculations. For instance, the value of a super-Gaussian integral is
\begin{equation}
  \int_{\mathbb R^{N|d}}d\boldsymbol\varphi\,e^{-\frac12\int d1\,d2\,\boldsymbol\varphi(1)^T\mathbb M(1,2)\boldsymbol\varphi(2)+\int d1\,\boldsymbol\phi(1)^T\boldsymbol\psi(1)}
  =(\operatorname{sdet}M)^{-\frac12}
  e^{\frac12\int d1\,d2\,\boldsymbol\psi(1)\mathbb M^{-1}(1,2)\boldsymbol\psi(2)}
\end{equation}
where $d\boldsymbol\phi$ contains the product of measures for each regular and
superspace component of $\boldsymbol\phi$ normalized by $(2\pi)^{-2^{d-2}}$. The inverse
of a superoperator can be constructed by representing it in the block matrix form
outlined above, inverting the block matrix, then mapping back onto the
corresponding basis elements.

\section{Quenched complexity of the spherical spin glasses}
\label{sec:quenched}

In Reference \cite{Kent-Dobias_2023_How} the quenched complexity of the
spherical spin glasses was derived under the restrictive assumption that the
correlation between the Hessian matrix and the gradient at different points in
space is negligible in the mean-field limit. Here, we see how the same
variational expression can be found without it. The quenched complexity is defined as the average of the logarithm of the number of stationary
points. We use replicas to reduce the logarithm to a power, which gives
\begin{equation}
  \Sigma'=\lim_{N\to\infty}\frac1N\overline{\log\mathcal N[H]}
  =\lim_{N\to\infty}\frac1N\lim_{n\to0}\frac\partial{\partial n}\overline{\mathcal N[H]^n}
\end{equation}
Following the same procedure as in Section \ref{sec:ssg} produces
\begin{equation}
  \overline{\mathcal N_\beta[H]^n}
  =\int_{\mathbb R^{n\times n|4\times 4}}d\mathbb Q
  \int_{\mathbb R^{n|4}}d\Lambda\,
  \,e^{N\mathcal S_\beta(\mathbb Q,\Lambda)}
\end{equation}
where the effective action is a function of an $n\times n$ matrix superoverlap
$\mathbb Q_{ab}(1,2)=\frac1N\boldsymbol\phi_a(1)\cdot\boldsymbol\phi_b(2)$ and
is given by
\begin{equation}
  \mathcal S_\beta(\mathbb Q,\Lambda)
  =\frac12\sum_{a=1}^n\int d1\,\Lambda_a(1)(1-\mathbb Q_{aa}(1,1))
  +\frac12\sum_{ab=1}^n\int d1\,d2\,\beta(1)\beta(2)f(\mathbb Q_{ab}(1,2))
  +\frac12\operatorname{sdet}\mathbb Q
\end{equation}
By assuming that at the relevant saddle point each element $\mathbb Q_{ab}$ is a minimally
\textsc{susy}-breaking operator and that each element of $\Lambda$ is a minimally \textsc{susy}-breaking vector, the action is expanded to
\begin{align}
  \mathcal S(Q_1,Q_\Phi,Q_\Pi,Q_\Theta,Q_\Phi,\lambda,\hat\lambda)
  =\frac12\sum_{a=1}^n\hat\lambda_a(1-Q^1_{aa})
  +\sum_{a=1}^n\lambda_a(Q^\Phi_{aa}-Q_{aa}^\Pi)
  \hspace{10em}
  \\ \notag
  +\frac12\sum_{ab}^n\left[
    \beta^2f(Q^1_{ab})
    +(Q^\Theta_{ab}-2\beta Q^\Pi_{ab})f'(Q^1_{ab})
    +[(Q^\Pi_{ab})^2-(Q^\Phi_{ab})^2+(Q^\Psi_{ab})^2]f''(Q^1_{ab})
  \right]
  \\ \notag
  +\frac12\log\det(Q_\Pi^2-Q_1Q_\Theta)
  -\frac12\log\det(Q_\Phi^2+Q_\Psi^2)
\end{align}
Consider the saddle-point conditions for $Q_\Psi$ and $Q_\Phi$, given by
\begin{align}
  0
  &=\frac{\partial\mathcal S}{\partial Q^\Phi_{ab}}
  =\lambda_a\delta_{ab}
  -Q^\Phi_{ab}f''(Q_{ab}^1)-\sum_c(Q_\Phi^2+Q_\Psi^2)^{-1}_{ac}Q^\Phi_{cb}
  \\
  0
  &=\frac{\partial\mathcal S}{\partial Q^\Psi_{ab}}
  =Q^\Psi_{ab}f''(Q_{ab}^1)-\sum_c(Q_\Phi^2+Q_\Psi^2)^{-1}_{ac}Q^\Psi_{cb}
\end{align}
These equations are always solved for diagonal
$Q^\Phi_{ab}=q_a^\Phi\delta_{ab}$ and $Q^\Psi_{ab}=q_a^\Psi\delta_{ab}$
regardless of the structure of $Q_1$. The values of the diagonal
that solve the above equations are
\begin{align}
  q_a^\Phi
  =\begin{cases}
    \frac1{2f''(1)}\left[\lambda_a-\operatorname{sgn}(\lambda_a)\sqrt{\lambda_a^2-4f''(1)}\right]
    \\
    \frac1{2f''(1)}\lambda_a
  \end{cases}
  &&
  q_a^\Psi
  =\begin{cases}
    0 & \lambda_a^2>4f''(1) \\
    \frac1{2f''(1)}\sqrt{4f''(1)-\lambda_a^2} & \lambda_a^2\leq4f''(1)
  \end{cases}
\end{align}
where we have used the fact that $Q^1_{aa}=1$ at the saddle point.
Substituting these solutions into the action to eliminate $Q_\Phi$ and $Q_\Psi$
and making the identification $Q_1=C$, $Q_\Theta=-D$, and $Q_\Pi=-R$, we have
\begin{align}
  \mathcal S(C,R,D,\lambda,\hat\lambda)
  =
  \sum_{a=1}^n\mathcal D(\lambda_a)
  +\frac12\sum_{a=1}^n\hat\lambda_a(1-C_{aa})
  +\sum_{a=1}^n\lambda_aR_{aa}
  +\frac12\log\det(CD+R^2)
  \hspace{3em}
  \\ \notag
  +\frac12\sum_{ab}^n\left[
    \beta^2f(C_{ab})
    +(2\beta R_{ab}-D_{ab})f'(C_{ab})
    +R_{ab}^2f''(C_{ab})
  \right]
\end{align}
where we have defined the function
\begin{align}
  \mathcal D(\lambda)
  &=\lambda q_\Phi+\frac12f''(1)\big(q_\Psi^2-q_\Phi^2\big)
  -\frac12\log\big(
    q_\Psi^2+q_\Phi^2
  \big)
  \\ \notag
  &=\begin{cases}
    \frac12+\log\frac{\lambda_\mathrm m}2+\frac{\lambda^2}{\lambda_\textrm m^2}-\left|\frac\lambda{\lambda_\mathrm m}\right|\sqrt{(\frac{\lambda}{\lambda_\mathrm m})^2-1}
    -\log\left(
      \left|\frac\lambda{\lambda_\mathrm m}\right|-\sqrt{(\frac{\lambda}{\lambda_\mathrm m})^2-1}
    \right)
    & \lambda^2>\lambda_\mathrm m^2 \\
    \frac12+\log\frac{\lambda_\mathrm m}2+\frac{\lambda^2}{\lambda_\mathrm m^2}
    & \lambda^2\leq\lambda_\mathrm m^2
  \end{cases}
\end{align}
for $\lambda_\mathrm m=\sqrt{4f''(1)}$. This is precisely the effective action
used in Reference \cite{Kent-Dobias_2023_How} conditioned on the radial
reaction by setting $\lambda_a=\mu$ by hand for all $a=1,\ldots,n$.

\paragraph{Acknowledgements}

JK-D is supported by FAPESP Young Investigator Grant No.~2024/11114-1. JK-D
also received support from the Simons Foundation Targeted Grant to ICTP-SAIFR.

\printbibliography[heading=bibintoc]

@article{Kurchan_1991_Replica,
 author = {Kurchan, Jorge},
 title = {Replica trick to calculate means of absolute values: applications to stochastic equations},
 journal = {Journal of Physics A: Mathematical and General},
 publisher = {IOP Publishing},
 year = {1991},
 month = {November},
 number = {21},
 volume = {24},
 pages = {4969--4979},
 url = {http://dx.doi.org/10.1088/0305-4470/24/21/011},
 doi = {10.1088/0305-4470/24/21/011},
 issn = {1361-6447}
}

@article{Parisi_2004_On,
 author = {Parisi, G and Rizzo, T},
 title = {On supersymmetry breaking in the computation of the complexity},
 journal = {Journal of Physics A: Mathematical and General},
 publisher = {IOP Publishing},
 year = {2004},
 month = {8},
 number = {33},
 volume = {37},
 pages = {7979--7992},
 url = {https://doi.org/10.1088%2F0305-4470%2F37%2F33%2F001},
 doi = {10.1088/0305-4470/37/33/001}
}

@article{Bray_2007_Statistics,
 author = {Bray, Alan J. and Dean, David S.},
 title = {Statistics of Critical Points of {Gaussian} Fields on Large-Dimensional Spaces},
 journal = {Physical Review Letters},
 publisher = {American Physical Society (APS)},
 year = {2007},
 month = {4},
 number = {15},
 volume = {98},
 pages = {150201},
 url = {https://doi.org/10.1103%2Fphysrevlett.98.150201},
 doi = {10.1103/physrevlett.98.150201}
}

@article{Castellani_2005_Spin-glass,
 author = {Castellani, Tommaso and Cavagna, Andrea},
 title = {Spin-glass theory for pedestrians},
 journal = {Journal of Statistical Mechanics: Theory and Experiment},
 publisher = {IOP Publishing},
 year = {2005},
 month = {5},
 number = {05},
 volume = {2005},
 pages = {P05012},
 url = {https://doi.org/10.1088%2F1742-5468%2F2005%2F05%2Fp05012},
 doi = {10.1088/1742-5468/2005/05/p05012}
}

@article{Fyodorov_2004_Complexity,
 author = {Fyodorov, Yan V.},
 title = {Complexity of Random Energy Landscapes, Glass Transition, and Absolute Value of the Spectral Determinant of Random Matrices},
 journal = {Physical Review Letters},
 publisher = {American Physical Society (APS)},
 year = {2004},
 month = {6},
 number = {24},
 volume = {92},
 pages = {240601},
 url = {https://doi.org/10.1103%2Fphysrevlett.92.240601},
 doi = {10.1103/physrevlett.92.240601}
}

@article{Urbani_2023_A,
 author = {Urbani, Pierfrancesco},
 title = {A continuous constraint satisfaction problem for the rigidity transition in confluent tissues},
 journal = {Journal of Physics A: Mathematical and Theoretical},
 publisher = {IOP Publishing},
 year = {2023},
 month = {February},
 number = {11},
 volume = {56},
 pages = {115003},
 url = {http://dx.doi.org/10.1088/1751-8121/acb742},
 doi = {10.1088/1751-8121/acb742},
 issn = {1751-8121}
}

@article{Kent-Dobias_2025_On,
 author = {Kent-Dobias, Jaron},
 title = {On the topology of solutions to random continuous constraint satisfaction problems},
 journal = {SciPost Physics},
 publisher = {Stichting SciPost},
 year = {2025},
 month = {May},
 number = {5},
 volume = {18},
 pages = {158},
 url = {http://dx.doi.org/10.21468/SciPostPhys.18.5.158},
 doi = {10.21468/scipostphys.18.5.158},
 issn = {2542-4653}
}

@article{Kent-Dobias_2024_Conditioning,
 author = {Kent-Dobias, Jaron},
 title = {Conditioning the complexity of random landscapes on marginal optima},
 journal = {Physical Review E},
 publisher = {American Physical Society (APS)},
 year = {2024},
 month = {December},
 number = {6},
 volume = {110},
 pages = {064148},
 url = {http://dx.doi.org/10.1103/PhysRevE.110.064148},
 doi = {10.1103/physreve.110.064148},
 issn = {2470-0053}
}

@article{Franz_2017_Universality,
 author = {Franz, Silvio and Parisi, Giorgio and Sevelev, Maxime and Urbani, Pierfrancesco and Zamponi, Francesco},
 title = {Universality of the {SAT-UNSAT} (jamming) threshold in non-convex continuous constraint satisfaction problems},
 journal = {SciPost Physics},
 publisher = {Stichting SciPost},
 year = {2017},
 month = {June},
 number = {3},
 volume = {2},
 pages = {019},
 url = {http://dx.doi.org/10.21468/SciPostPhys.2.3.019},
 doi = {10.21468/scipostphys.2.3.019},
 issn = {2542-4653}
}

@article{Folena_2020_Rethinking,
 author = {Folena, Giampaolo and Franz, Silvio and Ricci-Tersenghi, Federico},
 title = {Rethinking Mean-Field Glassy Dynamics and Its Relation with the Energy Landscape: The Surprising Case of the Spherical Mixed $p$-Spin Model},
 journal = {Physical Review X},
 publisher = {American Physical Society},
 year = {2020},
 month = {8},
 volume = {10},
 pages = {031045},
 url = {https://link.aps.org/doi/10.1103/PhysRevX.10.031045},
 doi = {10.1103/PhysRevX.10.031045},
 issue = {3},
 numpages = {26}
}

@article{Kent-Dobias_2023_How,
 author = {Kent-Dobias, Jaron and Kurchan, Jorge},
 title = {How to count in hierarchical landscapes: a full solution to mean-field complexity},
 journal = {Physical Review E},
 publisher = {American Physical Society (APS)},
 year = {2023},
 month = {6},
 number = {6},
 volume = {107},
 pages = {064111},
 url = {https://doi.org/10.1103/PhysRevE.107.064111},
 doi = {10.1103/PhysRevE.107.064111}
}

@article{Aspelmeier_2004_Complexity,
 author = {Aspelmeier, T. and Bray, A. J. and Moore, M. A.},
 title = {Complexity of {Ising} Spin Glasses},
 journal = {Physical Review Letters},
 publisher = {American Physical Society (APS)},
 year = {2004},
 month = {feb},
 number = {8},
 volume = {92},
 pages = {087203},
 url = {https://doi.org/10.1103%2Fphysrevlett.92.087203},
 doi = {10.1103/physrevlett.92.087203}
}

@article{Muller_2006_Marginal,
 author = {Müller, Markus and Leuzzi, Luca and Crisanti, Andrea},
 title = {Marginal states in mean-field glasses},
 journal = {Physical Review B},
 publisher = {American Physical Society (APS)},
 year = {2006},
 month = {10},
 number = {13},
 volume = {74},
 pages = {134431},
 url = {https://doi.org/10.1103%2Fphysrevb.74.134431},
 doi = {10.1103/physrevb.74.134431}
}

@article{Rizzo_2005_TAP,
 author = {Rizzo, Tommaso},
 title = {{TAP} complexity, the cavity method and supersymmetry},
 journal = {Journal of Physics A: Mathematical and General},
 publisher = {IOP Publishing},
 year = {2005},
 month = {March},
 number = {15},
 volume = {38},
 pages = {3287--3306},
 url = {http://dx.doi.org/10.1088/0305-4470/38/15/005},
 doi = {10.1088/0305-4470/38/15/005},
 issn = {1361-6447}
}

@article{Fyodorov_2019_A,
 author = {Fyodorov, Yan V.},
 title = {A Spin Glass Model for Reconstructing Nonlinearly Encrypted Signals Corrupted by Noise},
 journal = {Journal of Statistical Physics},
 publisher = {Springer Science and Business Media LLC},
 year = {2019},
 month = {1},
 number = {5},
 volume = {175},
 pages = {789--818},
 url = {https://doi.org/10.1007%2Fs10955-018-02217-9},
 doi = {10.1007/s10955-018-02217-9}
}

@article{Fyodorov_2020_Counting,
 author = {Fyodorov, Y. V. and Tublin, R.},
 title = {Counting Stationary Points of the Loss Function in the Simplest Constrained Least-square Optimization},
 journal = {Acta Physica Polonica B},
 publisher = {Jagiellonian University},
 year = {2020},
 number = {7},
 volume = {51},
 pages = {1663},
 url = {http://dx.doi.org/10.5506/APhysPolB.51.1663},
 doi = {10.5506/aphyspolb.51.1663},
 issn = {1509-5770}
}

@article{Fyodorov_2022_Optimization,
 author = {Fyodorov, Yan V and Tublin, Rashel},
 title = {Optimization landscape in the simplest constrained random least-square problem},
 journal = {Journal of Physics A: Mathematical and Theoretical},
 publisher = {IOP Publishing},
 year = {2022},
 month = {May},
 number = {24},
 volume = {55},
 pages = {244008},
 url = {http://dx.doi.org/10.1088/1751-8121/ac6d8e},
 doi = {10.1088/1751-8121/ac6d8e},
 issn = {1751-8121}
}

@phdthesis{Tublin_2022_A,
 author = {Tublin, Rashel},
 title = {A Few Results in Random Matrix Theory and Random Optimization},
 year = {2022},
 month = {September},
 url = {https://kclpure.kcl.ac.uk/portal/en/studentTheses/a-few-results-in-random-matrix-theory-and-random-optimization},
 school = {King's College London}
}

@article{Kamali_2023_Dynamical,
 author = {Kamali, Persia Jana and Urbani, Pierfrancesco},
 title = {Dynamical mean field theory for models of confluent tissues and beyond},
 journal = {SciPost Physics},
 publisher = {Stichting SciPost},
 year = {2023},
 month = {November},
 number = {5},
 volume = {15},
 pages = {219},
 url = {http://dx.doi.org/10.21468/SciPostPhys.15.5.219},
 doi = {10.21468/scipostphys.15.5.219},
 issn = {2542-4653}
}

@unpublished{Kamali_2023_Stochastic,
 author = {Kamali, Persia Jana and Urbani, Pierfrancesco},
 title = {Stochastic Gradient Descent outperforms Gradient Descent in recovering a high-dimensional signal in a glassy energy landscape},
 year = {2023},
 month = {sep},
 url = {http://arxiv.org/abs/2309.04788},
 note = {arXiv preprint},
 archiveprefix = {arXiv},
 eprint = {2309.04788},
 eprintclass = {cs.LG},
 eprinttype = {arxiv}
}

@unpublished{Urbani_2024_Statistical,
 author = {Urbani, Pierfrancesco},
 title = {Statistical physics of complex systems: glasses, spin glasses, continuous constraint satisfaction problems, high-dimensional inference and neural networks},
 year = {2024},
 month = {may},
 url = {http://arxiv.org/abs/2405.06384},
 note = {arXiv preprint},
 archiveprefix = {arXiv},
 eprint = {2405.06384},
 eprintclass = {cond-mat.dis-nn},
 eprinttype = {arxiv}
}

@unpublished{Montanari_2023_Solving,
 author = {Montanari, Andrea and Subag, Eliran},
 title = {Solving systems of Random Equations via First and Second-Order Optimization Algorithms},
 year = {2024},
 month = {dec},
 url = {http://arxiv.org/abs/2306.13326},
 note = {arXiv preprint},
 archiveprefix = {arXiv},
 eprint = {2306.13326},
 eprintclass = {math.PR},
 eprinttype = {arxiv}
}

@unpublished{Montanari_2024_On,
 author = {Montanari, Andrea and Subag, Eliran},
 title = {On {Smale}'s 17th problem over the reals},
 year = {2024},
 month = {may},
 url = {http://arxiv.org/abs/2405.01735},
 note = {arXiv preprint},
 archiveprefix = {arXiv},
 eprint = {2405.01735},
 eprintclass = {cs.DS},
 eprinttype = {arxiv}
}

@article{Urbani_2024_Quantum,
 author = {Urbani, Pierfrancesco},
 title = {Quantum exploration of high-dimensional canyon landscapes},
 journal = {Journal of Statistical Mechanics: Theory and Experiment},
 publisher = {IOP Publishing},
 year = {2024},
 month = {August},
 number = {8},
 volume = {2024},
 pages = {083301},
 url = {http://dx.doi.org/10.1088/1742-5468/ad0635},
 doi = {10.1088/1742-5468/ad0635},
 issn = {1742-5468}
}

@inproceedings{Montanari_2025_Dynamical,
 author = {Montanari, Andrea and Urbani, Pierfrancesco},
 title = {Dynamical Decoupling of Generalization and Overfitting in Large Two-Layer Networks},
 publisher = {Curran Associates, Inc.},
 year = {2025},
 volume = {38},
 pages = {26530--26608},
 url = {https://proceedings.neurips.cc/paper_files/paper/2025/file/261a154e3de61a0a8a1303c8a6c81ac7-Paper-Conference.pdf},
 booktitle = {Advances in Neural Information Processing Systems},
 editor = {Belgrave, D. and Zhang, C. and Lin, H. and Pascanu, R. and Koniusz, P. and Ghassemi, M. and Chen, N.}
}

@article{Gardner_1988_Optimal,
 author = {Gardner, E and Derrida, B},
 title = {Optimal storage properties of neural network models},
 journal = {Journal of Physics A: Mathematical and General},
 publisher = {IOP Publishing},
 year = {1988},
 month = {January},
 number = {1},
 volume = {21},
 pages = {271--284},
 url = {http://dx.doi.org/10.1088/0305-4470/21/1/031},
 doi = {10.1088/0305-4470/21/1/031},
 issn = {1361-6447}
}

@article{Gardner_1989_Three,
 author = {Gardner, E and Derrida, B},
 title = {Three unfinished works on the optimal storage capacity of networks},
 journal = {Journal of Physics A: Mathematical and General},
 publisher = {IOP Publishing},
 year = {1989},
 month = {June},
 number = {12},
 volume = {22},
 pages = {1983--1994},
 url = {http://dx.doi.org/10.1088/0305-4470/22/12/004},
 doi = {10.1088/0305-4470/22/12/004},
 issn = {1361-6447}
}

@article{Gardner_1988_The,
 author = {Gardner, E},
 title = {The space of interactions in neural network models},
 journal = {Journal of Physics A: Mathematical and General},
 publisher = {IOP Publishing},
 year = {1988},
 month = {January},
 number = {1},
 volume = {21},
 pages = {257--270},
 url = {http://dx.doi.org/10.1088/0305-4470/21/1/030},
 doi = {10.1088/0305-4470/21/1/030},
 issn = {1361-6447}
}

@article{Yamamura_2024_Geometric,
 author = {Yamamura, Atsushi and Mabuchi, Hideo and Ganguli, Surya},
 title = {Geometric Landscape Annealing as an Optimization Principle Underlying the Coherent {Ising} Machine},
 journal = {Physical Review X},
 publisher = {American Physical Society (APS)},
 year = {2024},
 month = {September},
 number = {3},
 volume = {14},
 pages = {031054},
 url = {http://dx.doi.org/10.1103/PhysRevX.14.031054},
 doi = {10.1103/physrevx.14.031054},
 issn = {2160-3308}
}

@article{Ghimenti_2026_Geometry,
 author = {Ghimenti, Federico and Sriram, Adithya and Yamamura, Atsushi and Mabuchi, Hideo and Ganguli, Surya},
 title = {Geometry and dynamics of annealed optimization in the coherent {Ising} machine with hidden and planted solutions},
 journal = {Physical Review E},
 publisher = {American Physical Society (APS)},
 year = {2026},
 month = {May},
 number = {5},
 volume = {113},
 pages = {054123},
 url = {http://dx.doi.org/10.1103/k73p-5k1w},
 doi = {10.1103/k73p-5k1w},
 issn = {2470-0053}
}

@article{Bray_1980_Metastable,
 author = {Bray, A J and Moore, M A},
 title = {Metastable states in spin glasses},
 journal = {Journal of Physics C: Solid State Physics},
 publisher = {IOP Publishing},
 year = {1980},
 month = {7},
 number = {19},
 volume = {13},
 pages = {L469--L476},
 url = {https://doi.org/10.1088%2F0022-3719%2F13%2F19%2F002},
 doi = {10.1088/0022-3719/13/19/002}
}

@article{Sherrington_1975_Solvable,
 author = {Sherrington, David and Kirkpatrick, Scott},
 title = {Solvable Model of a Spin-Glass},
 journal = {Physical Review Letters},
 publisher = {American Physical Society (APS)},
 year = {1975},
 month = {December},
 number = {26},
 volume = {35},
 pages = {1792--1796},
 url = {http://dx.doi.org/10.1103/PhysRevLett.35.1792},
 doi = {10.1103/physrevlett.35.1792},
 issn = {0031-9007}
}

@article{Thouless_1977_Solution,
 author = {Thouless, D. J. and Anderson, P. W. and Palmer, R. G.},
 title = {Solution of `Solvable model of a spin glass'},
 journal = {Philosophical Magazine},
 publisher = {Informa UK Limited},
 year = {1977},
 month = {March},
 number = {3},
 volume = {35},
 pages = {593--601},
 url = {http://dx.doi.org/10.1080/14786437708235992},
 doi = {10.1080/14786437708235992},
 issn = {0031-8086}
}

@article{Aspelmeier_2022_Free-energy,
 author = {Aspelmeier, T. and Moore, M. A.},
 title = {Free-energy barriers in the {Sherrington-Kirkpatrick} model},
 journal = {Physical Review E},
 publisher = {American Physical Society (APS)},
 year = {2022},
 month = {March},
 number = {3},
 volume = {105},
 pages = {034138},
 url = {http://dx.doi.org/10.1103/PhysRevE.105.034138},
 doi = {10.1103/physreve.105.034138},
 issn = {2470-0053}
}

@article{Cavagna_2003_On,
 author = {Cavagna, Andrea and Giardina, Irene and Parisi, Giorgio and Mézard, Marc},
 title = {On the formal equivalence of the {TAP} and thermodynamic methods in the {SK} model},
 journal = {Journal of Physics A: Mathematical and General},
 publisher = {IOP Publishing},
 year = {2003},
 month = {January},
 number = {5},
 volume = {36},
 pages = {1175--1194},
 url = {http://dx.doi.org/10.1088/0305-4470/36/5/301},
 doi = {10.1088/0305-4470/36/5/301},
 issn = {0305-4470}
}

@article{Annibale_2003_Supersymmetric,
 author = {Annibale, Alessia and Cavagna, Andrea and Giardina, Irene and Parisi, Giorgio},
 title = {Supersymmetric complexity in the {Sherrington-Kirkpatrick} model},
 journal = {Physical Review E},
 publisher = {American Physical Society (APS)},
 year = {2003},
 month = {12},
 number = {6},
 volume = {68},
 pages = {061103},
 url = {https://doi.org/10.1103%2Fphysreve.68.061103},
 doi = {10.1103/physreve.68.061103}
}

@article{Annibale_2003_The,
 author = {Annibale, Alessia and Cavagna, Andrea and Giardina, Irene and Parisi, Giorgio and Trevigne, Elisa},
 title = {The role of the {Becchi--Rouet--Stora--Tyutin} supersymmetry in the calculation of the complexity for the {Sherrington--Kirkpatrick} model},
 journal = {Journal of Physics A: Mathematical and General},
 publisher = {IOP Publishing},
 year = {2003},
 month = {10},
 number = {43},
 volume = {36},
 pages = {10937--10953},
 url = {https://doi.org/10.1088%2F0305-4470%2F36%2F43%2F018},
 doi = {10.1088/0305-4470/36/43/018}
}

@article{Annibale_2004_Coexistence,
 author = {Annibale, Alessia and Gualdi, Giulia and Cavagna, Andrea},
 title = {Coexistence of supersymmetric and supersymmetry-breaking states in spherical spin-glasses},
 journal = {Journal of Physics A: Mathematical and General},
 publisher = {IOP Publishing},
 year = {2004},
 month = {11},
 number = {47},
 volume = {37},
 pages = {11311--11320},
 url = {https://doi.org/10.1088%2F0305-4470%2F37%2F47%2F001},
 doi = {10.1088/0305-4470/37/47/001}
}

@article{Ros_2023_Quenched,
 author = {Ros, Valentina and Roy, Felix and Biroli, Giulio and Bunin, Guy},
 title = {Quenched complexity of equilibria for asymmetric generalized {Lotka–Volterra} equations},
 journal = {Journal of Physics A: Mathematical and Theoretical},
 publisher = {IOP Publishing},
 year = {2023},
 month = {7},
 number = {30},
 volume = {56},
 pages = {305003},
 url = {https://doi.org/10.1088%2F1751-8121%2Face00f},
 doi = {10.1088/1751-8121/ace00f}
}

@article{Kurchan_1992_Supersymmetry,
 author = {Kurchan, J.},
 title = {Supersymmetry in spin glass dynamics},
 journal = {Journal de Physique I},
 publisher = {EDP Sciences},
 year = {1992},
 month = {7},
 number = {7},
 volume = {2},
 pages = {1333--1352},
 url = {https://doi.org/10.1051%2Fjp1%3A1992214},
 doi = {10.1051/jp1:1992214}
}

@article{Pacco_2025_Triplets,
 author = {Pacco, Alessandro and Rosso, Alberto and Ros, Valentina},
 title = {Triplets of local minima in a high-dimensional random landscape: correlations, clustering, and memoryless activated jumps},
 journal = {Journal of Statistical Mechanics: Theory and Experiment},
 publisher = {IOP Publishing},
 year = {2025},
 month = {March},
 number = {3},
 volume = {2025},
 pages = {033302},
 url = {http://dx.doi.org/10.1088/1742-5468/adbe40},
 doi = {10.1088/1742-5468/adbe40},
 issn = {1742-5468}
}

@article{Lacroix-A-Chez-Toine_2022_Counting,
 author = {Lacroix-A-Chez-Toine, Bertrand and Fyodorov, Yan V},
 title = {Counting equilibria in a random non-gradient dynamics with heterogeneous relaxation rates},
 journal = {Journal of Physics A: Mathematical and Theoretical},
 publisher = {IOP Publishing},
 year = {2022},
 month = {March},
 number = {14},
 volume = {55},
 pages = {144001},
 url = {http://dx.doi.org/10.1088/1751-8121/ac564a},
 doi = {10.1088/1751-8121/ac564a},
 issn = {1751-8121}
}

@article{Fournier_2026_Nonreciprocal,
 author = {Fournier, Samantha J. and Pacco, Alessandro and Ros, Valentina and Urbani, Pierfrancesco},
 title = {Nonreciprocal interactions and high-dimensional chaos: Comparing dynamics and statistics of equilibria in a solvable class of models},
 journal = {Physical Review E},
 publisher = {American Physical Society (APS)},
 year = {2026},
 month = {March},
 volume = {113},
 pages = {044139},
 url = {http://dx.doi.org/10.1103/62sm-m7lw},
 doi = {10.1103/62sm-m7lw},
 issn = {2470-0053},
 issue = {4}
}

@article{Kac_1943_On,
 author = {Kac, M.},
 title = {On the average number of real roots of a random algebraic equation},
 journal = {Bulletin of the American Mathematical Society},
 publisher = {American Mathematical Society},
 year = {1943},
 month = {4},
 number = {4},
 volume = {49},
 pages = {314--320},
 url = {https://projecteuclid.org:443/euclid.bams/1183505112}
}

@article{Rice_1939_The,
 author = {Rice, S. O.},
 title = {The Distribution of the Maxima of a Random Curve},
 journal = {American Journal of Mathematics},
 publisher = {JSTOR},
 year = {1939},
 month = {4},
 number = {2},
 volume = {61},
 pages = {409},
 url = {https://doi.org/10.2307%2F2371510},
 doi = {10.2307/2371510}
}

@inproceedings{Maillard_2020_Landscape,
 author = {Maillard, Antoine and Ben Arous, Gérard and Biroli, Giulio},
 title = {Landscape Complexity for the Empirical Risk of Generalized Linear Models},
 publisher = {PMLR},
 year = {2020},
 month = {7},
 volume = {107},
 pages = {287--327},
 url = {https://proceedings.mlr.press/v107/maillard20a.html},
 booktitle = {Proceedings of The First Mathematical and Scientific Machine Learning Conference},
 editor = {Lu, Jianfeng and Ward, Rachel},
 series = {Proceedings of Machine Learning Research}
}

@article{Ros_2019_Complex,
 author = {Ros, Valentina and Ben Arous, Gérard and Biroli, Giulio and Cammarota, Chiara},
 title = {Complex Energy Landscapes in Spiked-Tensor and Simple Glassy Models: Ruggedness, Arrangements of Local Minima, and Phase Transitions},
 journal = {Physical Review X},
 publisher = {American Physical Society (APS)},
 year = {2019},
 month = {1},
 number = {1},
 volume = {9},
 pages = {011003},
 url = {https://doi.org/10.1103%2Fphysrevx.9.011003},
 doi = {10.1103/physrevx.9.011003}
}

@article{Crisanti_1992_The,
 author = {Crisanti, A. and Sommers, H.-J.},
 title = {The spherical $p$-spin interaction spin glass model: the statics},
 journal = {Zeitschrift für Physik B Condensed Matter},
 publisher = {Springer Science and Business Media LLC},
 year = {1992},
 month = {10},
 number = {3},
 volume = {87},
 pages = {341--354},
 url = {https://doi.org/10.1007%2Fbf01309287},
 doi = {10.1007/bf01309287}
}

@article{Kirkpatrick_1987_p-spin-interaction,
 author = {Kirkpatrick, T. R. and Thirumalai, D.},
 title = {$p$-spin-interaction spin-glass models: Connections with the structural glass problem},
 journal = {Physical Review B},
 publisher = {American Physical Society (APS)},
 year = {1987},
 month = {10},
 number = {10},
 volume = {36},
 pages = {5388--5397},
 url = {https://doi.org/10.1103%2Fphysrevb.36.5388},
 doi = {10.1103/physrevb.36.5388}
}

@article{Crisanti_2004_Spherical,
 author = {Crisanti, A. and Leuzzi, L.},
 title = {Spherical $2+p$ Spin-Glass Model: An Exactly Solvable Model for Glass to Spin-Glass Transition},
 journal = {Physical Review Letters},
 publisher = {American Physical Society (APS)},
 year = {2004},
 month = {11},
 number = {21},
 volume = {93},
 pages = {217203},
 url = {https://doi.org/10.1103%2Fphysrevlett.93.217203},
 doi = {10.1103/physrevlett.93.217203}
}

@article{Crisanti_2006_Spherical,
 author = {Crisanti, A. and Leuzzi, L.},
 title = {Spherical $2+p$ spin-glass model: An analytically solvable model with a glass-to-glass transition},
 journal = {Physical Review B},
 publisher = {American Physical Society (APS)},
 year = {2006},
 month = {1},
 number = {1},
 volume = {73},
 pages = {014412},
 url = {https://doi.org/10.1103%2Fphysrevb.73.014412},
 doi = {10.1103/physrevb.73.014412}
}

@article{Cavagna_1997_Structure,
 author = {Cavagna, Andrea and Giardina, Irene and Parisi, Giorgio},
 title = {Structure of metastable states in spin glasses by means of a three replica potential},
 journal = {Journal of Physics A: Mathematical and General},
 publisher = {IOP Publishing},
 year = {1997},
 month = {7},
 number = {13},
 volume = {30},
 pages = {4449--4466},
 url = {https://doi.org/10.1088%2F0305-4470%2F30%2F13%2F004},
 doi = {10.1088/0305-4470/30/13/004}
}

@article{Cavagna_1997_An,
 author = {Cavagna, Andrea and Giardina, Irene and Parisi, Giorgio},
 title = {An investigation of the hidden structure of states in a mean-field spin-glass model},
 journal = {Journal of Physics A: Mathematical and General},
 publisher = {IOP Publishing},
 year = {1997},
 month = {10},
 number = {20},
 volume = {30},
 pages = {7021--7038},
 url = {https://doi.org/10.1088%2F0305-4470%2F30%2F20%2F009},
 doi = {10.1088/0305-4470/30/20/009}
}

@article{Rosenblatt_1958_The,
 author = {Rosenblatt, F.},
 title = {The perceptron: A probabilistic model for information storage and organization in the brain.},
 journal = {Psychological Review},
 publisher = {American Psychological Association (APA)},
 year = {1958},
 number = {6},
 volume = {65},
 pages = {386--408},
 url = {http://dx.doi.org/10.1037/h0042519},
 doi = {10.1037/h0042519},
 issn = {0033-295X}
}

@unpublished{Stojnic_2013_Negative,
 author = {Stojnic, Mihailo},
 title = {Negative spherical perceptron},
 year = {2013},
 month = {jun},
 url = {http://arxiv.org/abs/1306.3980v1},
 eprint = {1306.3980v1},
 eprintclass = {math.PR},
 eprinttype = {arxiv}
}

@article{Annesi_2023_Star-shaped,
 author = {Annesi, Brandon Livio and Lauditi, Clarissa and Lucibello, Carlo and Malatesta, Enrico M. and Perugini, Gabriele and Pittorino, Fabrizio and Saglietti, Luca},
 title = {Star-Shaped Space of Solutions of the Spherical Negative Perceptron},
 journal = {Physical Review Letters},
 publisher = {American Physical Society (APS)},
 year = {2023},
 month = {11},
 number = {22},
 volume = {131},
 pages = {227301},
 url = {http://dx.doi.org/10.1103/PhysRevLett.131.227301},
 doi = {10.1103/physrevlett.131.227301},
 issn = {1079-7114}
}

@article{Annesi_2025_Exact,
 author = {Annesi, Brandon L. and Malatesta, Enrico M. and Zamponi, Francesco},
 title = {Exact full-{RSB} {SAT/UNSAT} transition in infinitely wide two-layer neural networks},
 journal = {SciPost Physics},
 publisher = {Stichting SciPost},
 year = {2025},
 month = {April},
 number = {4},
 volume = {18},
 pages = {118},
 url = {http://dx.doi.org/10.21468/SciPostPhys.18.4.118},
 doi = {10.21468/scipostphys.18.4.118},
 issn = {2542-4653}
}

@article{Kent-Dobias_2026_Structure,
 author = {Kent-Dobias, Jaron},
 title = {Structure of solutions to continuous constraint satisfaction problems through the statistics of wedged and inscribed spheres},
 journal = {Journal of Statistical Mechanics: Theory and Experiment},
 publisher = {IOP Publishing},
 year = {2026},
 month = {February},
 number = {2},
 volume = {2026},
 pages = {023301},
 url = {http://dx.doi.org/10.1088/1742-5468/ae4587},
 doi = {10.1088/1742-5468/ae4587},
 issn = {1742-5468}
}

@article{Rieger_1992_The,
 author = {Rieger, H.},
 title = {The number of solutions of the {Thouless-Anderson-Palmer} equations for $p$-spin-interaction spin glasses},
 journal = {Physical Review B},
 publisher = {American Physical Society (APS)},
 year = {1992},
 month = {12},
 number = {22},
 volume = {46},
 pages = {14655--14661},
 url = {https://doi.org/10.1103%2Fphysrevb.46.14655},
 doi = {10.1103/physrevb.46.14655}
}

@book{DeWitt_1992_Supermanifolds,
 author = {DeWitt, Bryce S.},
 title = {Supermanifolds},
 publisher = {Cambridge University Press},
 year = {1992},
 address = {Cambridge ; New York},
 edition = {2nd ed},
 isbn = {9780521413206 9780521423779},
 series = {Cambridge monographs on mathematical physics}
}

@article{Tsironis_2025_Landscape,
 author = {Tsironis, Theodoros G. and Moustakas, Aris L.},
 title = {Landscape complexity for the empirical risk of generalized linear models: Discrimination between structured data},
 journal = {Physical Review E},
 publisher = {American Physical Society (APS)},
 year = {2025},
 month = {December},
 number = {6},
 volume = {112},
 pages = {065307},
 url = {http://dx.doi.org/10.1103/3mbj-xkgk},
 doi = {10.1103/3mbj-xkgk},
 issn = {2470-0053}
}

\end{document}